\documentclass[acmsmall]{acmart}

\AtBeginDocument{%
  \providecommand\BibTeX{{%
    \normalfont B\kern-0.5em{\scshape i\kern-0.25em b}\kern-0.8em\TeX}}}

\setcopyright{acmcopyright}
\copyrightyear{2022}
\acmYear{2022}
\acmJournal{TRETS}
\acmVolume{1}
\acmNumber{1}
\acmArticle{1}
\acmMonth{1}

\usepackage{caption}
\usepackage{subcaption}
\usepackage{float}
\usepackage{amsmath}
\usepackage{amsfonts}
\usepackage{graphicx}
\usepackage{textcomp}
\usepackage{balance}
\usepackage{xcolor}
\usepackage{algorithm}
\usepackage[noend]{algpseudocode}
\usepackage{mathtools}
\usepackage{multirow}
\usepackage{calc}
\usepackage{pifont}
\usepackage{soul}
\usepackage{siunitx}
\newcommand{\cmark}{\ding{51}}%
\newcommand{\xmark}{\ding{55}}%

\newcommand{\toolName}{SASA}

\newcommand{\rev}[1]{\textcolor{black}{#1}}
\newcommand{\revsnd}[1]{\textcolor{black}{#1}}

\usepackage{enumitem}

\setenumerate[1]{itemsep=0.5pt,partopsep=0pt,parsep=\parskip, topsep=2pt, leftmargin=12pt}

\usepackage{listings}
\definecolor{dkgreen}{rgb}{0,0.6,0}
\definecolor{gray}{rgb}{0.5,0.5,0.5}
\definecolor{mauve}{rgb}{0.58,0,0.82}

\lstdefinestyle{interfaces}{
  float=tp,
  floatplacement=tbp,
}

\begin{document}
%\input{Sections/revision}
%%
%% The "title" command has an optional parameter,
%% allowing the author to define a "short title" to be used in page headers.
\title{{\toolName: A Scalable and Automatic Stencil Acceleration Framework for Optimized Hybrid Spatial and Temporal Parallelism on HBM-based FPGAs}}

%%
%% The "author" command and its associated commands are used to define
%% the authors and their affiliations.
%% Of note is the shared affiliation of the first two authors, and the
%% "authornote" and "authornotemark" commands
%% used to denote shared contribution to the research.
% Zhenman: Be careful to not include , at the end of streetaddress
\author{Xingyu Tian}
\email{xingyu_tian@sfu.ca}
\affiliation{%
  \institution{School of Engineering Science, Simon Fraser University}
  \streetaddress{8888 University Dr}
  \city{Burnaby}
  \state{BC}
  \country{Canada}
  \postcode{V5A1S6}
}

\author{Zhifan Ye}
\email{yezhifan@mail.ustc.edu.cn}
\authornote{The work was done when Zhifan was a research intern at Simon Fraser University.}
\affiliation{%
  \institution{School of the Gifted Young, University of Science and Technology of China}
  \streetaddress{No. 96 Jinzhai Road}
  \city{Hefei}
  \state{Anhui}
  \country{China}
  \postcode{230026}
}

\author{Alec Lu}
\email{alec_lu@sfu.ca}
\affiliation{%
  \institution{School of Engineering Science, Simon Fraser University}
  \streetaddress{8888 University Dr}
  \city{Burnaby}
  \state{BC}
  \country{Canada}
  \postcode{V5A1S6}
}

\author{Licheng Guo}
\email{lcguo@ucla.edu}
\author{Yuze Chi}
\email{chiyuze@cs.ucla.edu}
\affiliation{%
  \institution{Computer Science Department, University of California, Los Angeles}
  \streetaddress{404 Westwood Plaza}
  \city{Los Angeles}
  \state{California}
  \country{United States}
  \postcode{90095}
}

\author{Zhenman Fang}
\email{zhenman@sfu.ca}
\affiliation{%
  \institution{School of Engineering Science, Simon Fraser University}
  \streetaddress{8888 University Dr}
  \city{Burnaby}
  \state{BC}
  \country{Canada}
  \postcode{V5A1S6}
}

%%
%% By default, the full list of authors will be used in the page
%% headers. Often, this list is too long, and will overlap
%% other information printed in the page headers. This command allows
%% the author to define a more concise list
%% of authors' names for this purpose.

%%
%% The abstract is a short summary of the work to be presented in the
%% article.
% \begin{abstract}
%   A clear and well-documented \LaTeX\ document is presented as an
%   article formatted for publication by ACM in a conference proceedings
%   or journal publication. Based on the ``acmart'' document class, this
%   article presents and explains many of the common variations, as well
%   as many of the formatting elements an author may use in the
%   preparation of the documentation of their work.
% \end{abstract}

\begin{abstract}
%approximately 100 words

% Alec's Edit Start - 100 words
%Stencil computation is crucial in many demanding applications, such as image processing and cellular automata. Previous studies have explored various approaches to accelerate stencil kernels on FPGAs; however, it is non-trivial to design an automated acceleration framework.
%In this paper, we present~\toolName, a scalable stencil kernel acceleration framework that can automatically balance both spatial and temporal parallelisms while efficiently leveraging the logic resources and the high bandwidth memory (HBM) on modern FPGAs. Compared to the state-of-the-art designs, our framework can automatically generate the optimized designs and bring up to \al{XXX performance speedup} on the Xilinx Alveo U280 datacenter FPGA.
% Alec's Edit End

%FCCM: no more than 250 words

Stencil computation is one of the fundamental computing patterns in many application domains such as scientific computing and image processing. While there are promising studies that accelerate stencils on FPGAs, there lacks an automated acceleration framework to systematically explore both spatial and temporal parallelisms for iterative stencils that could be either computation-bound or memory-bound. In this paper, we present \toolName, a scalable and automatic stencil acceleration framework on modern HBM-based FPGAs. \toolName~takes the high-level stencil DSL and FPGA platform as inputs, automatically exploits the best spatial and temporal parallelism configuration based on our accurate analytical model, and generates the optimized FPGA design with the best parallelism configuration in \rev{TAPA} high-level synthesis C++ as well as its corresponding host code. Compared to state-of-the-art automatic stencil acceleration framework SODA that only exploits temporal parallelism, \toolName~achieves an average speedup of 
 \rev{3.74$\times$
and up to 
15.73$\times$}
speedup on the HBM-based Xilinx Alveo U280 FPGA board for a wide range of stencil kernels.

\end{abstract}

% \keywords{FPGAs; HLS; ...}

%%
%% The code below is generated by the tool at http://dl.acm.org/ccs.cfm.
%% Please copy and paste the code instead of the example below.
%%


%%
%% Keywords. The author(s) should pick words that accurately describe
%% the work being presented. Separate the keywords with commas.

%%
%% This command processes the author and affiliation and title
%% information and builds the first part of the formatted document.
% \input{Sections/revision}
\newpage

\maketitle

%\vspace{-0.1in}
\section{Introduction}\label{sec:intro}

Stencil computation is one of the most widely used computing patterns in many important application domains, such as scientific computing, image processing, and cellular automata ~\cite{CFD14, ImageProcessing13, CellularAutomata18, CellularAutomata21}. Due to its importance, stencil kernels have been well studied and accelerated on multicore CPUs, GPUs, and FPGAs~\cite{CPUGPUStencil10, GPUStencil12, CPUStencil20, GPUAutoStencil20, SODA18,SpatialTemporyBlocking18,HighOrderFPGAStencil18,MultiFPGAStencil19, StencilHBM20,StencilComputeReuse20,ShuoWangDAC17}. Among these approaches, FPGA acceleration~\cite{SODA18,SpatialTemporyBlocking18,HighOrderFPGAStencil18,MultiFPGAStencil19, StencilHBM20,StencilComputeReuse20, ShuoWangDAC17, TemporalICCAD17, TemporalTACO16, HDLMultiFPGA21, NERO20, HighLevelStencil21} is getting increasing attention due to its high performance, low power consumption, and high flexibility for customization. For example, in the automatic stencil acceleration framework SODA~\cite{SODA18}, it designed an optimized dataflow architecture with optimal data reuse and achieved up to 3.28$\times$ speedup on an FPGA over a 24-thread CPU.

%high computation demand, previous efforts have proposed optimized designs that leverage multicore CPUs and/or GPUs to accelerate stencil computations for various of application. With the image processing resolution increasing beyond 4K and the need for higher simulation accuracy, there is an ever-increasing demand of computing resource and memory bandwidth to further accelerate stencil computations.

%On the other hand, driven by the significant slowdown of CPU performance scaling and the high power consumption of GPUs, recently, researchers and the industry have explored high-performance energy efficient FPGA accelerator designs to accelerate stencil computations~\cite{SODA18,SpatialTemporyBlocking18,HighOrderFPGAStencil18,MultiFPGAStencil19, StencilHBM20,StencilComputeReuse20}. SODA~\cite{SODA18} proposes a automated framework which minimizes the on-chip reuse buffer size. It can reach 3.28x speedup over 24-thread CPU. Another work on~\cite{StencilHBM20} investigate the effectiveness of spatial and temporal blocking which can reach up to 760 and 375 GFLOP/s on an Intel Arria 10 FPGA for 2D and 3D stencil compuatation.
% \al{Give description of high level ideas of previous works, and give some good performance.} 
% SODA and HBM_paper1 good performance speedup numbers.

\begin{figure}[!htb]
    \vspace{-0.05in}
    \centering
    \begin{subfigure}[t]{0.48\textwidth}
        \centering
        \includegraphics[width=\textwidth]{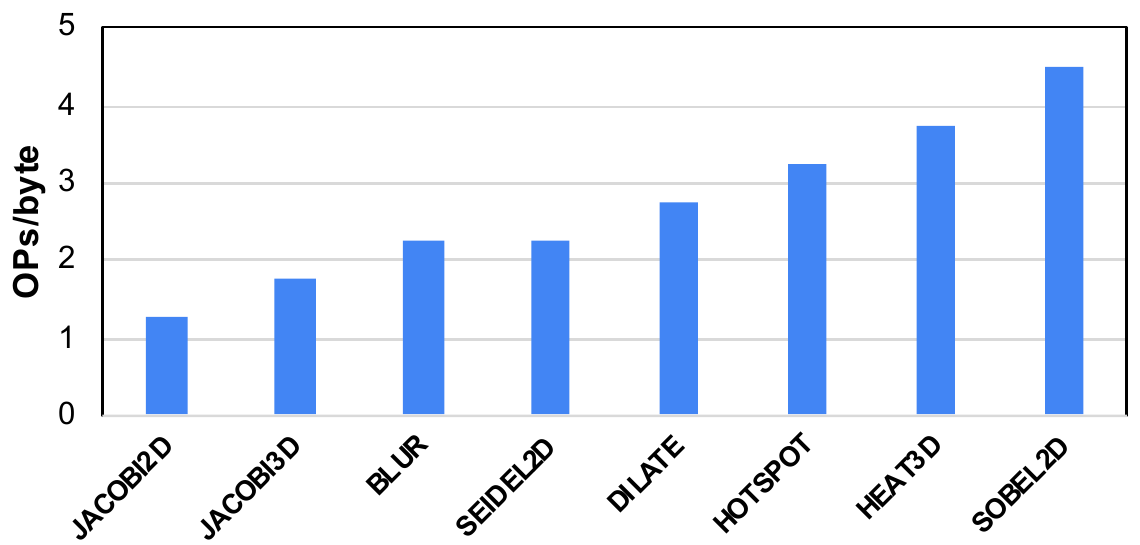}
        \caption{Computation intensity of different stencil kernels with the number of iterations = 1}
        \label{fig:intensityA}
    \end{subfigure}
    \hspace{+0.1in}
    \begin{subfigure}[t]{0.48\textwidth}
        \centering
        \includegraphics[width=\textwidth]{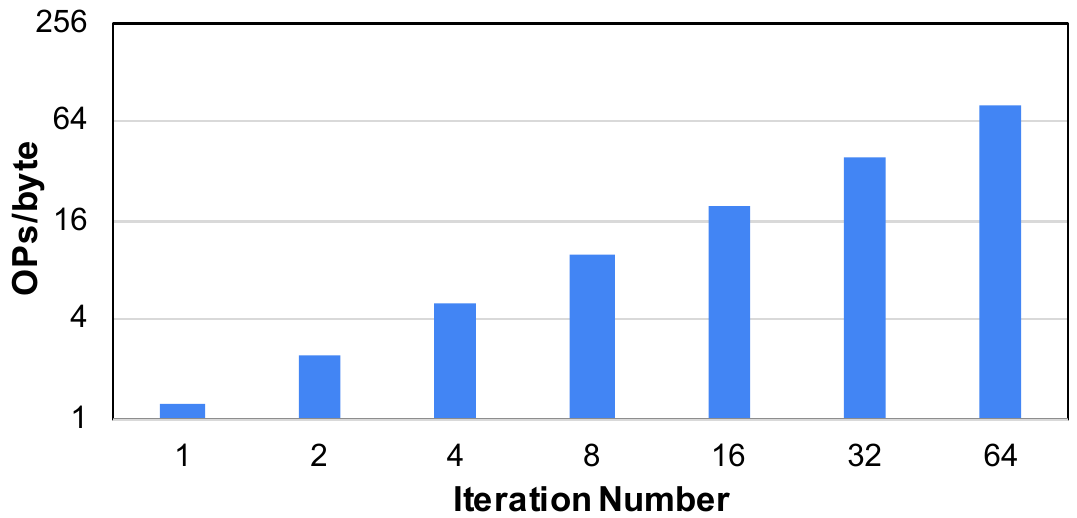}
        \caption{Computation intensity of the JACOBI2D stencil kernel with different numbers of iterations}
        \label{fig:intensityB}
    \end{subfigure}
    \vspace{-0.1in}
    \caption{Computation intensity (number of algorithmic operations per byte of off-chip memory access, i.e., OPs/byte) comparison for different stencil kernels and different numbers of iterations.} %A high computation intensity indicates computation-bound and a low one indicates memory-bound.}
    \vspace{-0.05in}
    \label{fig:intensity}
\end{figure}

However, one important factor that is often overlooked in prior studies is that, the stencil computation can be either computation-bound or memory-bound, depending on the stencil operations in the kernel and the number of iterations in the stencil kernel. To demonstrate this, we have measured the computation intensity, defined as the number of algorithmic operations divided by the number of bytes for off-chip memory accesses (OPs/byte), for a wide range of stencil kernels (detailed experimental setup in Section~\ref{sec:setup}). 
The measurement is based on the assumption of the optimal data reuse, i.e., every byte of data only needs to be accessed from off-chip memory once.
As shown in Figure~\ref{fig:intensityA}, the computation intensity varies between different stencil kernels, ranging from 1.25 to 4.5. Moreover, as shown in Figure~\ref{fig:intensityB}, the computation intensity increases linearly with the number of iterations. A high computation intensity indicates that the stencil kernel is computation-bound, while a low one indicates the stencil kernel is memory-bound.

Such observations suggest that different types of parallelism optimizations are needed for a stencil kernel to achieve the best performance on an FPGA. \rev{In general, there is a broad range of iteration numbers for stencil applications. For non-iterative stencil kernels, the iteration number is considered as one. Some iterative stencil kernels have a small iteration number (e.g., up to four in \cite{small-iterations}), while others could have a large iteration number~\cite{SODA18}.} 
Depending on the stencil kernel and its number of iterations, one may need to either 1) parallelize the computation along the iteration dimension (called \emph{temporal parallelism}), or 2) parallelize the memory access along the data dimension (called \emph{spatial parallelism}), or 3) combine both parallelisms together (called \emph{hybrid parallelism}). Moreover, it is nontrivial to program FPGAs to realize the best parallelism for the stencil kernels, especially for domain experts who program in high-level languages. Ideally, domain experts would only need to program in a simple stencil domain-specific language (DSL), and a tool would automatically compile the DSL to a highly efficient stencil accelerator on an FPGA and choose the best parallelism (and the optimal data reuse). Unfortunately, as summarized in Table~\ref{tbl:feature-compare} and Section~\ref{subsec:fpga-acc}, none of the prior studies satisfy all these requirements.

% \begin{figure}[h]
%     \centerline{
%     \includegraphics[width=0.5\textwidth]{Fig/Intensity.pdf}}
%     \vspace{-0.05in}
%     \caption{Computation Intensity Comparison}
%     \label{fig:intensity}
% \end{figure}

In this paper, we design and implement~\toolName, a DSL-based, \underline{s}calable, and \underline{a}utomatic  \underline{s}tencil  \underline{a}cceleration framework on modern HBM-based FPGAs. 
%To support both temporal and spatial parallelism and scale the accelerator design to different modern FPGAs, 
To support different types of parallelisms for stencil kernels,~\toolName~takes a scalable multi-PE (processing element) approach. For the single PE design, we take a similar design to SODA~\cite{SODA18}, which explores fine-grained data parallelism that matches the data streaming speed from a single memory bank and achieves the optimal data reuse within a single stencil iteration. Moreover, we further optimize the single PE design by utilizing the \emph{coalesced reuse buffers} (i.e., widened and shortened FIFOs) to reduce its resource utilization and reduce the high fan-out for better timing. For computation-bound stencil kernels, we scale the number of PEs to explore temporal parallelism between stencil iterations and use the same coalesced reuse buffer technique to dataflow between multiple PEs and exploit data reuse between stencil iterations. 
%We also use line buffers between stencil iterations to support the temporal parallelism and exploit data reuse as summarized in Section~\ref{temporal_parallelism}. 
For memory-bound stencil kernels, we scale the number of PEs to explore the coarse-grained spatial parallelism to better utilize the available off-chip bandwidth of multiple HBM banks on modern FPGAs. %as summarized in Section~\ref{spatial_parallelism}. 
Moreover, we support the combination of temporal and spatial parallelisms to get benefits from both sides. %A detailed presentation of these different types of parallelisms 

To bridge the programming gap, we support a simple stencil DSL so that end-users can easily develop their stencil algorithm and get hardware acceleration on FPGAs. Given the stencil DSL and FPGA platform as inputs, \toolName~can automatically generate a scalable stencil accelerator design in \rev{TAPA~\cite{tapa} high-level synthesis (HLS) C++ and its corresponding host code}. The generated stencil design automatically chooses the best temporal and spatial parallelism based on our accurate analytical model. \rev{Moreover, the open source TAPA framework~\cite{tapa,Autobridge21,guo2022rapidstream} invokes Vitis HLS to compile our generated TAPA HLS code in parallel, applies coase-grained floorplanning and pipelining to improve the timing closure.} Experimental results for a wide range of stencil kernels and iterations confirm the effectiveness of \toolName. Compared to state-of-the-art automatic stencil acceleration framework SODA~\cite{SODA18} that only explores temporal parallelism, \toolName~explores the optimized hybrid spatial and temporal parallelism and achieves an average speedup of \rev{3.74$\times$
and up to 
15.73$\times$} speedup on the HBM-based Xilinx Alveo U280 FPGA board. 

In summary, this paper makes the following contributions:

\begin{itemize}[leftmargin=*,topsep=0pt]
    \item Scalable stencil accelerator design optimizations, including coalesced reuse buffers to further improve the resource usage of the already well-optimized dataflow stencil design~\cite{SODA18} that exploits the temporal parallelism, and two design alternatives---redundant computation without communication vs. \rev{border streaming} for fast border communication---to exploit the spatial parallelism.

    \item An accurate analytical model, which has less than 5\% performance prediction error, to choose the best parallelism configuration for a given iterative stencil kernel, based on whether it is computation-bound or memory-bound.
    
    %We formulate performance models for accelerator designs that leverage spatial, temporal, and hybrid parallelism, based on key factors such as, number of iteration, number of stages and number of kernels. According to these models, we tune the hybrid spatial and temporal parallelism of our design depending on whether the stencil kernel is computation-bound or memory-bound.
    
    \item An end-to-end automation framework that takes the high-level stencil DSL and FPGA platform as inputs, and automatically generates the optimized FPGA design with the best parallelism configuration on that FPGA.
    
    %optimized Xilinx HLS code that can efficiently scale resource utilization and bandwidth on modern HBM-based FPGAs.
\end{itemize}

% \begin{itemize}[leftmargin=*]
%     \item \textbf{Scalable stencil accelerator design and exploration:}
%     We introduce novel design optimizations, coalesced fifo, based on a well-optimized streaming stencil design that exploits the temporal parallelism to further reduce the FPGA BRAM resource usage by 1.9\%-35.9\%. To further leverage the spatial parallelism in stencil designs, we explore two implementation approaches: redundant computation and kernel-to-kernel streaming.
    
%     \item \textbf{Accurate analytical model to choose the best parallelism:}
%     We formulate performance models for accelerator designs that leverage spatial, temporal, and hybrid parallelism, based on key factors such as, number of iteration, number of stages and number of kernels. According to these models, we tune the hybrid spatial and temporal parallelism of our design depending on whether the stencil kernel is computation-bound or memory-bound.
    
%     \item \textbf{End-to-end automation framework:}
%     We provide a high-level DSL that takes users' stencil kernel configurations, and automatically generates optimized Xilinx HLS code that can efficiently scale resource utilization and bandwidth on modern HBM-based FPGAs.
% \end{itemize}

%\zf{I will revisit this paragraph}

The rest of the paper is organized as follows. 
Section~\ref{sec:background} introduces the stencil computation pattern and presents the previous studies and their limitations in accelerating stencil kernels on FPGAs.
Section~\ref{sec:design} presents the scalable stencil accelerator architecture design of~\toolName, and its various types parallelism optimizations.
%for the efficient acceleration of stencil kernels with different numbers of iterations. 
Section~\ref{sec:automation} describes our end-to-end automation framework, including the high-level stencil DSL, the analytical performance models for our accelerator design, the code generator and automation tool flow. 
Section~\ref{sec:results} evaluates the performance of \toolName~on a comprehensive set of stencil benchmarks with different numbers of iterations, compares the performance of different parallelism optimizations, and demonstrates that~\toolName{} achieves an average speedup of \rev{3.74$\times$
and up to 
15.73$\times$} speedup over state-of-the-art automatic stencil acceleration framework SODA~\cite{SODA18}.
Finally, Section~\ref{sec:concl} concludes this paper and discusses the future work. 

%\vspace{-0.1in}
\section{Background and Related Work} \label{sec:background}

In this section, we first introduce the stencil computation pattern. Then, we discuss related literature on FPGA acceleration for stencil computations and their limitations. Finally, we describe the goal of our paper.

\subsection{Stencil Computation}

\begin{figure}[h]
    \vspace{-0.2in}
    \centerline{
    \includegraphics[width=0.8\textwidth]{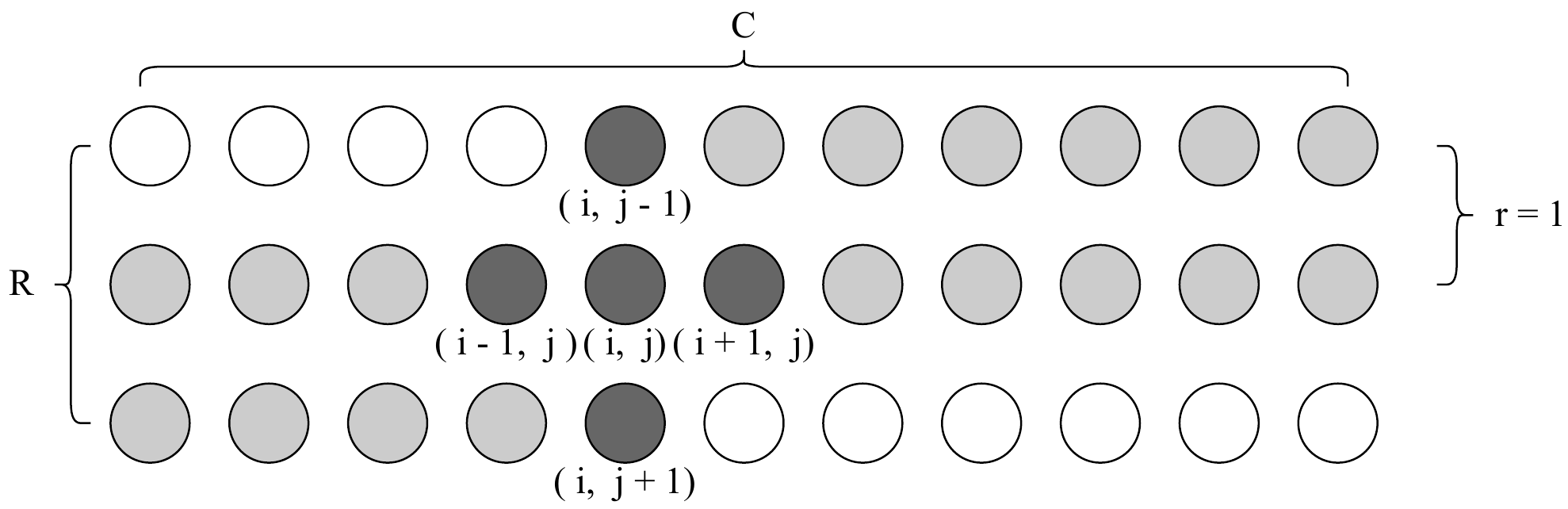}}
    \vspace{-0.05in}
    \caption{\rev{Stencil access pattern of JACOBI2D: R and C are the number of input rows and columns, r is the stencil radius size.}} %\zf{Subfig (a) needs an update, comments sent on Skype}\xy{Updated now}}
    \vspace{-0.1in}
    \label{fig:stencil_example}
\end{figure}

Stencil computation usually operates on a multidimensional array and updates each data cell using its neighbor cells in a fixed pattern. Listing~\ref{list:example} \rev{and Figure~\ref{fig:stencil_example}} show an example of the JACOBI2D stencil kernel, which is a 5-point, 2-dimensional stencil that computes and updates each data cell (i.e., $output[i][j]$) with the values from itself (i.e., $input[i][j]$) and its four neighbor cells (i.e., $input[i][j-1]$, $input[i-1][j]$, $input[i][j+1]$, and $input[i+1][j]$). \rev{Its stencil kernel radius size is 1, which is defined as the distance between the center cell and its furthest neighbor cell.} In practice, such a stencil kernel will be executed multiple iterations; in the next iteration, the output array from the previous iteration becomes the input, while the input array from the previous iteration becomes the output. As discussed earlier in the introduction, depending on the stencil operations and the number of iterations, the stencil kernel could be either computation-bound or memory-bound, and would require a different parallelism optimization to achieve the best performance.

%\vspace{0.2in}
\begin{lstlisting}[frame=single, numbers=none,  basicstyle=\small, caption = A 5-point stencil JACOBI2D kernel, label={list:example}, language=C++]
void jacobi2d (float input[R][C], float output[R][C]) {
  for (int i = 1; i < R - 1; ++i)
    for (int j = 1; j < C - 1; ++j)
      output[i][j] = (input[i][j-1] + input[i-1][j] + input[i][j] + input[i][j+1] + input[i+1][j]) / 5;
}
\end{lstlisting}

%\zf{I will revisit Sec 2.2 and 2.3 after Alec addresses my comments}

%\zf{Related work needs more work: 1) we need to cite more papers, classify them into these categories, and discuss their limitations. Take a look at the related work section of SODA and some other recent papers. I remember in recent couple of years, there are also some new papers on stencil acceleration. We should get this done for the FPL submission. 2) We need more quantitative comparison to those using hybrid parallelisms. Let's do this after the FPL submission.}

\subsection{FPGA Acceleration for Stencil Computation} \label{subsec:fpga-acc}

Previous research efforts in FPGA-based stencil computing generally target the following aspects: 1) improving the on-chip data reuse and reducing the FPGA on-chip memory usage for the stencil computation, 2) exploring temporal and/or spatial parallelisms in the stencil accelerator designs, including leveraging the HBM bandwidth on modern FPGAs to further extend the spatial parallelism, and 3) facilitating automatic stencil design generation. Some of these studies also need pre-processing on the host CPU to enable their optimizations. Next we discuss the prior studies considering these aspects. Some of the recent studies are also summarized in Table~\ref{tbl:feature-compare} for comparison and illustrating the novelty of our work.

% Table to show the feature coverage in previous work and ours
\begin{table}[!htb]
%\vspace{-0.05in}
\caption{Comparison of stencil acceleration frameworks}
%\vspace{-0.05in}
\label{tbl:feature-compare}
\centering
\begin{tabular}{|c|c|c|c|c|}
\hline
\multicolumn{1}{|l|}{} &
  \begin{tabular}[c]{@{}c@{}}Multi-PE\\ parallelism\end{tabular} &
  \begin{tabular}[c]{@{}c@{}}Pre-processing\\ free\end{tabular} &
  \begin{tabular}[c]{@{}c@{}}Automatic\\ optimization\end{tabular} &
  \begin{tabular}[c]{@{}c@{}}On-chip\\ data reuse\end{tabular} \\ \hline
~\cite{TemporalICCAD17, TemporalTACO16} & temporal & \cmark & \cmark & streaming\\ \hline
~\cite{SODA18}                          & temporal & \cmark & \cmark & streaming\\ \hline
~\cite{HDLMultiFPGA21}                  & temporal & \cmark & \xmark & streaming\\ \hline
~\cite{MultiFPGAStencil19}              & temporal & \cmark & \xmark & streaming\\ \hline
\rev{~\cite{SpatialTemporyBlocking18}}        & \rev{temporal}   & \xmark & \xmark & streaming\\ \hline
~\cite{ShuoWangDAC17}                   & hybrid   & \cmark & \xmark & buffering\\ \hline
~\cite{NERO20}                          & hybrid   & \cmark & \xmark & buffering\\ \hline
~\cite{StencilHBM20}                    & hybrid   & \xmark & \xmark & buffering\\ \hline
~\cite{HighLevelStencil21}              & hybrid   & \xmark & \xmark & streaming\\ \hline
Ours                                    & hybrid   & \cmark & \cmark & streaming\\ \hline
\end{tabular} 
\end{table}

In effort to improve on-chip data reuse and reduce the FPGA on-chip memory usage for the stencil computation, Chi et al. presented SODA~\cite{SODA18} and proposed the optimal streaming solution to minimize the reuse buffer size and leveraged microarchitectural design optimizations to also minimize the external memory access. Therefore, we also build our baseline design based on SODA. 
Other methods such as a sliding-window design is used in~\cite{SlidingWindow}, which requires maintaining only a small on-chip buffer to reduce the BRAM usage, but introduces additional off-chip communication overhead.
Further, another graph-theory based implementation proposed in~\cite{GraphBasedFPGA18} can derive the minimum memory partition factor for the on-chip memory banks, but such a design only supports a limited set of stencil kernels.

Various studies have explored accelerating the iterative stencil computation through different types of parallelisms.
Namely, acceleration through \textbf{temporal parallelism} has been well studied and explored in most previous iterative stencil kernel accelerator designs on FPGA~\cite{TemporalICCAD17, TemporalTACO16, ShuoWangDAC17, SpatialTemporyBlocking18, HighOrderFPGAStencil18, SODA18, MultiFPGAStencil19, HDLMultiFPGA21, StencilHBM20, NERO20, HighLevelStencil21, StencilSoC18}.
For example, in~\cite{TemporalICCAD17, TemporalTACO16}, Natale and Cattaneo et al. designed a dataflow architecture that executes multiple stencil computing iterations as multiple temporal stages. However, it lacks exploiting the fine-grained spatial parallelism during the processing of a single iteration stage.
In SODA~\cite{SODA18}, Chi et al. also presented a streaming-based accelerator design that exploits the temporal parallelism in the iterative stencil acceleration and proposed microarchitectural design optimizations to minimize the reuse buffer size and external memory access. 
For these designs~\cite{TemporalICCAD17, TemporalTACO16, SODA18}, there is no data pre-processing requirement; and they also provide an automatic design optimization framework to explore design space and optimize designs based on accurate analytical performance models. 
Further, Hasitha et al.~\cite{MultiFPGAStencil19} and Reggiani et al.~\cite{HDLMultiFPGA21} explored scaling the temporal parallelism of their accelerator design across multiple FPGAs without requiring any redundant computations. However, their design is only efficient for the computation-bound stencil kernel, where it can benefit from the great amount of data reuse.
A common limitation these temporally accelerated designs share is that they do not exploit any coarse-grained spatial parallelism, which would lead to under-optimized performance when the stencil kernel has a low number of iterations and is memory-bound.

To leverage both temporal and spatial parallelisms, previous studies have exploited \textbf{hybrid parallelism} in their designs~\cite{ShuoWangDAC17, HighOrderFPGAStencil18,  HighLevelStencil21, NERO20, StencilHBM20}.
Unlike the streaming-based designs, some of these designs require data buffering and typically have a significant on-chip memory utilization requirement. For example, the designs in~\cite{ShuoWangDAC17, NERO20} need to load multiple tiles of data on chip for its PEs to execute in parallel; another approach presented in~\cite{StencilHBM20} requires a single but rather larger on-chip buffer.
To further exploit the spatial parallelism by leveraging the HBM bandwidth on modern FPGAs, previous work also explored utilizing multiple memory banks of an HBM in their multi-PE design~\cite{HighLevelStencil21} and single-PE design~\cite{StencilHBM20}. However, their implementations require data pre-processing on the host CPU side to allow the parallel memory access and to exploit the efficient burst access from the FPGA off-chip memory. 
Another common limitation of these work is the lack of a design automation framework to facilitate the automatic design generation and mitigate the long design exploration process. 
% More recently, Licht et al. developed StencilFlow~\cite{StencilFlow21} to accelerate stencil computations with large stencil data in a non-iterative fashion on the distributed spatial computing systems. It maximizes the data reuse between different stencil programs through dataflow analysis, and supports datapath vectorization to increase spatial parallelism. Furthermore, they include a code generator to automatically generate the accelerator design. However, the design lacks in the automatic design optimization since the users would still need to manually choose the vectorization factor of the design to achieve the best parallelism and off-chip memory bandwidth utilization on a given platform. Also it does not consider leveraging the HBM bandwidth on the modern FPGAs as we do in this work.

\subsection{Goal of This Paper}
The goal of this paper is to develop an automatic stencil acceleration framework to incorporate the all the features as summarized in Table~\ref{tbl:feature-compare}. 
First, a streaming-based design similar to SODA~\cite{SODA18} is used in~\toolName{} for achieving the optimal data reuse; we also additionally include coalesced reuse buffers to further reduce resource utilization in our design.
Second,~\toolName{} leverages a hybrid multi-PE design architecture, exploiting both spatial and temporal parallelisms (presented in Section~\ref{sec:design}).
Furthermore,~\toolName{} utilizes the HBM memory on modern FPGAs to achieve higher throughput, yet unlike previous HBM-based stencil designs, our tool does not require any data pre-processing on the host CPU side.
And lastly, in terms of design automation,~\toolName{} enables end-users to define their stencil operations through a DSL and will automatically compile and optimize the accelerator designs with the best parallelism configuration based an analytical model (detailed in Section~\ref{sec:automation}).

%\vspace{-0.1in}
\section{Scalable Stencil Accelerator Design with Hybrid Temporal and Spatial Parallelism}\label{sec:design}

% Talk about the overall architecture (figure needed) of our scalable stencil accelerator design first. 

% 1. Single PE: briefly talk about the baseline PE design that is optimized for a single HBM bank, refer the audience to SODA for more details. Then focus on our own optimization: coalesced FIFO

% 2. Temporal parallelism optimization: briefly talk about the idea, refer the audience to SODA for more details.

% 3. Spatial parallelism optimization: talk more details about redundant computation approach and kernel-to-kernel streaming approach. 

% Leave code generator to Section 4.
%  
%\zf{Please briefly summarize the novel contributions in this paragraph, and why exploring each kind of parallelism. See my writing in intro}

In this section, we explore different types of parallelism optimizations in~\toolName, as summarized in Figure~\ref{fig:overall_structure_temporal}, ~\ref{fig:overall_structure_spatial} and ~\ref{fig:overall_structure_hybrid}. It is nontrivial to choose the best parallelism for a stencil kernel, since the most appropriate parallelism varies, depending on the stencil kernel and its number of iterations.
First, in Section~\ref{subsec:single-pe}, we present our single PE design, which is based on the streaming-based architecture proposed in SODA~\cite{SODA18}; and we describe our \emph{coalesced reuse buffer} design optimization for further reducing the resource utilization of the design. 
Next, in Section~\ref{subsec:temporal_parallelism}, we briefly describe our temporal parallelism design for computation-bound stencils with high number of iterations, which is similar to SODA but uses our \emph{coalesced reuse buffer} optimization.
After that, in Section~\ref{subsec:spatial_parallelism}, we introduce our spatial parallelism design for memory-bound stencil kernels and discuss two design alternatives to implement it.
Finally, in Section~\ref{subsec:hybrid_parallelism}, we explore hybrid parallelism, which exploits benefits of both spatial and temporal parallelisms in our multi-PE designs.

\subsection{Single PE Optimization} \label{subsec:single-pe}
\begin{figure}[!t]
    \centerline{
    \includegraphics[width=0.85\textwidth]{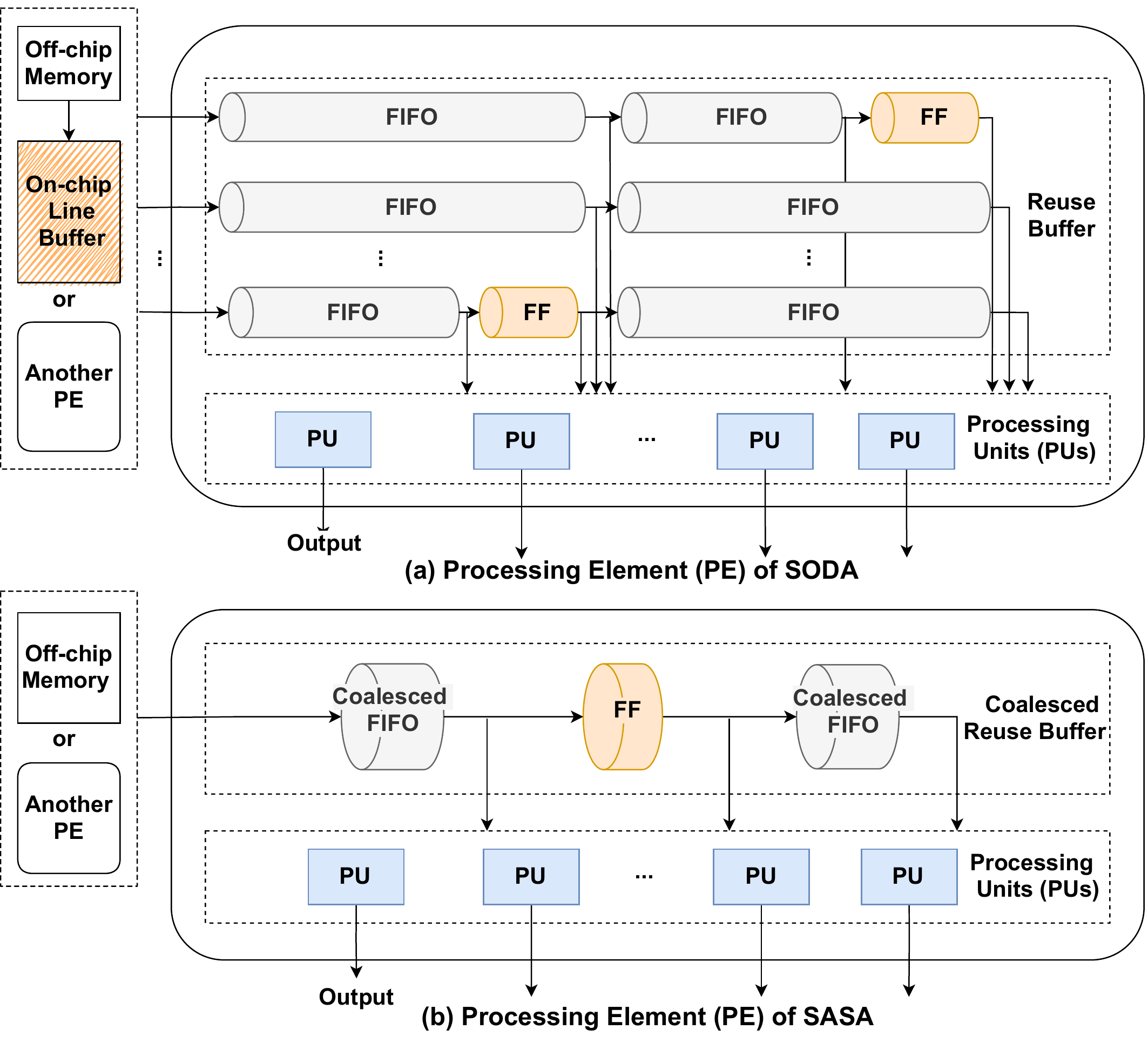}}
    %\vspace{-0.05in}
    \caption{Single processing element (PE) architecture based on SODA~\cite{SODA18}, with optimization of coalesced reuse buffers to reduce on-chip BRAM usage.} %\zf{Subfig (a) needs an update, comments sent on Skype}\xy{Updated now}}
    \label{fig:fifo}
\end{figure}

% Alec's edit start
% 1. We use SODA as our baseline design.
% 2. Why we use SODA as our baseline design? 
% 3. What does the baseline design look like?
% 4. Drawbacks of baseline (SODA) design
% 5. What are our design optimizations and why they improve the design?
% 6. How are these PEs used to realize temporal parallelism in our design?
As mentioned in Section II, our single PE design is based on SODA's design~\cite{SODA18}, since it achieves the optimal reuse buffer size and off-chip memory access requirement. 
Figure~\ref{fig:fifo} (a) shows an architecture overview of SODA's single PE design. For the input data that stream from the off-chip memory, SODA exploits the memory coalescing optimization to stream 512-bit wide data every cycle and stores it in an on-chip line buffer using BRAM. Then, it distributes this buffered data into reuse buffer channels composed of FIFOs and FFs, where each has a data width that matches the size of each stencil data cell (e.g., 32-bit for the \emph{float} data type). The data from these reuse buffer channels are then forwarded to the parallel processing units (PUs) for exploiting the fine-grained (spatial) parallelism and generate the output results. Each PU computes and updates for one data cell in the stencil, as shown in the JACOBI2D example in Listing~\ref{list:example}. The degree of fine-grained parallelism (i.e., the number of PUs) is set to saturate the off-chip bandwidth of a single memory bank and ensure the design executes in a dataflow fashion. For example, for a single HBM bank that uses 512-bit wide AXI interface and a data cell type of \emph{float}, the number of PUs can be derived as $512\ bits\ /\ (8 bits/byte)\ /\ sizeof(float)\ bytes = 16$. For the detailed microarchitecture design of the baseline PE, we refer the audience to the SODA paper~\cite{SODA18}.

However, based on our experiments, SODA's distributed reuse buffer channel implementation can be further optimized to reduce the on-chip BRAM usage. For such, in~\toolName, we propose an alternative implementation that removes the on-chip line buffer for storing the input data, and coalesces all the narrow distributed reuse buffers into a single wide coalesced reuse buffer as shown in Figure~\ref{fig:fifo} (b), to further reduce the BRAM usage. \rev{With memory coalescing, the input data it reads in from off-chip memory is typically 512-bit wide. Without coalesced FIFOs, it needs an on-chip line buffer to store such 512-bit wide data that is read in a AXI burst mode. And then it distributes such wide data from the on-chip line buffer onto multiple narrow (32-bit wide for floating data type) FIFOs, as shown in Figure~\ref{fig:fifo} (a). With coalesced FIFO, we stream in 512-bit data and write them into the 512-bit wide FIFO (i.e., coalesced FIFO) directly. Thus, we can get rid of the extra on-chip line buffer. Each cycle we also read one 512-bit data from each coalesced FIFO, divide it into mulitple 32-bit registers, and feed them to the parallel PUs.}
Another benefit of our optimization is that it helps reducing the number of fan-outs from SODA's line buffer design, and thus allows the design to achieve a higher operating frequency when further scaling out to multiple PEs.

\subsection{Temporal Parallelism Optimization} \label{subsec:temporal_parallelism}

\begin{figure}[!t]
    \centerline{
    \includegraphics[width=0.85\textwidth]{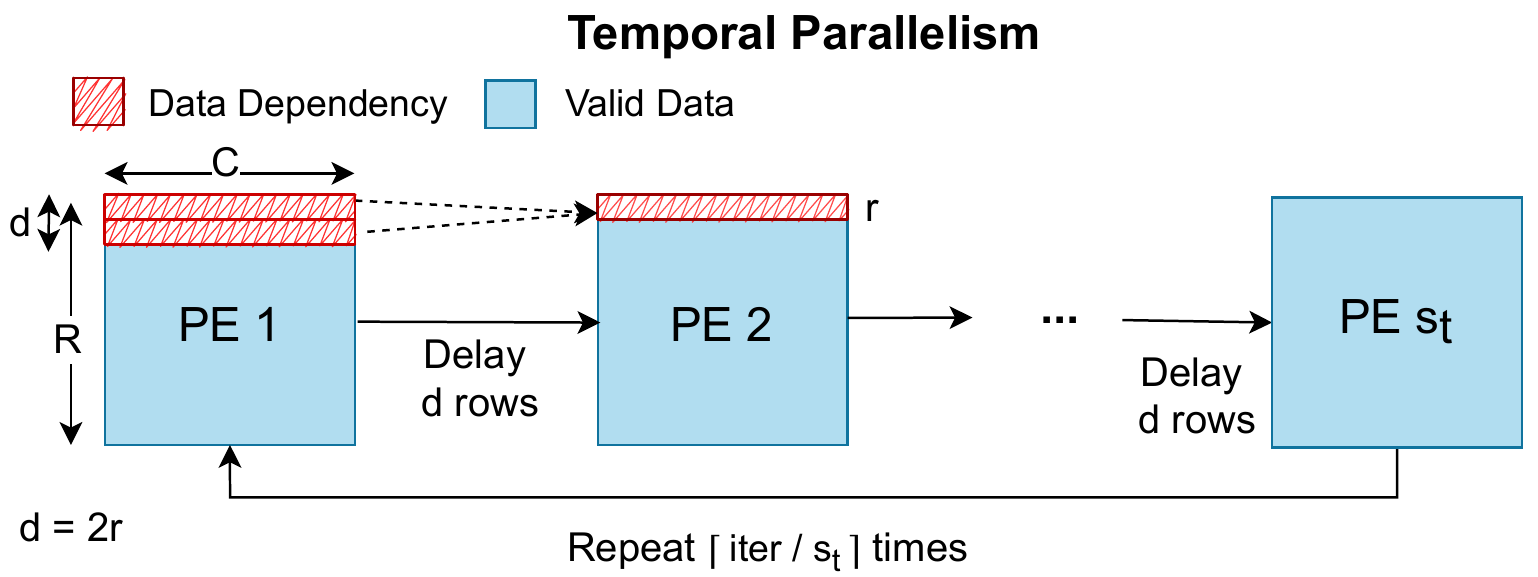}}
    %\vspace{-0.05in}
    \caption{\rev{Temporal parallelism among the stencil iterations. $d$ stands for delay between two temporal stages; $r$ stands for stencil radius size; $R$ and $C$ stand for the row and column size;  $s_t$ stands for the number of temporal stages; $iter$ stands for the number of iterations. These are summarized in Table~\ref{tab:configuration} as well.}} 
    \label{fig:overall_structure_temporal}
\end{figure}

In order to exploit the temporal parallelism, we instantiate multiple of our single PEs in a cascaded pipeline fashion as shown in Figure~\ref{fig:overall_structure_temporal}, which is similar to SODA's temporal parallelism design. The difference is that we use the coalesced reuse buffers to connect multiple PEs.
The input data is only read once from the off-chip memory and the output result is also written once back to the off-chip memory after processing N iterations of the stencil computation. Each PE handles one iteration of the stencil processing. 
For computation-bound stencil kernel designs with high number of iterations, it is more efficient to leverage the temporal parallelism since it allows for a higher level of data reuse across processing multiple consecutive stencil iterations in a pipelined fashion on the FPGA and does not require a huge amount of off-chip memory bandwidth. However, when the number of iterations becomes low, it will be hard to leverage the benefits of this temporal parallelism.

% Alec's edit end

\subsection{Spatial Parallelism Optimization} \label{subsec:spatial_parallelism}
% \begin{figure*}[!t]
%     \centering
%     \includegraphics[width=\textwidth]{Fig/AcceleratorStructure_new.pdf}
%     \caption{Five different types of parallelisms to scale the number of stencil PEs in \toolName: a) temporal parallelism among the stencil iterations; b) spatial parallelism with redundant computation; c) spatial parallelism with kernel-to-kernel streaming; d) hybrid parallelism with redundant computation; e) hybrid parallelism with kernel-to-kernel streaming. \zf{this figure needs an update, comments sent on Skype}}
%     %\vspace{-0.05in}
%     \label{fig:overall_structure}
% \end{figure*}

\begin{figure}[!t]
    \centerline{
    \includegraphics[width=0.75\textwidth]{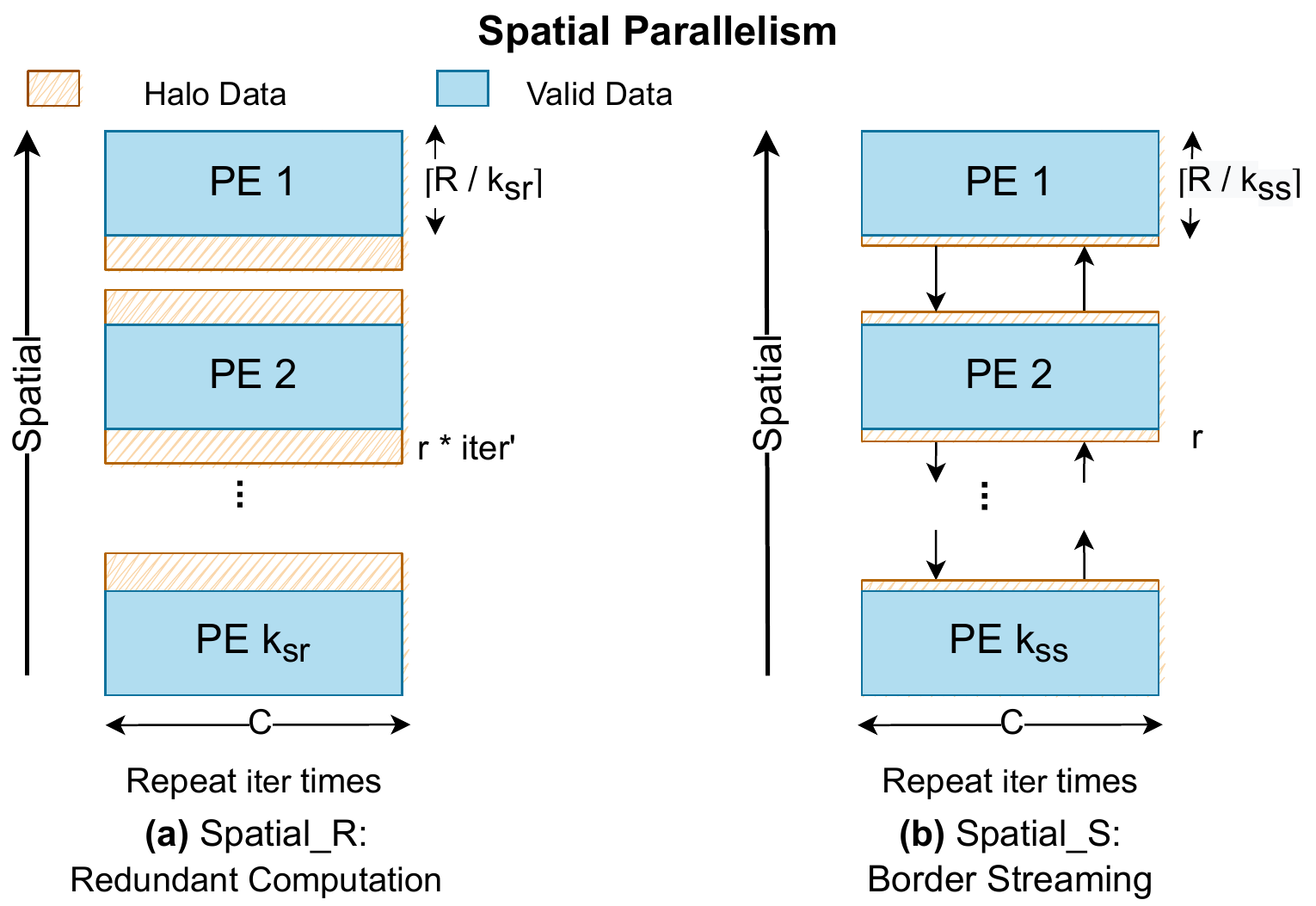}}
    %\vspace{-0.05in}
    \caption{\rev{Spatial parallelism with redundant computation and border streaming. $R$ and $C$ stands for the row and column size; $r$ stands for stencil radius size; $k_{sr}$ stands for the number of PEs in spatial parallelism with redundant computation; $k_{ss}$ stands for for the number of PEs spatial parallelism with border streaming. These are summarized in Table~\ref{tab:configuration} as well.}} 
    \label{fig:overall_structure_spatial}
\end{figure}

% Alec's edit start
% 1. What's the main idea behind spatial parallelism in stencil designs?
% 2. What are we implementation objectives?
% 3. What are the caveats behind leveraging the spatial parallelism?
% 4. What are our implementation approaches?
For stencil kernels with low computation intensity and low number of iterations, parallelizing the memory access along the spatial (data) dimension is more efficient, compared to leveraging the temporal parallelism.
To fully exploit the spatial parallelism during a single iteration of the stencil computation, first we need to evenly partition the input data and store them onto different HBM banks to allow for more parallel memory access. Note that here we are just simply partitioning the input data vertically by the rows, so there is no data pre-processing overhead. After that, we can then instantiate multiple spatially parallel PEs to distribute the workload for coarse-grained parallel computation and memory access.
Due to the dependency of the halo data, which is the boarder data between partitions, synchronization could be required at the end of each stencil iteration to maintain correctness of the output results. 

To address the halo synchronization issue when leveraging the spatial parallelism, Figure~\ref{fig:overall_structure_spatial} (a) and~\ref{fig:overall_structure_spatial} (b) present the two approaches in our stencil accelerator design:

\begin{enumerate}
    \item \textbf{Redundant Computation}: In order to reduce the memory transfer overhead during the data synchronization, one way is to avoid data synchronization. As shown in Figure~\ref{fig:overall_structure_spatial} (a), input data is partitioned into multiple tiles and each PE processes one tile. Each PE needs to read additional halo data from neighbouring tiles at the start, then performs the computation of all iterations without synchronization. The halo size is decided by the number of iterations and the stencil algorithm itself.
    
    \item \textbf{\rev{Border Streaming}}: Another way is to exchange the halo data between neighbor PEs via the border streaming technique. As shown in Figure~\ref{fig:overall_structure_spatial} (b), each PE only computes its own input tile without redundant computations for extra halo data. Instead, it exchanges the required halo data with the neighbouring PEs at the end of every iteration. To support efficient halo data exchange, it exchanges such data via on-chip streaming. %Here the kernel refers to the FPGA kernel that the host code can call upon and each kernel includes only one PE. 
    Compared to redundant computation, this approach uses slightly more on-chip resource (e.g., LUTs and FFs) to implement border streaming interfaces, but can reduce the computation overhead.
\end{enumerate}

\subsection{Hybrid Parallelism Optimization} \label{subsec:hybrid_parallelism}

\begin{figure}[!t]
    \centerline{
    \includegraphics[width=0.85\textwidth]{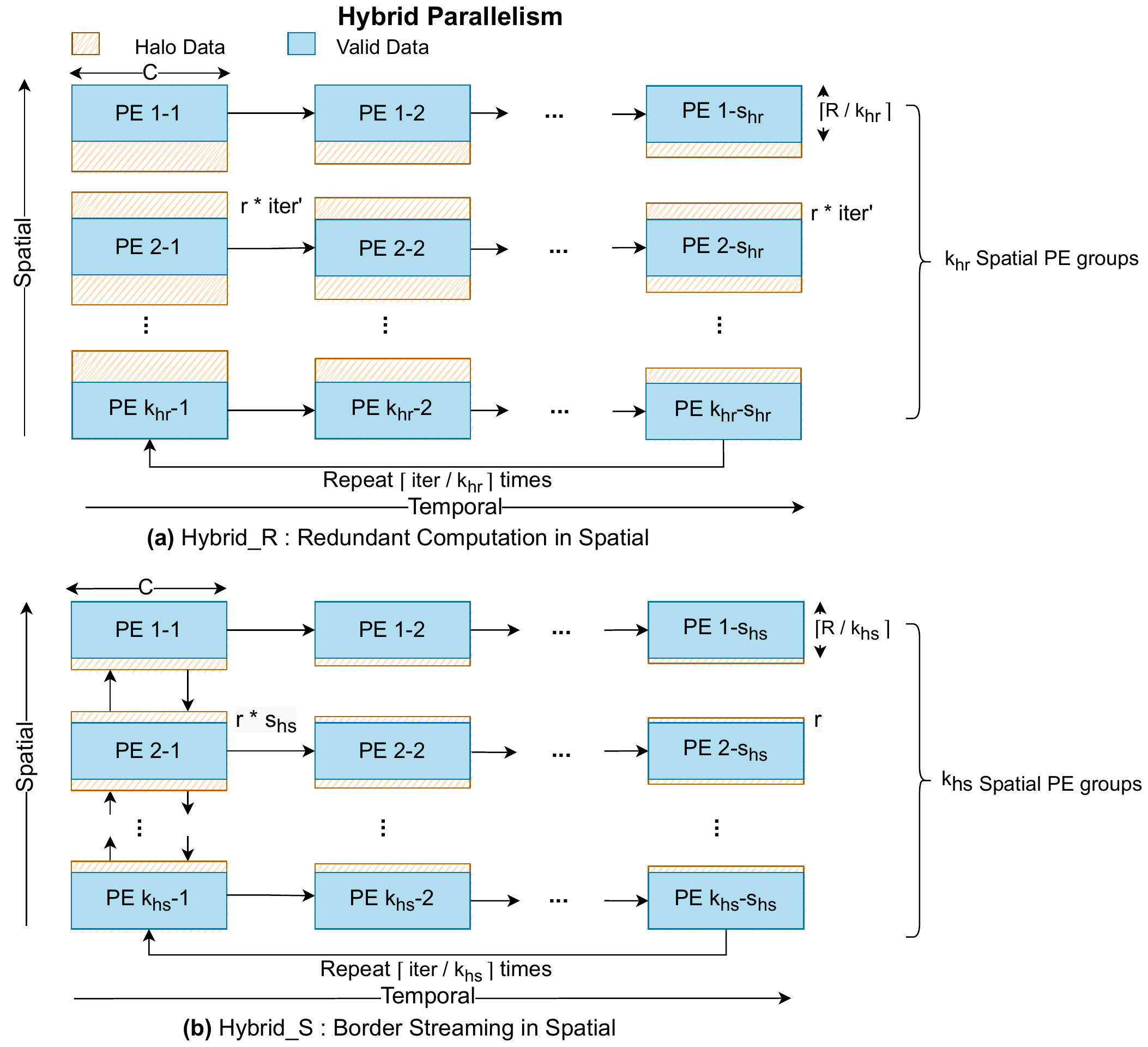}}
    %\vspace{-0.05in}
    \caption{\rev{Hybrid parallelism with redundant computation and border streaming. $R$ and $C$ stands for the row and column size; $r$ stands for stencil radius size; $k_{hr}$ and $s_{hr}$ stand for the degree of spatial parallelism and temporal parallelism respectively in hybrid parallelism with redundant computation; $k_{hs}$ and $s_{hs}$ \revsnd{stand for} the degree of spatial parallelism and temporal parallelism respectively in hybrid parallelism with border streaming. These are summarized in Table~\ref{tab:configuration} as well. %\zf{Please change "Kernels" to "Spatial PE Groups" in the figure"}
    }}% \zf{change upper case K to lower case k}} 
    \label{fig:overall_structure_hybrid}
\end{figure}

There are limitations for both temporal parallelism and spatial parallelism. As previously discussed in Section ~\ref{sec:intro}, the performance bottleneck varies with the algorithmic intensity and number of iterations of a stencil kernel. This is because when the stencil iteration number is high and computation intensity is high (i.e., computation-bound), the major performance improvement comes from the parallel processing across multiple consecutive stencil iterations in a pipelined fashion with high on-chip data reuse on the FPGA (i.e., temporal parallelism).
Conversely, for the memory-bound stencil kernels with a low iteration number, the performance gain comes from the parallel memory access within each stencil iteration, and the spatial parallelism can efficiently parallelize the computation across the data dimension.

In our hybrid parallelism approach, both temporal and spatial parallelisms are exploited to better support efficient acceleration of the arbitrary stencil operations. 
In terms of the design architecture, we integrate the temporal parallelism and explore the two variants of the spatial parallelism optimizations: 1) $Hybrid\_R$ (temporal with the $Spatial\_R$ parallelisms) and 2) $Hybrid\_S$ (temporal with the $Spatial\_S$ parallelisms), as shown in Figure~\ref{fig:overall_structure_hybrid} (a) and (b), respectively.

%\zf{Needs more description of how hybrid\_R and hybrid\_S work. I will revisit it after you two finish}

\begin{enumerate}
    \item $\textbf{Hybrid\_R}$: To integrate temporal parallelism with $Spatial\_R$ spatial parallelism, we instantiate multiple ($k_{hr}$) \rev{spatial PE groups} to concurrently process different partitions of the input data without any synchronization requirement as described in Section~\ref{subsec:spatial_parallelism}. Within each \rev{spatial PE group}, we apply temporal parallelism to concurrently process multiple stencil iterations using multiple ($s_{hr}$) PEs in a dataflow fashion, as shown in Figure~\ref{fig:overall_structure_hybrid} (a). To avoid halo synchronization, PEs in the earlier stages need to compute increasingly more halo data than those PEs in the later stages. In total, there are $k_{hr} \times s_{hr}$ PEs running concurrently, processing $s_{hr}$ stencil iterations at a time. The whole design has to be executed multiple rounds to finish the entire stencil iterations.

    % we first leverage the design architecture as $Spatial\_R$ where multiple spatial PEs are instantiated and assigned to process different partitions of the input data without any synchronization requirement as described in Section~\ref{subsec:spatial_parallelism}.
    % Then we apply temporal parallelism for each spatial PE by appending multiple temporal stages to it as shown in Figure~\ref{fig:overall_structure} (d). 
    % To realize the design, we implement $k_{hr}$ number of the spatial kernels, which contains $s_{hr}$ number of temporal PEs. Note the same number of temporal PEs is used in all spatial kernels.
    
    \item $\textbf{Hybrid\_S}$: To integrate temporal parallelism with $Spatial\_S$ spatial parallelism, we adopt a similar approach as the $Hybrid\_R$ design. We denote the number of \rev{spatial PE groups} as $k\_{hs}$ for spatial parallelism, and the number of temporal stages within each \rev{spatial PE group} as $s\_{hs}$.
    The main difference lays in the additional synchronization step to update the halo region data after processing each iteration/temporal stage.
    %as described in Section~\ref{subsubsec:perf_spatial}. 
    If we simply replicate the $Spatial\_S$ design by $s_{hs}$ number of temporal stages, the corresponding number of border streaming connections will also increase, and thus may cause overhead in the placement and routing, as well as design frequency degradation.
    As an optimization, in our design, only the spatial PEs in the first temporal stages have the border streaming connections between themselves and perform the halo data exchange. Instead of only exchanging one $halo$ rows of data, they exchange all required $halo \times s\_{hs}$ rows of data required by PEs for all following $s\_{hs}$ temporal stages. For the remaining temporal stages, no more synchronization is required. The whole design has to be executed multiple rounds to finish the entire stencil iterations, and only at the beginning of each round, there is halo data exchange required.
    
%    Before the computation starts, these PEs can first broadcast and receive all the halo region data required throughout all the iterations among themselves to eliminate the inter-iteration synchronization overhead.
\end{enumerate}

% In $Hybrid\_R$, each kernel consists of multiple PEs and processes one tile of input data with additional halo data. First PE of each kernel will read data from its corresponding off-chip memory bank streamingly and multiple PEs work in a dataflow pattern like temporal parallelism. As shown in Figure~\ref{fig:overall_structure}d, each row of PEs consist one kernel. Then middle kernel reads the most halo data since it requires halo data from both top and bottom sides. The required halo data size is decided by the number of remaining iterations and the stencil algorithm. Thus the required halo data will decrease after each PE. The last PE will write the output data back to off-chip memory bank. The kernel will be executed multiple times until all iterations are processed.

% In $Hybrid\_S$, multiple PEs within the same kernel will work in dataflow fashion as the one in $Hybrid\_R$ but the required halo size is decided by the number of stages. The first PE of each kernel will read the tile of input data from corresponding off-chip memory bank and share halo data with neighbouring kernel at the start of each kernel run. The halo data is only need to cover the computation of the remaining PEs in the same kernel and its size is decided by stencil algorithms and number of stages within single kernel. The last PE of each kernel will write the output data to off-chip memory bank. When the iteration number is sufficiently large(i.e., larger than the number of stages), the halo size of $Hybrid\_S$ will be smaller than the one of $Hybrid\_R$.
%\vspace{-0.1in}
\section{Automation Framework for \toolName} \label{sec:automation}
\label{Automation}
In this section, we discuss the automation perspectives of \toolName~and its fundamental components. First, we describe a domain-specific language (DSL), which is similar to the one used in SODA~\cite{SODA18}, for domain experts to easily define their stencil computation settings. Then, we introduce the analytical model for estimating the design performance under different types of parallelisms according to the design parameters. Finally, we present the entire work flow of our code generator that incorporates our analytical model to automatically determine the best parallelism configuration, compiles the DSL to the corresponding optimized stencil design in \rev{TAPA high-level synthesis (HLS) C++~\cite{tapa}, and generates the corresponding TAPA host code in C++~\cite{tapa}.}

\subsection{Stencil DSL}
\label{subsec:dsl}

\vspace{0.1in}
\begin{minipage}{\linewidth}
\begin{lstlisting}[frame=single, numbers=none, basicstyle=\small, caption = A 5-point stencil JACOBI2D kernel description in~\toolName{} DSL, label={list:dsl}]
(*\bfseries kernel*): JACOBI2D
(*\bfseries iteration*): 4
(*\bfseries input*) float: in_1(9720, 1024)
(*\bfseries output*) float: out_1(0,0) = ( in_1(0,1) + in_1(1,0) + in_1(0,0) + in_1(0,-1) + in_1(-1,0) ) / 5
\end{lstlisting}
\end{minipage}
% \vspace{0.1in}
\begin{minipage}{\linewidth}
\begin{lstlisting}[frame=single, numbers=none, basicstyle=\small, caption = \rev{A 9-point stencil HOTSPOT kernel description in~\toolName{} DSL with two inputs}, label={list:hotspot}]
(*\bfseries kernel*): HOTSPOT
(*\bfseries iteration*): 64
(*\bfseries input*) float: in_1(9720, 1024)
(*\bfseries input*) float: in_2(9720, 1024)
(*\bfseries output*) float: out_1(0,0) = 1.296 * ((in_2(-1,0) + in_2(1,0) - in_2(0,0) + in_2(0,0)) * 0.949219 + in_1(-1,0) + (in_2(0,-1) + in_2(0,1) - in_2(0,0) + in_2(0,0)) * 0.010535 + (80 - in_2(0,0)) * 0.00000514403)
\end{lstlisting}
\end{minipage}
% \vspace{0.1in}
\begin{minipage}{\linewidth}
\begin{lstlisting}[frame=single, numbers=none, basicstyle=\small, caption =\rev{A description of two combined stencil kernels in~\toolName{} DSL}, label={list:multiple}]
(*\bfseries kernel*): BLUR-JACOBI2D
(*\bfseries iteration*): 4
(*\bfseries input*) float: in(9720, 1024)
(*\bfseries local*) float: temp(0,0) = (in(-1,0) + in(-1,1) + in(-1,2) + in(0,0) + in(0,1) + in(0,2) + in(1,0) + in(1,1) + in(1,2)) / 9
(*\bfseries output*) float: out(0,0) = (temp(0,1) + temp(1,0) + temp(0,0) + temp(0,-1) + temp(-1,0)) / 5
\end{lstlisting}
\end{minipage}

To allow domain experts to easily define any arbitrary stencil computing workload at a high abstraction level,~\toolName{} provides a stencil domain-specific language (DSL) similar to SODA~\cite{SODA18}. 
Here We present a few stencil kernel samples using \toolName~DSL: Listing~\ref{list:dsl} shows the description of a 5-point, 2-dimensional JACOBI2D stencil kernel; \rev{Listing~\ref{list:hotspot} shows the description of a 9-point, 2 dimensional HOTSPOT stencil kernel handling two input data; and Listing~\ref{list:multiple} shows the description of two combined stencil loops.}
\begin{enumerate}
    \item The text following the \textbf{kernel} keyword specifies the name of the stencil kernel, which is also used as the name of the top-level function in HLS for the FPGA kernel. 
    \item The number after the \textbf{iteration} keyword specifies the number of iterations that the stencil kernel will be executed. 
    \item For \textbf{input} keyword, first, the data type of each stencil cell is specified, followed by the name and dimension of the input data array. 
    \item Similarly, for the \textbf{output} keyword, the data type is first specified. Then, users should specify the name of the output data and the formula to compute and update one output data cell. 
    \item \rev{Multiple inputs and outputs and multiple stencil loops are supported.}
    \item \rev{The \textbf{local} keyword is used to define the intermediate data between multiple stencil loops.}
    % \item Although not shown here, \toolName~also supports stencil processing on multiple inputs and multiple outputs, which is not supported in SODA~\cite{SODA18}.
\end{enumerate}

%\zf{I will revisit Sec 4.2 after Alec finishes}

\subsection{Analytical Performance Model} \label{subsec:analytical_model}
In this section, to guide the design automation, we build an analytical performance model for our accelerator framework with the temporal, spatial, and hybrid parallelism optimizations presented in Section~\ref{sec:design}. Such a comprehensive model is not present in previous studies. Table~\ref{tab:para} shows the description of the parameters used in our analytical model to determine the latency $L$ of designs with different parallelisms. Parameters such as the number of input rows and columns ($R$ and $C$), number of stencil iterations ($iter$), and stencil radius size ($r$) can be extracted from the input stencil DSL. Note that our analytical model only models a two-dimensional stencil. As will be presented in Section~\ref{subsec:automation}, our code generator will transform a multidimensional array specified in the stencil DSL into a two-dimensional array. Parameters such as the delay between two temporal stages ($d$) and size of halo region for one iteration ($halo$) can be directly derived from the input $r$, i.e., $d=halo=2\times r$. For other parameters such as the number of PUs inside each PE ($U$), \rev{the degree of spatial parallelism} ($k$ with different subscripts), and \rev{the degree of temporal parallelism} ($s$ with different subscripts), our automation tool flow (Section~\ref{subsec:automation}\revsnd{)} will automatically choose the best configurations.

\begin{table}[h]
  \caption{Description of analytical model parameters}
  %\vspace{-0.1in}
  \label{tab:para}
\begin{tabular}{|lc|l|}
\hline
\multicolumn{2}{|l|}{Parameter} & Definition \\ \hline
\multicolumn{1}{|l|}{Output} & $L$ & Overall execution latency \\ \hline
\multicolumn{1}{|l|}{\multirow{4}{*}{Input}} & $R$ & Number of input rows \\ \cline{2-3} 
\multicolumn{1}{|l|}{} & $C$ & Number of input columns \\ \cline{2-3} 
\multicolumn{1}{|l|}{} & $iter$ & Number of stencil iterations \\ \cline{2-3} 
\multicolumn{1}{|l|}{} & $r$ & Stencil radius size \\ \hline
\multicolumn{1}{|l|}{\multirow{2}{*}{Derived}} & $d$ & Delay between two temporal stages ($d=2\times r$) \\ \cline{2-3} 
\multicolumn{1}{|l|}{} & $halo$ & Size of halo region for one iteration ($halo=2\times r$) \\ \hline
\multicolumn{1}{|l|}{\multirow{3}{*}{\begin{tabular}[c]{@{}l@{}}\toolName\\automatic\\ exploration\end{tabular}}} & $U$ & Unroll factor along column dimension, i.e., number of PUs per PE \\ \cline{2-3} 
\multicolumn{1}{|l|}{} & $k$ & \rev{Degree of spatial parallelism} \\ \cline{2-3} 
\multicolumn{1}{|l|}{} & $s$ & \rev{Degree of temporal parallelism} \\ \hline
\multicolumn{1}{|l|}{\multirow{5}{*}{Subscript}} & subscript $t$ & Temporal parallelism \\ \cline{2-3} 
\multicolumn{1}{|l|}{} & subscript $sr$ & Spatial parallelism with redundant computation \\ \cline{2-3} 
\multicolumn{1}{|l|}{} & subscript $ss$ & Spatial parallelism with border streaming \\ \cline{2-3} 
\multicolumn{1}{|l|}{} & subscript $hr$ & Hybrid parallelism with redundant computation \\ \cline{2-3} 
\multicolumn{1}{|l|}{} & subscript $hs$ & Hybrid parallelism with border streaming \\ \hline
\end{tabular}
\end{table}

For each PE, the latency to execute one stencil iteration is determined by the dimension (i.e., number of rows ($R$) and columns ($C$)) of the input data and the number of PUs inside each PE (i.e., $U$) of the design, which is $\lceil \frac{R \times  C}{U}\rceil$. Next we describe our analytical models to compute the latency for each parallelism configuration for the multi-PE design.

% \begin{figure}[!h]
%     \centerline{
%     \includegraphics[width=0.6\textwidth]{Fig/Temporal.pdf}}
%     %\vspace{-0.05in}
%     \caption{}
%     \label{fig:temporal}
% \end{figure}

\textbf{\rev{Resource Bound and Memory Bound}}.
\rev{The maximum number of PEs that can be implemented is limited by both on-chip hardware resource and available off-chip memory banks (i.e., bandwidth). For the limitation of on-chip resource, we have:}

\begin{equation}\label{eq:pe_res}
    \#PE_{res} = \frac{\alpha \times total\_FPGA\_resource}{resource\_per\_PE}
\end{equation}
\rev{where $\alpha$ is the FPGA resource utilization constraint ratio and is initially set as 75\%, since typical design that uses more than 75\% of the FPGA resource becomes very difficult to pass the placement and routing.
}

\rev{For the constraint of off-chip memory banks, the number of spatial PEs is bounded as:}
\begin{equation}\label{eq:pe_bw}
    \#PE_{bw} = \frac{\#total\_off\_chip\_mem\_banks}{\#off\_chip\_mem\_banks\_per\_spatial\_PE}
\end{equation}
\rev{where $\#off\_chip\_mem\_banks\_per\_spatial\_PE$ is defined by the inputs and outputs number of stencil algorithm. Then, we can determine the maximum PE number based on FPGA platform specification and hardware resource constraints:}
\begin{equation}\label{eq:pe_max}
    Max\ \#PE = min \left(\#PE_{res},\ \#PE_{bw}\times s\right)
\end{equation}
\rev{where $s$ is the number of temporal stages (i.e., the degree of temporal parallelism) in each spatial PE group and these temporal stages do not require extra bandwidth.} 

\textbf{Temporal Parallelism}. 
As shown in Figure~\ref{fig:overall_structure_temporal}, we exploit the temporal parallelism in our design by cascading $s_t$ number of PEs to execute in a dataflow fashion. This also means $s_t$ iterations of the stencil computation is processed concurrently as input data get streamed through our design. To process an iterative stencil computation with $iter$ iterations, our design should be executed $\lceil iter/s_t \rceil$ times.
%As shown in Figure~\ref{fig:overall_structure_temporal}, 
to compute any single output in PE $i$ (except the first PE), data across two stencil radius size (2r) are required from the previous PE $i-1$. Thus, there is a delay $d=2r$ rows between any two neighbor stages. The last PE $s_t$ has to wait $d \times (s_t-1) \times  C$ cycles to start the execution.  Therefore, we determine overall latency of the temporal parallelism design as:
\begin{equation}\label{eq:temporal_model}
    L_t = \left\lceil \frac{(R + d \times (s_t-1)) \times  C}{U}\right\rceil \times \left\lceil \frac{iter}{s_t} \right\rceil,\ s_t \leq\#PE_{res}
\end{equation}
%Note that there is only one \rev{degree of spatial parallelism} % FPGA kernel 
\rev{In this case and $s_t$ is limited by $\#PE_{res}$, i.e., the available computing resource.}

\textbf{Spatial Parallelism}. 
In the context of spatial parallelism as shown in Figure~\ref{fig:overall_structure_spatial} (a) and (b), we have two different design implementations: redundant computation ($Spatial\_R$) and border streaming ($Spatial\_S$).
In both of these implementations, the computation of a single stencil iteration is distributed across multiple parallel spatial PEs. Every single PE processes $\lceil R/k_{sr} \rceil$ or $\lceil R/k_{ss} \rceil$ rows of the input data, plus some halo region rows. The design has to be executed $iter$ times. Note in the spatial parallelism, we put one PE inside each FPGA spatial PE grouop and the number of FPGA spatial PE groups equals to the number of PEs.
% data determines the iteration latency of the overall design.

For $Spatial\_R$, the latency can determined as
\begin{equation}\label{eq:spatial_R_model}
    L_{sr} = \left \lceil\frac{(\lceil\frac{R}{k_{sr}}\rceil + halo \times iter') \times  C}{U} \right \rceil \times iter,\ k_{sr} \leq Max\ \#PE
\end{equation}
where $halo \times iter'$ represents the halo data size gradually decreases over the processing iteration (i.e., $iter'$) as explained in Section~\ref{subsec:spatial_parallelism}. On average, $iter' = iter/2$.

As for $Spatial\_S$, we calculate its latency in Equation~\ref{eq:spatial_S_model}, since all the PEs synchronize with their neighboring PEs for a fixed number of $halo$ rows after each stencil iteration. 
\begin{equation}\label{eq:spatial_S_model}
    L_{ss} = \left \lceil \frac{(\lceil\frac{R}{k_{ss}}\rceil + halo) \times C}{U} \right \rceil \times iter,\ k_{ss} \leq Max\ \#PE
\end{equation}

\rev{For both spatial parallelisms, they are limited by both the computing resource and memory bandwidth, i.e., $Max\ \#PE$.}
% Alec: maybe move this to Sec 3.3 and/or 3.4
% The major difference between these two implementations lays in the handling approach for the halo region data. 
% In $Spatial\_R$, the halo region for each PE is defined based on both the iteration number and the stencil kernel; and its size gradually decreases over the processing time or iteration. This is because the PEs need to redundantly compute additional halo region data necessary for the subsequent iterations. We choose this approach to avoid data coalition across PEs after processing each iteration. To achieve a high HBM bandwidth, we partition our input data and store them onto different HBM banks, and so each PE can access its own portion of the input data independently from different HBM banks to extend the spatial parallelism. On the other hand, in the $Spatial\_S$ approach, no redundant halo data computation is required. Instead, adjacent PEs need to exchange the halo region data after processing each stencil iteration.

\textbf{Hybrid Parallelism}. 
As shown in Figure~\ref{fig:overall_structure_hybrid} (a) and (b), when combining spatial and temporal parallelisms, there are $k_{hr}$ (or $k_{hs}$) FPGA spatial PE groups running concurrently, each FPGA spatial PE group processing $\lceil R/k_{hr} \rceil$ (or $\lceil R/k_{hs} \rceil$) rows of input data. Within each FPGA spatial PE group, there are $s_{hr}$ (or $s_{hs}$)  temporal stages running concurrently in a dataflow fashion, processing $s_{hr}$ (or $s_{hs}$) stencil iterations at a time. Therefore, our design with hybrid parallelism has to execute $\lceil iter/s_{hr} \rceil$ (or $\lceil iter/s_{hs} \rceil$) times. In total, there are $k_{hr} \times s_{hr}$ (or $k_{hs} \times s_{hs}$) PEs running concurrently in the design.

For $Hybrid\_R$, which is the combination of temporal parallelism and spatial parallelism with redundant computation, for one round of execution, all PEs within each FPGA spatial PE group $i$ complete exactly at the same time. The reason is that the prior PE $i, j-1$ needs to redundantly compute $halo$ more rows of data than the next PE $i, j$, while the next PE $i, j$ needs to wait $d$ rows of data from the prior PE $i, j-1$, where $halo = d = 2r$. Therefore, we derive Equation~\ref{eq:hybrid_r} for the latency of $Hybrid\_R$:
%the temporal parallelism with $Spatial\_R$  based Equation~\ref{eq:temporal_model} and ~\ref{eq:spatial_R_model} to compute the latency. The number of spatial kernels and temporal stages are reflected as $k_{hr}$ and $s_{hr}$, respectively.
% With combined spatial and temporal parallelism, the latency will be influence by both halo region and number of stages. Compared with pure spatial parallelism, hybrid parallelism with redundant computation has the same size halo region. Besides, each stage can process less data than previous one since halo region decrease after each stage and this can offset the delay between stages. Thus, all stages can complete at the same time and the latency $L_{hs}$ is,
\begin{equation} \label{eq:hybrid_r}
    L_{hr} = \lceil \frac{(\lceil\frac{R}{k_{hr}}\rceil + halo \times iter') \times C}{U} \rceil \times \lceil \frac{iter}{s_{hr}} \rceil,\ k_{hr} \leq PE_{bw},\ k_{hr} \times s_{hr} \leq\ Max\ \#PE
\end{equation}
where the first term represents the latency to execute one round of our hybrid parallelism design, and $halo \times iter'$ represents the halo data size gradually decreases over the processing rounds. On average, $iter' = iter/2$. \rev{In this case, the degree of spatial parallelism is limited by the memory bandwidth, and the total degree of parallelism is limited by both the computing resource and memory bandwidth.}

For $Hybrid\_S$, which is the combination of temporal parallelism and spatial parallelism with border computation, similarly, for one round of execution, all PEs within each FPGA spatial PE group $i$ complete exactly at the same time. However, in this design, these PEs only need an extra latency for $halo$ more rows within each round. Between different rounds, PEs in the first temporal stage exchange $halo$ data with each other using border streaming. Therefore, we derive its latency as below:
%which is the combina the hybrid parallelism combining temporal parallelism with $Spatial\_S$ (), we compute its latency with Equation~\ref{eq:hybrid_s} based on Equation~\ref{eq:temporal_model} and ~\ref{eq:spatial_S_model}. Since there are $s_{hs}$ temporal stage in the design, the same number of halo region row data need to be transferred during the synchronization step between temporal stages. 
\begin{equation}  \label{eq:hybrid_s}
    L_{hs} = \lceil\frac{ (\lceil\frac{R}{k_{hs}}\rceil + halo \times s_{hs}) \times C}{U} \rceil \times \lceil \frac{iter}{s_{hs}} \rceil,\ k_{hs} \leq PE_{bw},\ k_{hs} \times s_{hs} \leq\ Max\ \#PE
\end{equation}
\rev{In this case, the degree of spatial parallelism is limited by the memory bandwidth, and the total degree of parallelism is limited by both the computing resource and memory bandwidth.}

\textbf{Automatic Parallelism Optimization}.
In order to automatically determine the optimal parallelism, the automation tool would need to find the parallelism with the minimum latency as:
\begin{equation} \label{eq:optimal}
    L_{optimal} = \min(L_t, L_{sr}, L_{ss}, L_{hs}, L_{hs})
\end{equation}

Examining our analytical performance model at a high level and assuming maximum PE number is the same across different parallelisms and $R$, $C$, $U$ and $r$ are fixed during running time, we summarize the following observations:

\begin{enumerate}
    \item In spatial parallelism, $L_{sr}$ grows with $iter$ slightly more than linearly, while $L_{ss}$ grows with $iter$ exactly linearly. It shows that both solutions provide proximate performance when the iteration number is relatively small. But as the iteration number increases, border streaming will outperform redundant computation. For the two hybrid parallelism alternatives, $L_{hr}$ and $L_{hs}$ have the same relation as the one between $L_{sr}$ and $L_{ss}$.
    \item Comparing spatial parallelism with temporal parallelism, when $iter$ is large enough and $iter$ is divisible by $s_t$, temporal parallelism could achieve a similar performance to spatial parallelism, as the value of $s_t$ would be set to the same as $k_{sr}$ and $k_{ss}$. In addition, temporal parallelism requires much less amount of off-chip bandwidth. Howerver, when $iter$ is small enough, the largest value of $s_t$ is the same as $iter$, while $k_{sr}$ and $k_{ss}$ can be much larger than $iter$ by exploiting off-chip memory bandwidth (especially on HBM-based FPGAs). This will bring significant performance degradation for temporal parallelism. In this case, hybrid parallelism can further improve the performance with less bandwidth requirement. Finally, when $iter$ is not divisible by $s_t$ (or $s_{hr}$, or $s_{hs}$), $L_t$ (or $L_{hr}$, or $L_{hs}$) will also suffer some overhead to process the whole input data with some PEs idle in the last round. 
\end{enumerate}

% \begin{figure}[!h]
%     \centerline{
%     \includegraphics[width=0.9\textwidth]{Fig/tapa_flow.pdf}}
%     %\vspace{-0.05in}
%     \caption{Overall Framwork of TAPA}
%     \label{fig:tapa}
% \end{figure}

\subsection{Code Generator and Automation Tool Flow} \label{subsec:automation}

\begin{figure}[!t]
    \centerline{
    \includegraphics[width=0.9\textwidth]{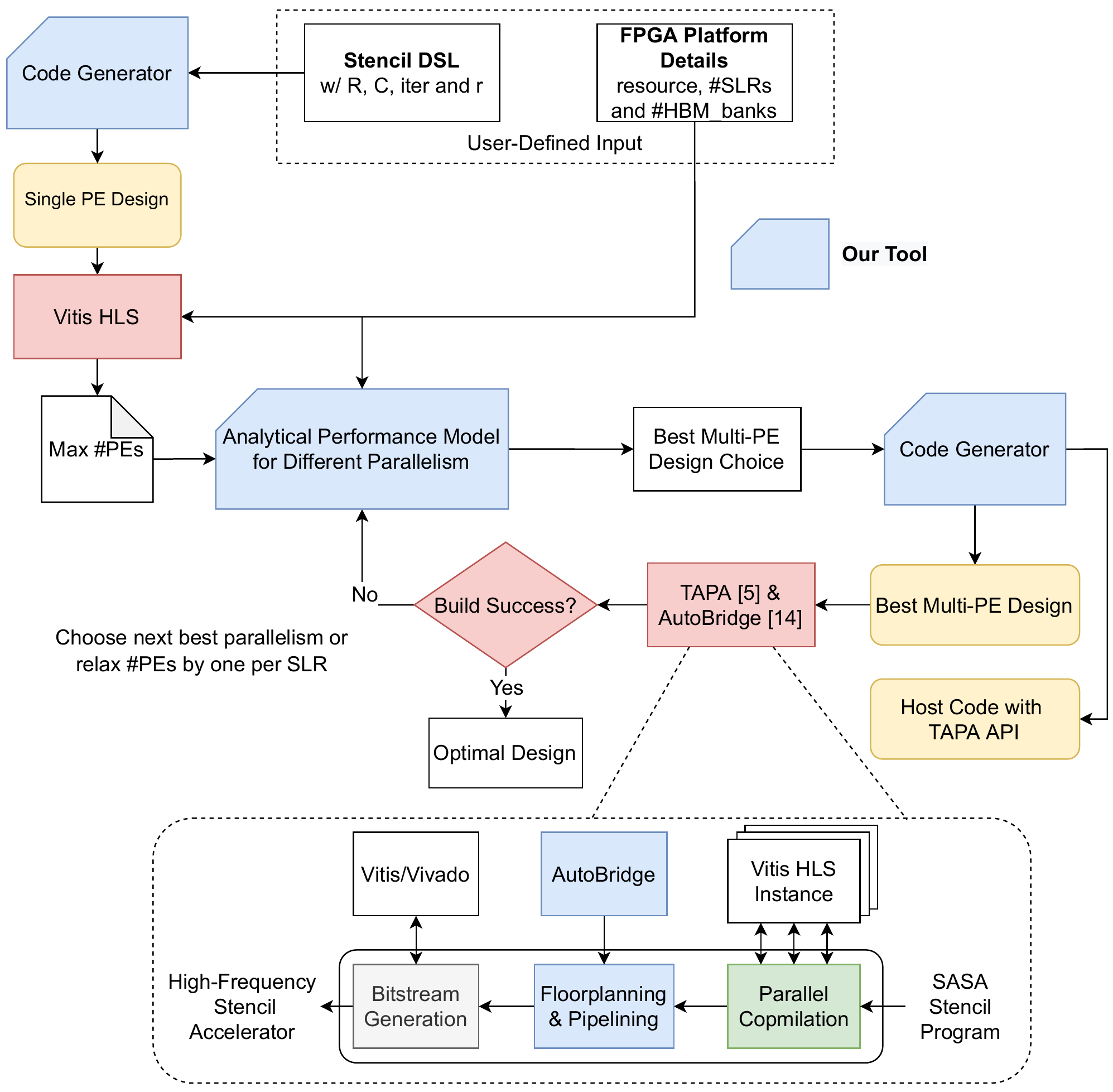}}
    %\vspace{-0.05in}
    \caption{Overall automation tool flow in \toolName}
    \label{fig:flowchart}
\end{figure}

Figure~\ref{fig:flowchart} shows an overview of the automation flow for \toolName. It takes a stencil DSL and FPGA platform information as input, and automatically generates optimized FPGA accelerator design with the best parallelism optimization as the output. \rev{To address the timing closure issue (and conduct a fairer comparison between different parallelism implementations), we have integrated the open source TAPA/AutoBridge framework [5,14] into our SASA framework to build our generated multi-PE design. TAPA/AutoBridge is a high-performance fast-compiling HLS framework that is fully compatible with the Xilinx Vitis/Vivado workflow. It takes in task-parallel program in Vitis HLS syntax with additional TAPA APIs. It has three major advantages. First, it supports easier programming of task-parallel dataflow programs in C++, without the need of the more complex OpenCL approach (using multiple OpenCL kernels) to support task parallelism. Our SASA code generator automatically converts the stencil DSL to the TAPA HLS and host code. Second, it replaces the resource-inefficient AXI interface with a lightweight streaming interface to access off-chip memory. The standard AXI interface always buffers data in BRAM and consumes a significant amount of resources on the bottom die of the HBM-based U280 FPGA (with multiple AXI interfaces), which often causes place-and-route congestion and timing violation timing congestion on the bottom die. With the lightweight streaming interfaces, it saves resources for actual computations and reduces place-and-route congestion and timing violation timing congestion on the bottom die. Third, it automatically applies coarse-grained floorplanning and pipelining optimizations to improve the timing closure for dataflow programs, and can often greatly improve the design build success rate and the final design frequency. With this integration, we are able to generate high-frequency stencil accelerators.}

The detailed steps inside the \toolName~automation flow are described as below.

%For a given FPGA platform and the stencil parameters (input dimension, iteration number, and stencil operation) provided by the end-users.
\begin{enumerate}
    % Pythong textx-> ast -> model ->
    \item Our code generator first parses the user programmed stencil DSL for a given stencil application and generates the optimized single PE design in Vitis HLS C++ code. To do this, our code generator uses a Python based meta-language specification, $textX$\cite{textX}, which uses meta-model to define a DSL. With our pre-defined meta-model grammar as illustrated in Section~\ref{subsec:dsl}, our compiler parses the DSL, generates the abstract syntax tree (AST), and extracts the user-defined stencil configurations. The stencil configurations include the number of input rows ($R$), the number of input columns ($C$), the number of stencil iterations ($iter$), and the stencil radius size ($r$). 
    Note that for a multidimensional array specified in the DSL, our code generator flattens all the dimensions except the first dimension into one dimension. \rev{Take 3D stencil input size 256$\times$16$\times$16 as an example, we buffer two rows of 16$\times$16 data in the row buffer like a 2D stencil kernel. The difference between 2D and 3D stencil accelerator designs is that they read data from different locations of the row buffer when updating each cell.}
    Then, it coverts the AST into a model consisting of Python objects. The code generator further interprets the model to analyze the data dependency in each statement between the input(s) and output(s). After that, the code generator generates the Vitis HLS C++ code for a single PE design presented in Section~\ref{subsec:single-pe}, based on the user-defined configurations (i.e., $R$, $C$, $iter$, $r$) extracted from the DSL. The unroll factor (i.e., the number of PUs inside each PE in Figure~\ref{fig:fifo}), $U$ (e.g., 16), is chosen based on the AXI interface width (e.g., 512-bit) of a single memory bank and the size of each stencil data cell (e.g., 32-bit) to saturate the off-chip bandwidth.
    
    %to match the streaming flow which equals to the number of PE in Figure~\ref{fig:fifo}. %The stencil radius, $r$ will be decided by the stencil pattern.
    \item To determine the maximum number of PEs that can be instantiated on the FPGA platform, we first estimate the resource utilization of the single-PE design generated from the code generator block, by running Vitis HLS~\cite{vitis} synthesis. Then, combined with the FPGA platform specification and hardware utilization constraints, we determine the maximum PE number \rev{as described in Equation~\ref{eq:pe_res},~\ref{eq:pe_bw} and~\ref{eq:pe_max}}.
    % \begin{equation}
    % \begin{aligned}
    % Max\ \#PEs &= min\left(\frac{\alpha \times total\_FPGA\_resource}{resource\_per\_PE}\ ,\   \frac{\#total\_off\_chip\_mem\_banks}{\#off\_chip\_mem\_banks\_per\_PE}\right)
    % \end{aligned}
    % \label{eq:max-pe-num}
    % %\vspace{-0.02in}
    % \end{equation}
    \item Once the $\#PE_{res}$, $\#PE_{bw}$, and $Max\ \#PEs$ are determined, we explore different temporal and spatial parallelism configurations of the multi-PE design based on the analytical performance model presented in Section~\ref{subsec:analytical_model}, and choose the optimal design choice such that it achieves the least execution latency, based on Equations~\ref{eq:temporal_model} to~\ref{eq:optimal}. For temporal parallelism, we set $s_t = \#PE_{res}$ in Equation~\ref{eq:temporal_model}. For the spatial parallelism alternatives, we set $k_{sr} = k_{ss} = Max\ \#PEs$ in in Equations~\ref{eq:spatial_R_model} and \ref{eq:spatial_S_model}. For the two hybrid parallelism implementations, we \rev{explore all combinations of $(k_{hr}, s_{hr})$ and $(k_{hs}, s_{hs})$ that meets}
    $k_{hr} \times s_{hr} = k_{hs} \times s_{hs} = Max\ \#PEs$ 
    \rev{, $k_{hr} \leq PE_{bw}$, and  $k_{hs} \leq PE_{bw}$}.
    in Equations~\ref{eq:hybrid_r} and \ref{eq:hybrid_s}. To simplify the floorplanning, we limit the number of FPGA spatial PE groups $k_{hr}$ and $k_{hs}$ to be a multiple of $\#SLRs$, so that we have a very small number of $(k_{hr}, s_{hr})$ and $(k_{hs}, s_{hs})$ pairs to explore. Our analytic model will select the best multi-PE design choice with the best parallelism. \rev{When multiple parallelisms achieve a similar performance, we choose the most resource-efficient one. For example,  $Spatial\_S$ and $Hybrid\_S$ are the two best choices among many configurations, then our model will choose $Hybrid\_S$ as it uses fewer HBM banks.}
    %Moreover, to simplify the placement of these PEs on the multiple SLRs (super logic regions, i.e., FPGA dies) on modern FPGAs, we limit the $Max\ \#PEs$ to be a multiple of $\#SLRs$ (number of SLRs), so that each SLR hosts an equal number of PEs. 
    %and the $max \#PEs$ is the product between the spatial kernel number $k$ and the temporal stage number $s$.
    \item Once the best multi-PE design choice is selected, our code generator will automatically generate the corresponding multi-PE accelerator design in TAPA HLS C++~\cite{tapa}, based on the multi-PE architecture presented in Section~\ref{sec:design} and single-PE design generated in step 1. Moreover, we will also automatically generate the corresponding host code \rev{with TAPA API} to manage this FPGA kernel, which includes common FPGA device setup, host buffer allocation, data communication between the host and the FPGA, and signal to start the FPGA kernel execution.
    \item Finally, we build the optimal design from our code generator using Xilinx Vitis 2020.2 tool to generate the final FPGA bitstream and host executable. If the design is successfully built and meets the frequency requirement, it will be output as the optimal design. Otherwise, our automation tool will first attempt to build the next best parallelism design with the same number of PEs. If none of those designs can pass the requirement, our tool will lower the number of PEs by the number of SLRs (i.e., $Max\ \#PEs = Max\ \#PEs - \#SLRs$) and repeat steps 3 to 5 until the design can be successfully built.
\end{enumerate}

Code generator is one fundamental block of our automation framework, which is utilized at two different stages of the automation flow. 
First, after the stencil DSL is parsed and interpreted, the code generator needs to automate the generation of a single-PE design. At this point, only the datapath logic is defined based on the stencil operation, and the fine-grained data parallelism is set to match the off-chip memory bandwidth to enable the dataflow computing requirement.
The second function of the code generator is to automate the multi-PE binding code generation when the number of PEs and the optimal design parallelism settings have been chosen by our analytical performance model. This time the code generator will return a software driver code to run on the host CPU and an optimized stencil accelerator design to deploy on the chosen FPGA platform.

In summary, with our automation framework \toolName, for a given FPGA platform, users can easily define the stencil computing parameters (i.e., input and output data dimension, iteration number, and stencil operation) through a high-level DSL. To automate the design space exploration, we derive analytical models for all five types of parallelisms shown in Figure~\ref{fig:overall_structure_temporal}, ~\ref{fig:overall_structure_spatial} and ~\ref{fig:overall_structure_hybrid}. As a result, our automation framework supports arbitrary stencil workload and can generate performance portable accelerator designs with the optimized parallelism across different HBM-based FPGAs. 
%\vspace{-0.1in}
\section{Experimental Results}\label{sec:results}

%\zf{Please add an overview paragraph here}
In this section, we conduct a comprehensive evaluation of our proposed framework \toolName~and compare it to state-of-the-art automatic stencil acceleration framework SODA~\cite{SODA18}, which only exploits temporal parallelism.
%over single PE optimization, performance model accuracy and comparison of different parallelisms. 
First, we introduce the experiment setup of our evaluation. Second, we present the improvement of our single PE optimization over SODA. Finally, we compare different parallelism optimizations, and discuss the results of the best parallelism configuration. 
%Finally, ...\xy{to be updated}

\subsection{Experimental Setup} \label{sec:setup}

We evaluate a wide range of stencil benchmarks including: 
\rev{
\begin{enumerate}
    \item JACOBI2D/3D from from SODA testbench. They are a 2D 5-point stencil kernel and a 3D 7-point stencil kernel, respectively. They are used in linear algebra algorithms to find the solution for linear equations.
    \item BLUR from SODA testbench~\cite{SODA18}. It is a 2D 9-point stencil kernel. It is commonly used for edge smoothing and noise removing in image processing domains.
    \item SEIDEL2D from SODA testbench. It is a 2D 9-point stencil kernel and used in linear algebra to solve a system or linear equations.
    \item DILATE from the Rodinia-HLS benchmark suite~\cite{RodiniaHLS}. It is a 2D 13-point stencil kernel and used to detect and track leukocyte of blood vessel in biomedical research.
    \item HOTSPOT from the Rodinia-HLS benchmark suite. It is a 2D 5-point stencil kernel with two inputs and one output. It is used to estimate processor temperature based on power grid and temperature of the corresponding area.
    \item HEAT3D from SODA testbench. It is a 3D 7-point stencil kernel and used for heat diffusion simulation. 
    \item SOBEL2D from SODA testbench. It is a 2D 9-point stencil kernel and used for image processing, particularly for edge detection.
\end{enumerate}
}
We use four different input sizes, \rev{256 $\times$ 256, 720 $\times$ 1024,} 9720 $\times$ 1024 \rev{and 4096 $\times$ 4096,} when evaluating all the 2-dimensional stencil benchmarks; and use \rev{256 $\times$ 16 $\times$ 16, 720 $\times$ 32 $\times$ 32,} 9720 $\times$ 32 $\times$ 32 \rev{and 4096 $\times$ 64 $\times$ 64,} input sizes for the 3-dimensional stencil benchmarks. Furthermore, we sweep the iteration number from 1 to 64 at a power of 2 increment.
% ~\al{Please confirm the following sentence: Note that when the iteration number is 1, all parallelism will have the same throughput.}
Note that when the iteration number is 1, spatial parallelism and hybrid parallelism will be the same and have the same throughput. 
%\zf{Looks we need to present results for iteration number is 1}. 
All these stencil kernels are written in the stencil DSL as illustrated in Section~\ref{subsec:dsl}.

We evaluate \toolName~on Xilinx Alevo U280 datacenter FPGA board with 32 HBM2 banks~\cite{alveoU280}. First, \toolName~compiles the stencil DSL into the optimized FPGA design in Xilinx Vitis HLS C++ \rev{with TAPA APIs~\cite{tapa}} and the corresponding host code. Then it uses \rev{AutoBridge~\cite{Autobridge21} to do the floorplanning and pipelining optimizations for our design and} Vitis 2020.2~\cite{vitis} to build the generated design to run on the U280 FPGA.
We set 225 MHz as the target frequency of our designs since all of them use 512-bit wide streaming connections and can already fully utilize the effective bandwidth from each HBM memory bank. This is because on the U280 FPGAs, the two HBM stacks operate at 450 MHz, and are connected to 32 hardened AXI ports in width of 256-bit. 
Therefore, to achieve the HBM memory bandwidth using a 512-bit AXI port, the kernel frequency needs to be above $\SI{450}{\MHz}\ \times \  $256-bit / 512-bit $= \SI{225}{\MHz}$.
And the theoretical peak bandwidth of a single 512-bit AXI port accessing a single HBM bank is 512 bits/cycle $\times\ \SI{225}{\MHz}$ / 8 bits-per-byte = $\SI{14.4}{GB/s}$.

% In our design, we choose 512 bits instead of data width(i.e. 32 bits for float) as port width to utilize the bandwidth of interface. With larger bit width, the required frequency can be lower. 
% On Alveo U280 FPGA board, two HBM stacks runs at 450 MHz. There are 32 hardened AXI ports connecting to HBM banks with fixed 256 bits width. 
% The peak throughput of single AXI port connected to HBM is $256 \times \SI{450}{\MHz} / 8 bits/byte = \SI{14.4}{\giga\byte/s}$. With 512-bit port width, the kernel frequency must be above $\SI{14.4}{\giga\byte/s} \times 8 / 512 = \SI{225}{\MHz} $ to fully utilize the HBM bandwidth.

\begin{figure}[!t]
    \begin{subfigure}[b]{0.48\textwidth}
        \centering
        \includegraphics[width=\textwidth]{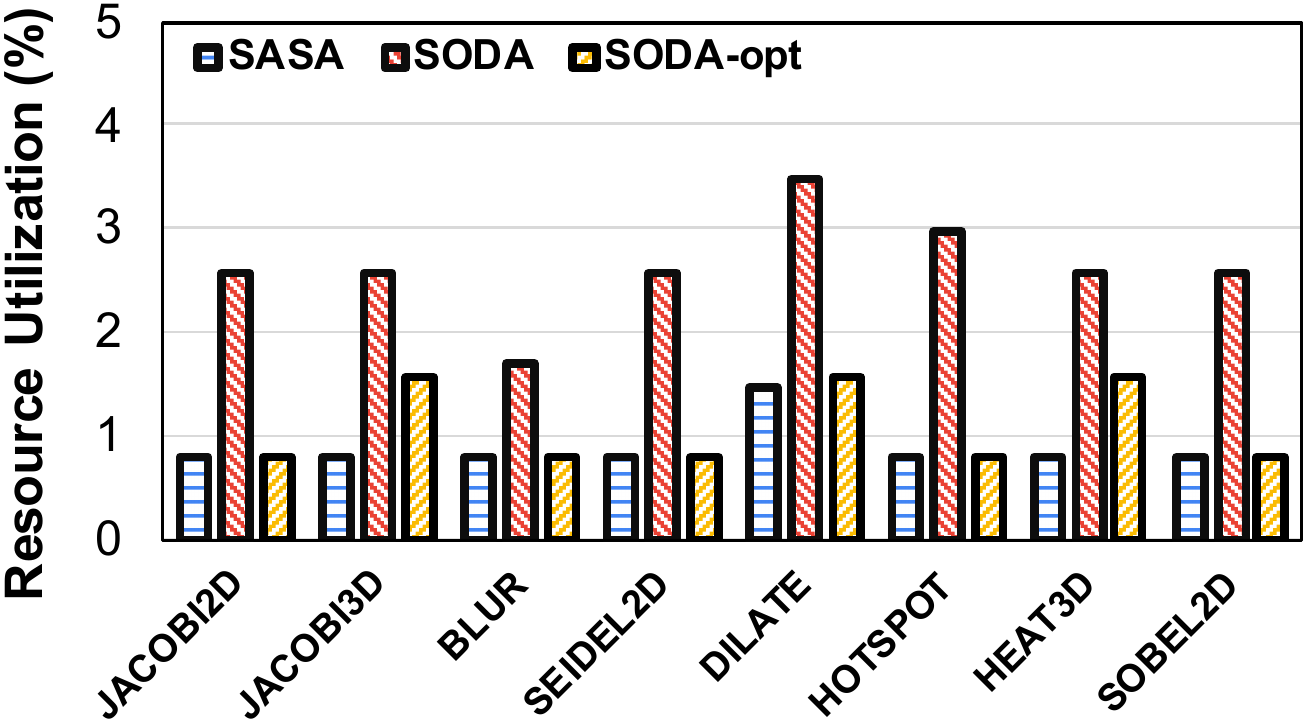}
        \caption{BRAM utilization}
        \label{fig:res_bram}
    \end{subfigure}
    \hspace{+0.1in}
    \begin{subfigure}[b]{0.48\textwidth}
        \centering
        \includegraphics[width=\textwidth]{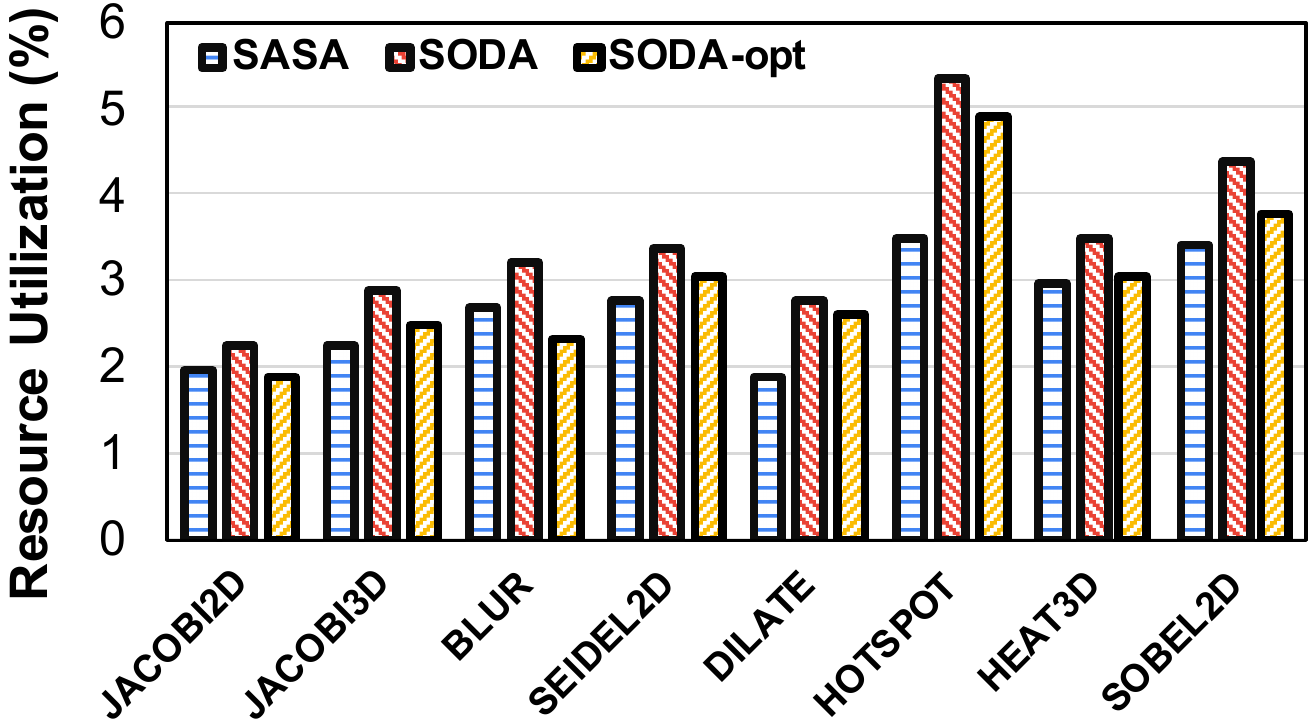}
        \caption{FF utilization}
        \label{fig:res_ff}
    \end{subfigure}
    \begin{subfigure}[b]{0.48\textwidth}
        \centering
        \includegraphics[width=\textwidth]{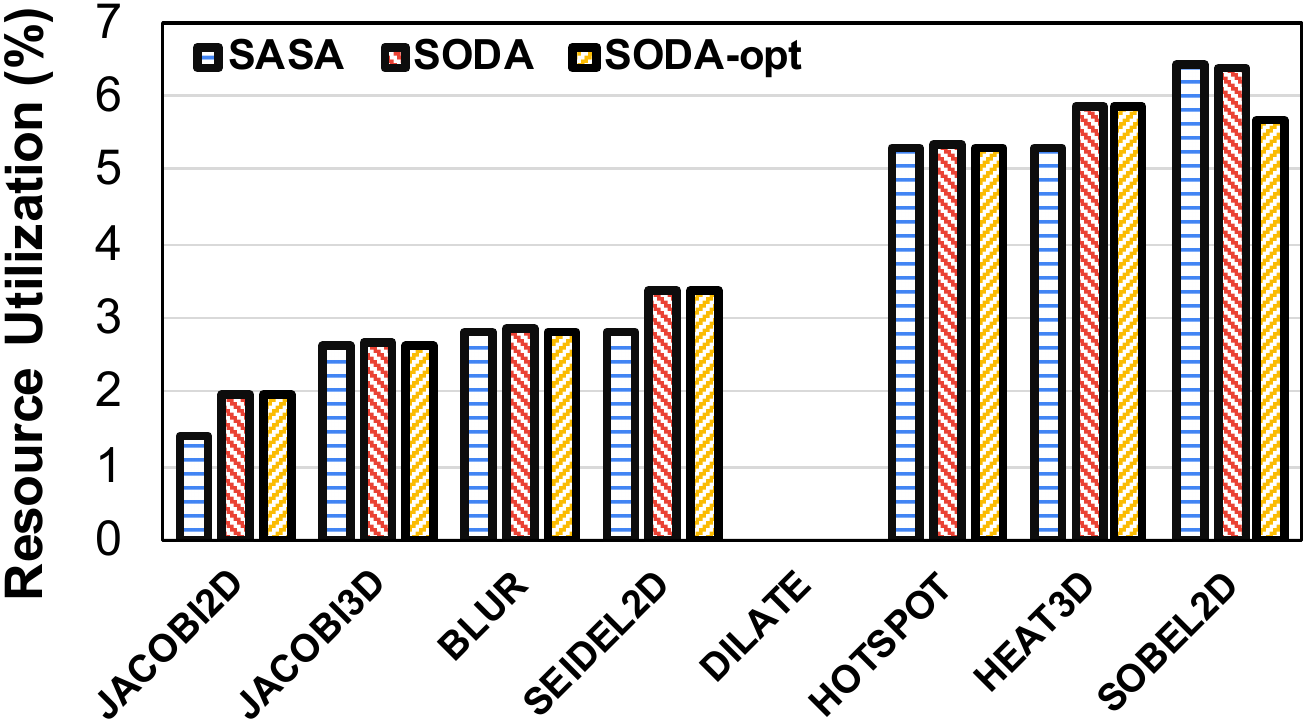}
        \caption{DSP utilization}
        \label{fig:res_dsp}
    \end{subfigure}
    \hspace{+0.1in}
    \begin{subfigure}[b]{0.48\textwidth}
        \centering
        \includegraphics[width=\textwidth]{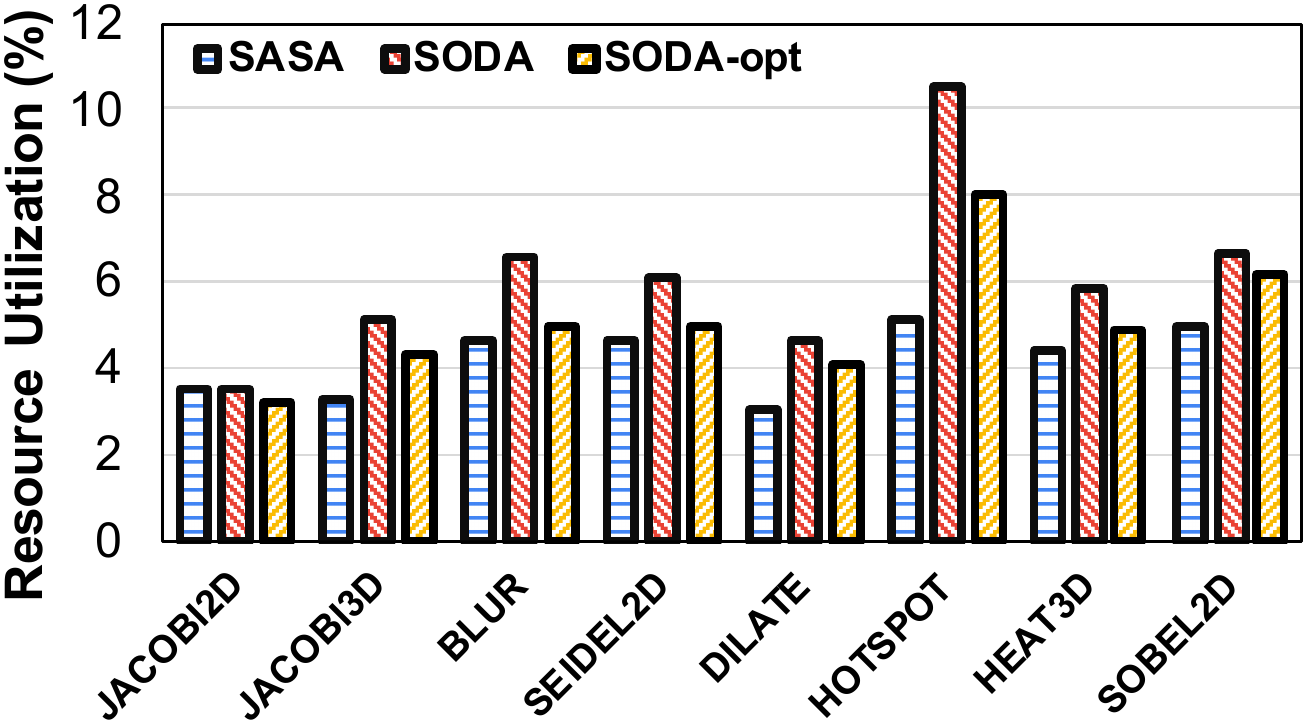}
        \caption{LUT utilization}
        \label{fig:res_lut}
    \end{subfigure}
    \caption{\rev{Resource utilization of a single PE saturating one HBM bank bandwidth on Alveo U280 for the input size of 9720$\times$1024 and 9720$\times$32$\times$32}} %\zf{Update this figure, enlarge font size, can make the height of the figure larger} \xy{updated now}}
    \label{fig:resource}
\end{figure}

\subsection{Results for Single PE Optimization} \label{subsec:result_singlePE_parallisim}

To demonstrate the quality of our design, we first evaluate our optimized single PE design, which accelerates one stencil iteration using the optimized streaming access from one HBM bank. From the performance perspective, it saturates the bandwidth of a single HBM bank by placing 16 parallel PUs (processing units) inside each PE to execute in a fully streaming fashion. Therefore, it achieves the optimal performance given one HBM bank, which is the same as SODA~\cite{SODA18} that uses the optimal data reuse size and memory access requirement. 

Next, we focus on the comparison of its resource consumption. 
Figure~\ref{fig:resource} shows an overall resource utilization comparison with \rev{original SODA and optimized SODA (i.e., SODA-opt) that is integrated with TAPA/AutoBridge~\cite{tapa,Autobridge21} for a fair comparison}, including BRAM, FF, DSP and LUT consumption.
%\zf{Please double check the numbers and add a comparison to SODA-opt.}
Compared to the original SODA, the major benefit of our design comes from removing the on-chip line buffer by introducing the coalesced reuse buffer design. It brings a \rev{4.3\%-69.8\%} reduction in the BRAM utilization compared to the previous SODA design.
Consequently, the BRAM reduction further reduces the FFs and LUTs consumption of the design by \rev{12.9\%-34.8\% and 1.8\%-51.7\%,} respectively.
Since both of SODA and our design use the same fine-grained parallelism and place 16 PUs inside each PE (i.e., loop unroll factor $U=16$), we both achieve the same DSP utilization. Note that DILATE only has boolean logic operations and thus does not utilize any DSP resource.

\rev{For the majority of benchmarks, both our implementation and the optimized SODA achieve a similar amount of resources.}

\subsection{Results for Different Multi-PE Parallelisms} \label{subsec:result_multiPE_parallisim}

In this subsection, we first validate the accuracy our analytical performance model. Then we evaluate the performance trend of temporal parallelism, two spatial parallelisms, and two hybrid parallelisms, respectively, when the number of iterations changes. Finally, we compare the performance between temporal, spatial, and hybrid parallelisms and summarize the best parallelism configurations. All results for different parallelisms are summarized in Figure~\ref{fig:throughput_a} to~\ref{fig:throughput_h}, which are measured using the common throughput metric GCell/s (i.e., how many billion of stencil data cells it can process per second). 

% Make sure \#iteration is incorporated in the analysis.
\subsubsection{Performance Model Accuracy}
\begin{figure}[!t]
    \centerline{
    \includegraphics[width=0.7\textwidth]{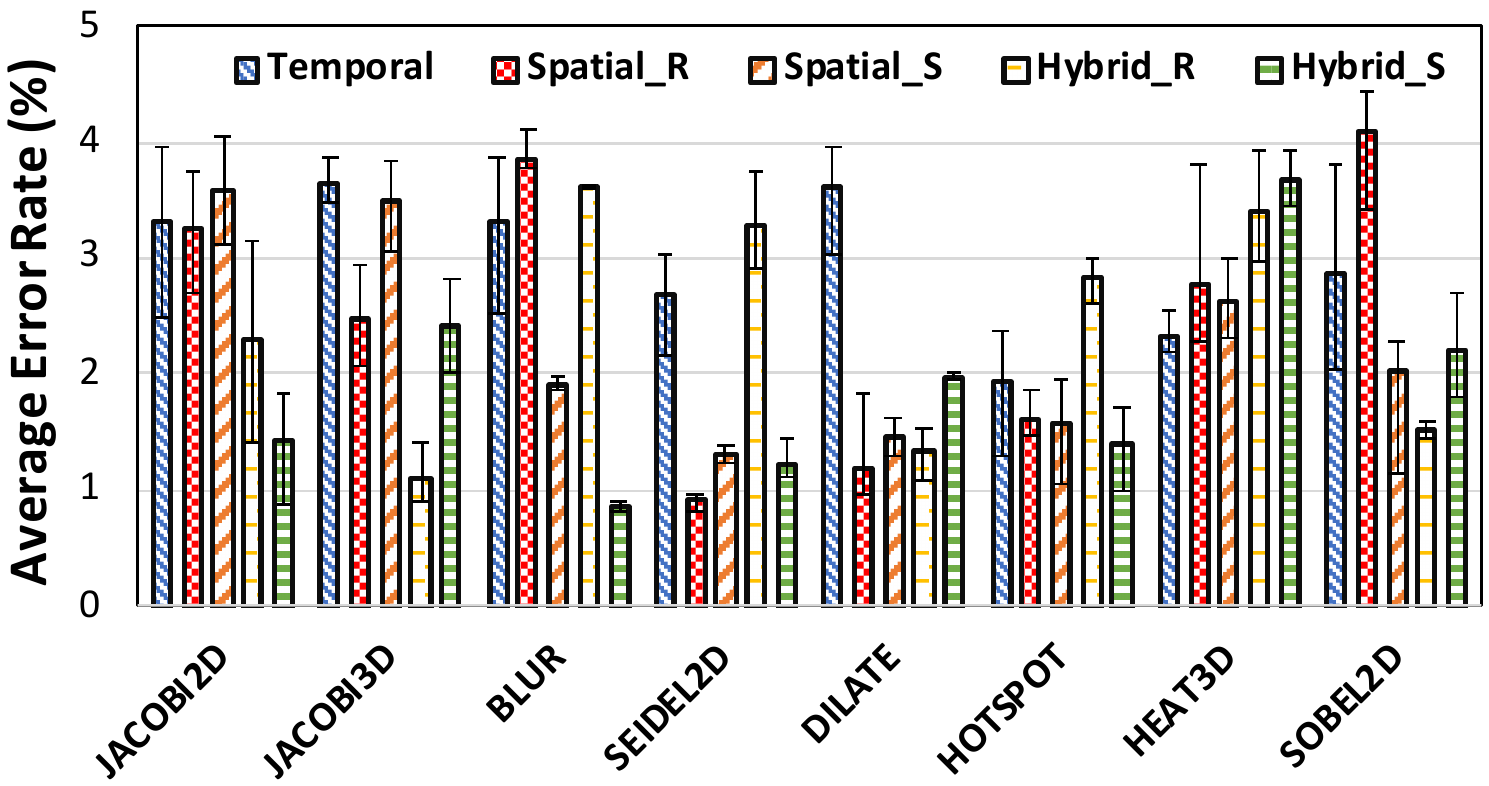}}
    \vspace{-0.05in}
    \caption{\rev{Accuracy of our analytical performance model}} %\zf{Update this figure: BLUR error rate} \xy{Updated now}}
    \label{fig:accuracy}
\end{figure}

\begin{figure}[h]
    \centering
    \begin{subfigure}[b]{0.447\textwidth}
        \centering
        \includegraphics[width=\textwidth]{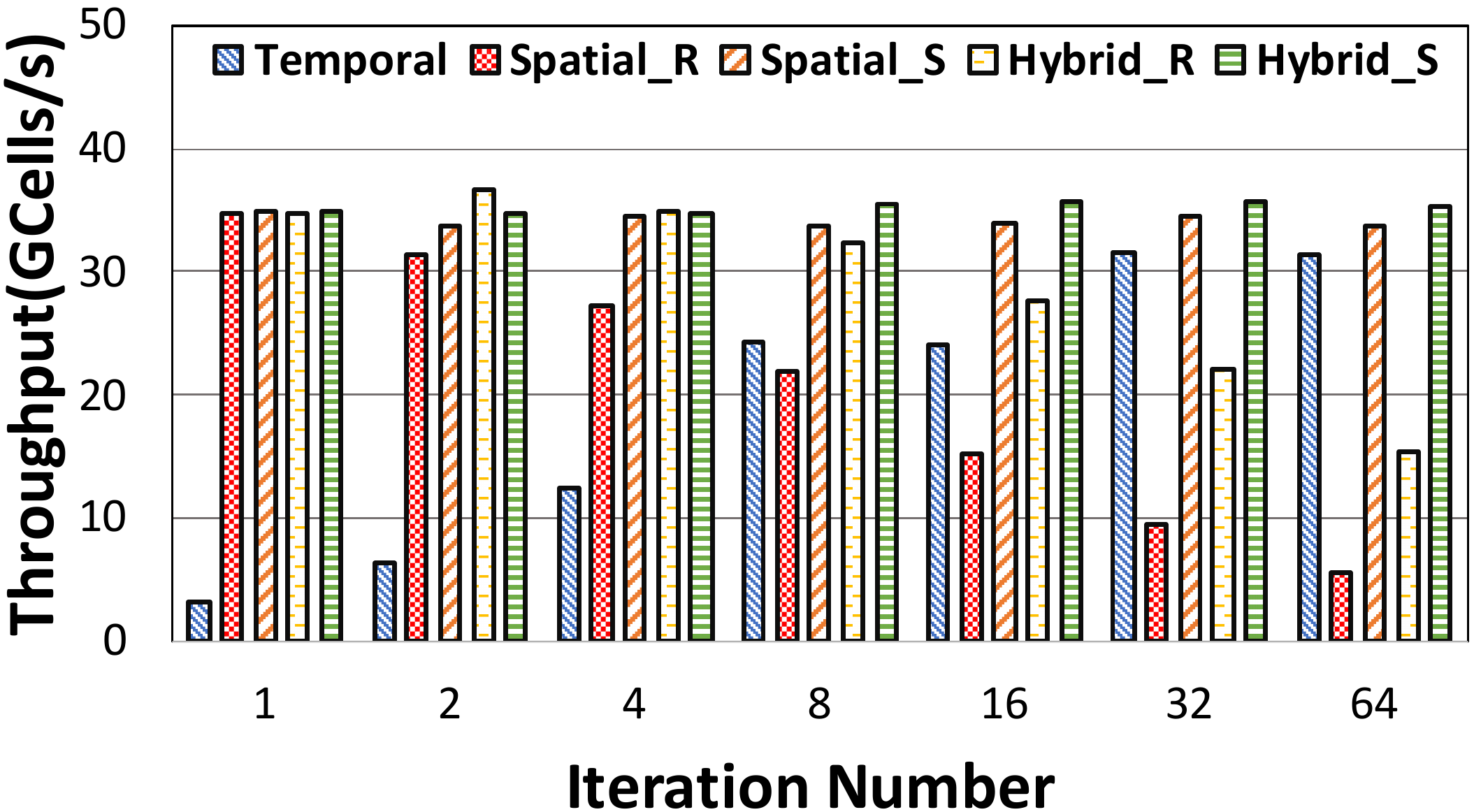}
        \caption{Throughput of BLUR $256 \times 256$}
        \label{fig:c_256}
    \end{subfigure}
    \begin{subfigure}[b]{0.447\textwidth}
        \centering
        \includegraphics[width=\textwidth]{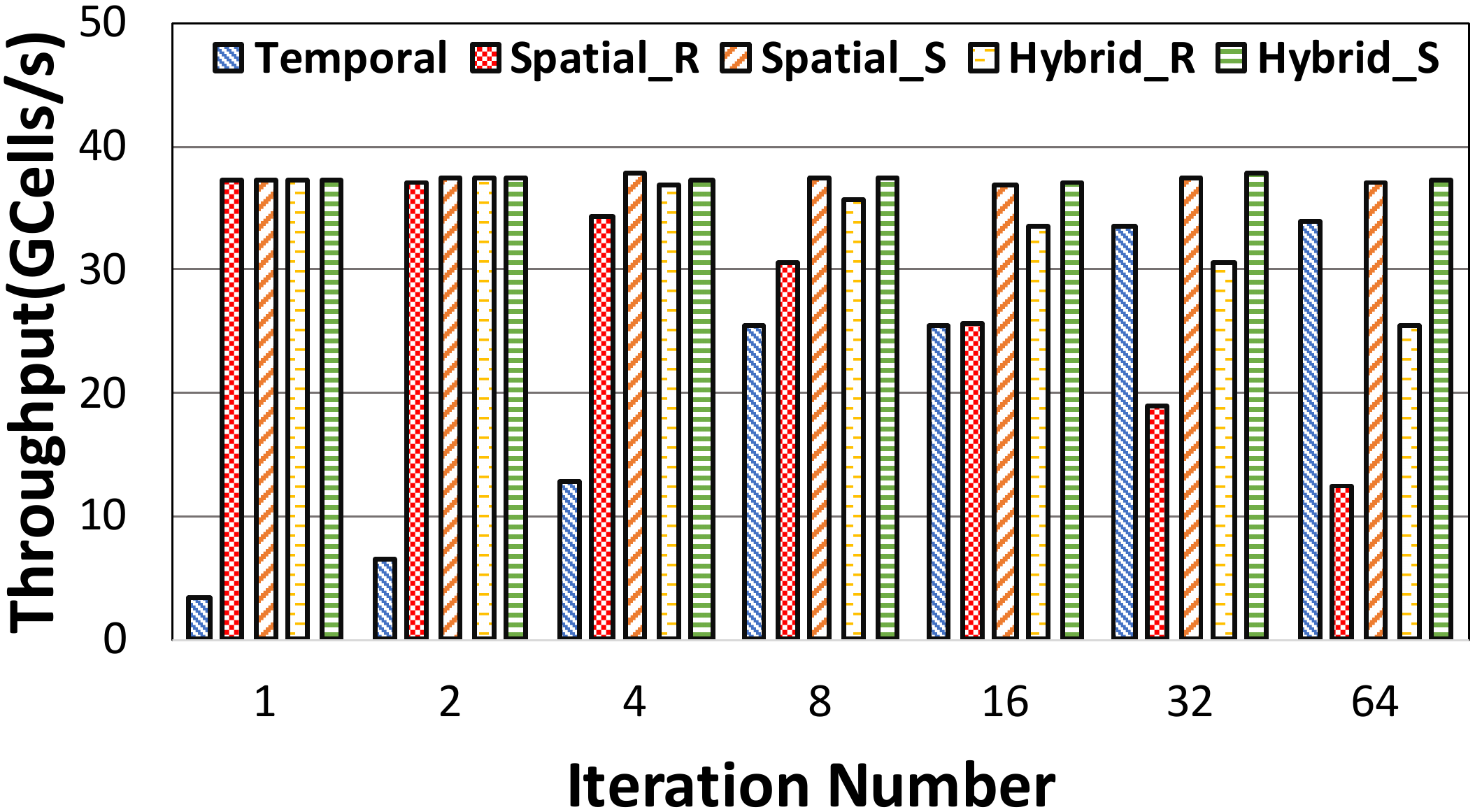}
        \caption{Throughput of BLUR $720 \times 1024$}
        \label{fig:c_720}
    \end{subfigure}
    \begin{subfigure}[b]{0.447\textwidth}
        \centering
        \includegraphics[width=\textwidth]{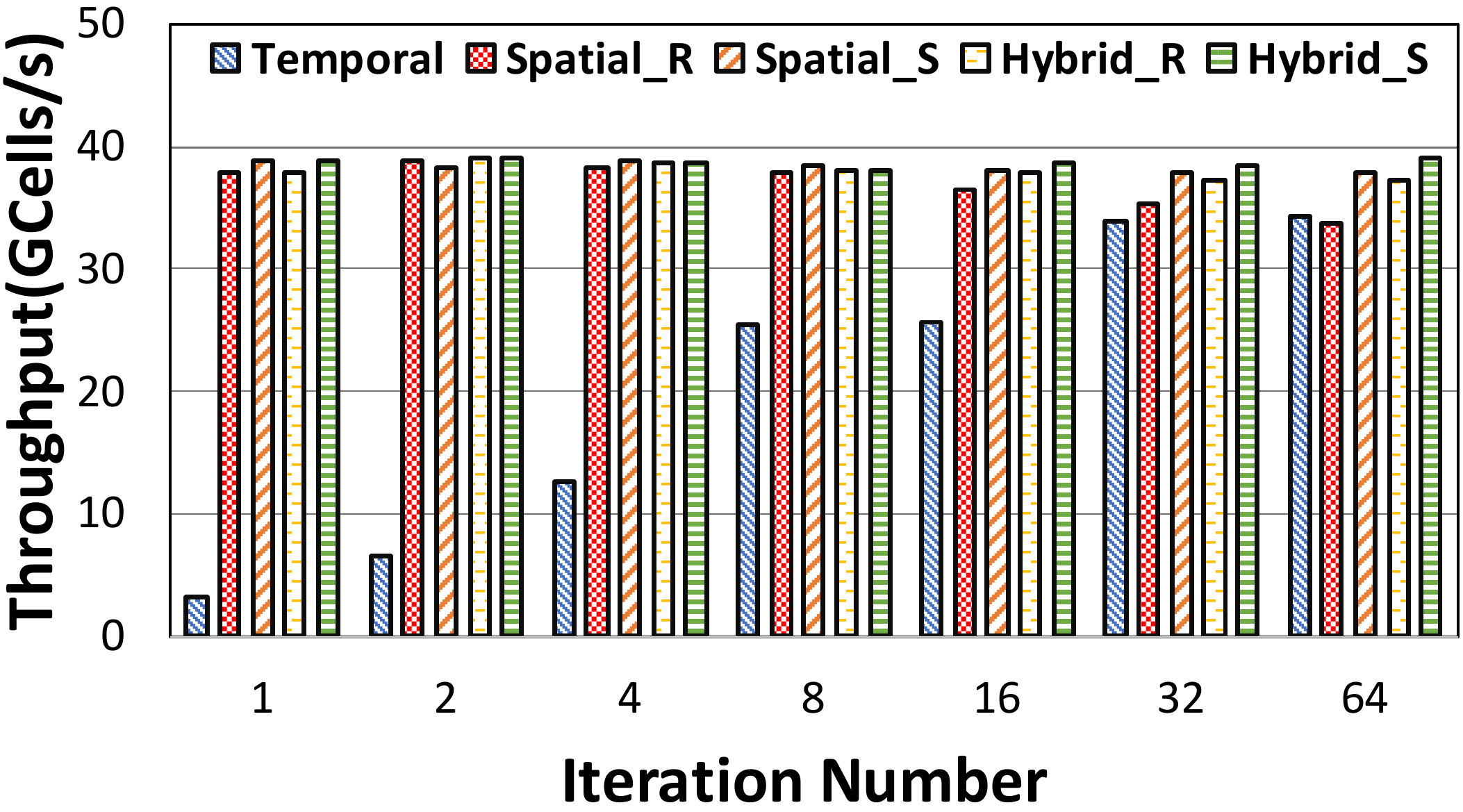}
        \caption{Throughput of BLUR $9720 \times 1024$}
        \label{fig:c_9720}
    \end{subfigure}
    \begin{subfigure}[b]{0.447\textwidth}
        \centering
        \includegraphics[width=\textwidth]{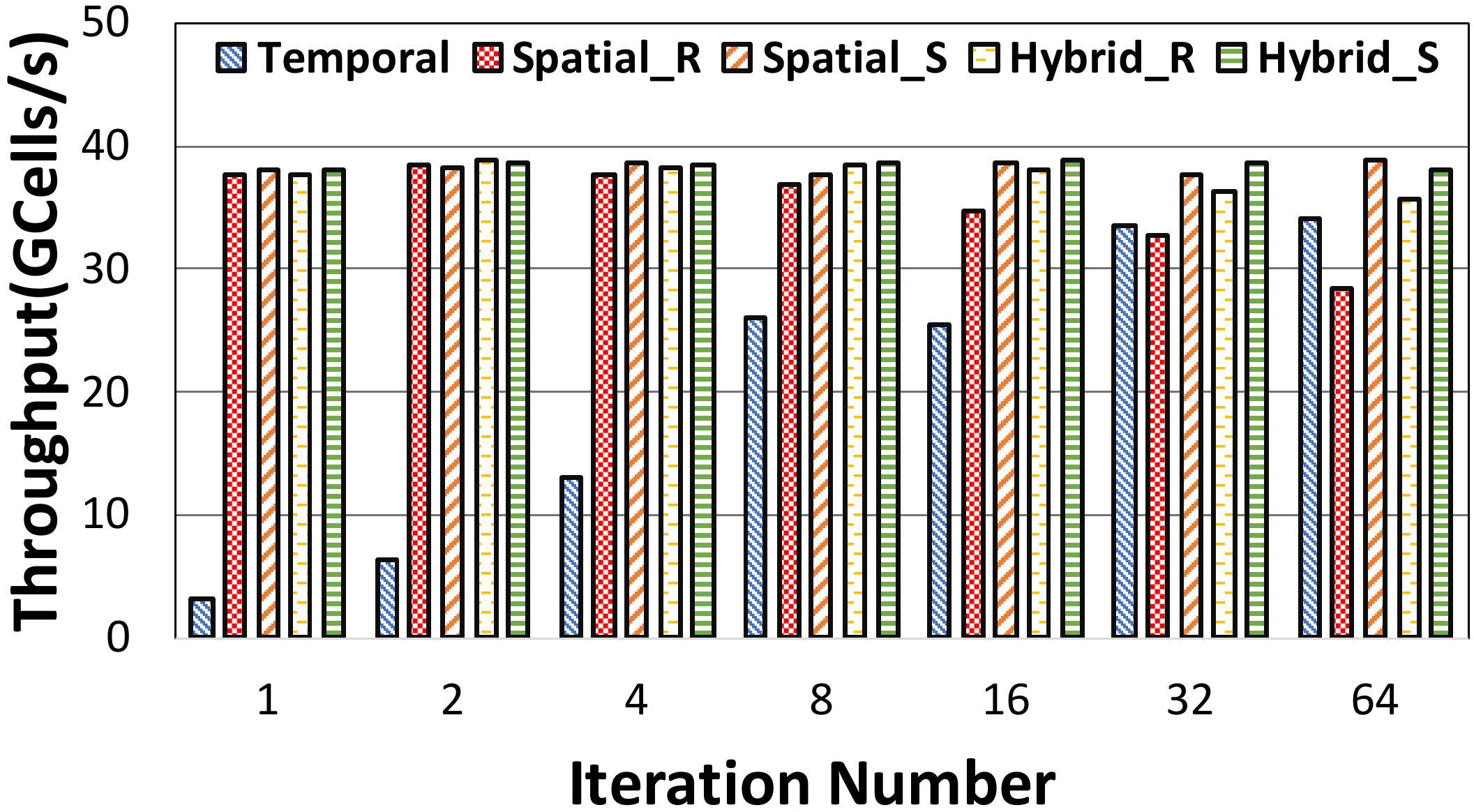}
        \caption{Throughput of BLUR $4096 \times 4096$}
        \label{fig:c_4096}
    \end{subfigure}
    \caption{\rev{Throughput (GCell/s) comparison of different parallelism optimizations for BLUR with the number of iterations changing from 1 to 64} 
    \label{fig:throughput_a}}
\end{figure}

\begin{figure}[h]
    \centering
    \begin{subfigure}[b]{0.448\textwidth}
        \centering
        \includegraphics[width=\textwidth]{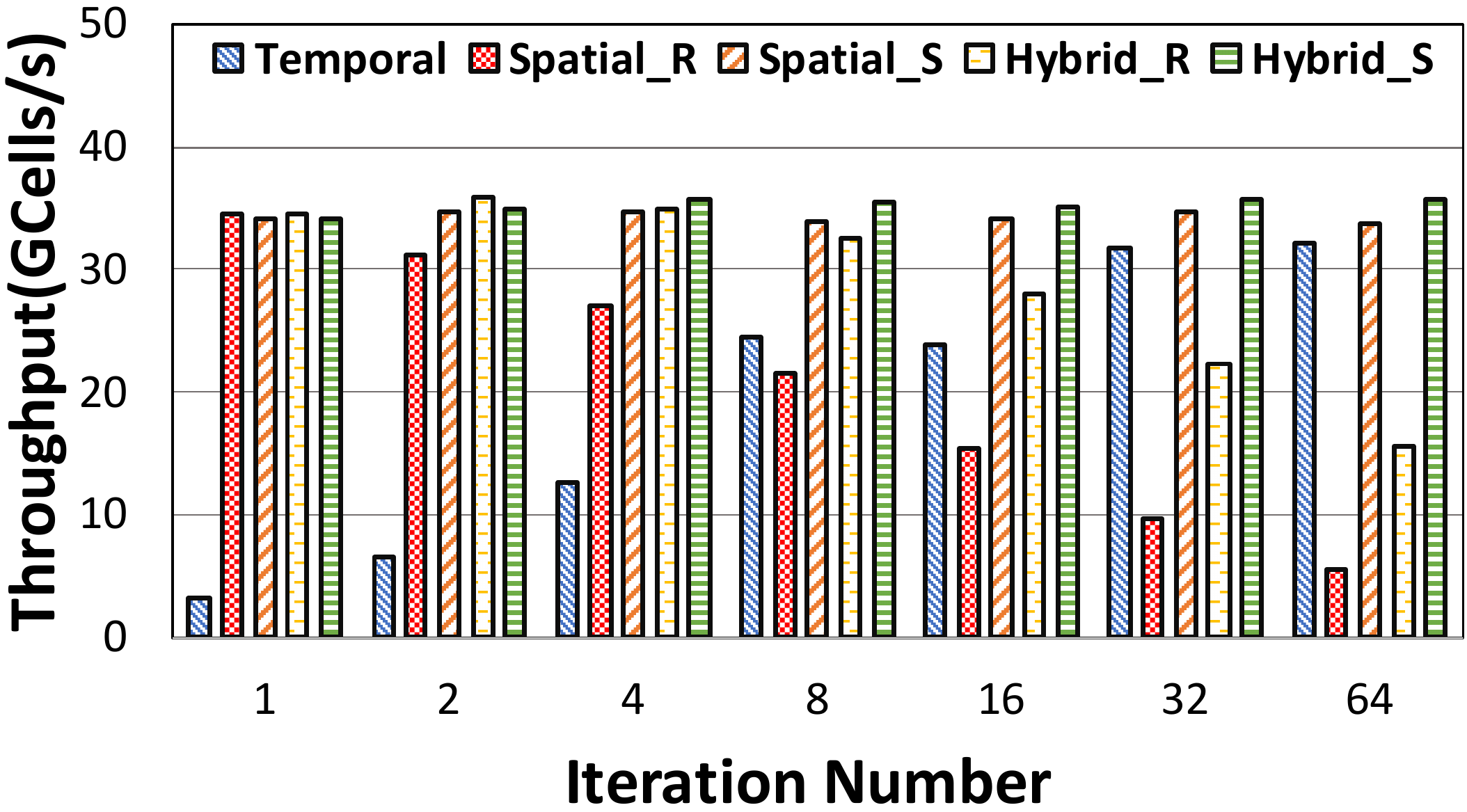}
        \caption{Throughput of SEIDEL2D $256 \times 256$}
        \label{fig:d_256}
    \end{subfigure}
    \begin{subfigure}[b]{0.448\textwidth}
        \centering
        \includegraphics[width=\textwidth]{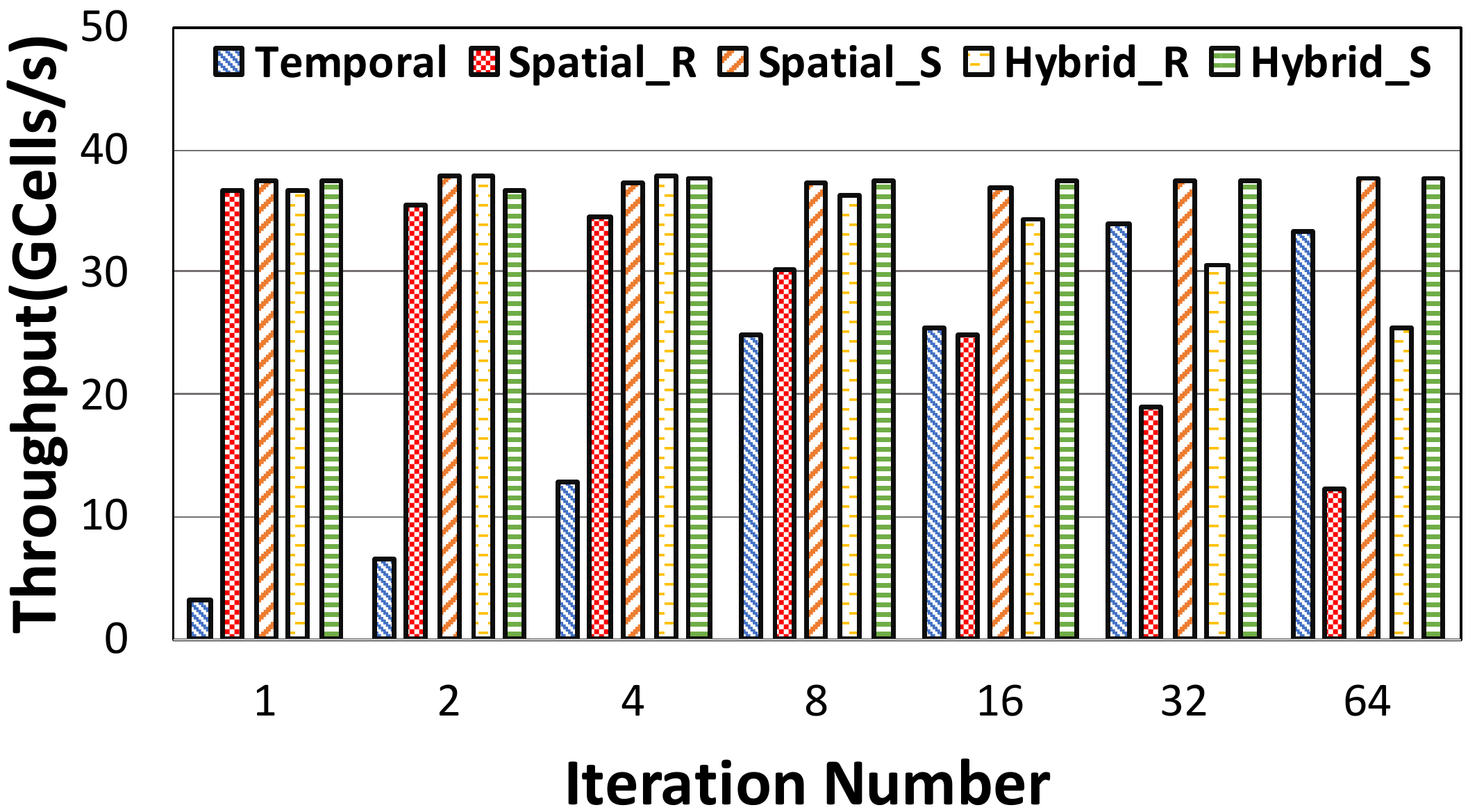}
        \caption{Throughput of SEIDEL2D $720 \times 1024$}
        \label{fig:d_720}
    \end{subfigure}
    \begin{subfigure}[b]{0.448\textwidth}
        \centering
        \includegraphics[width=\textwidth]{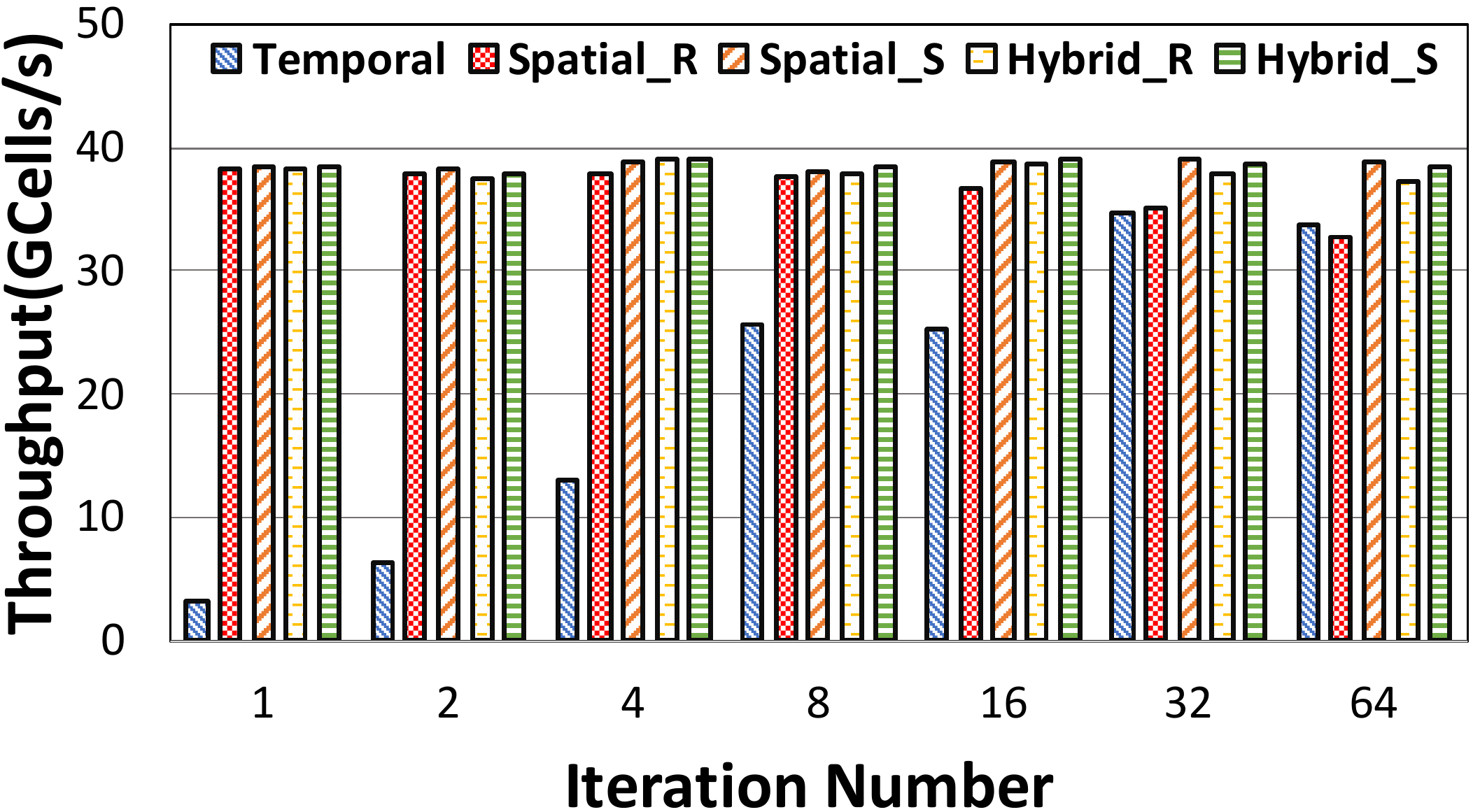}
        \caption{Throughput of SEIDEL2D $9720 \times 1024$}
        \label{fig:d_9720}
    \end{subfigure}
    \begin{subfigure}[b]{0.448\textwidth}
        \centering
        \includegraphics[width=\textwidth]{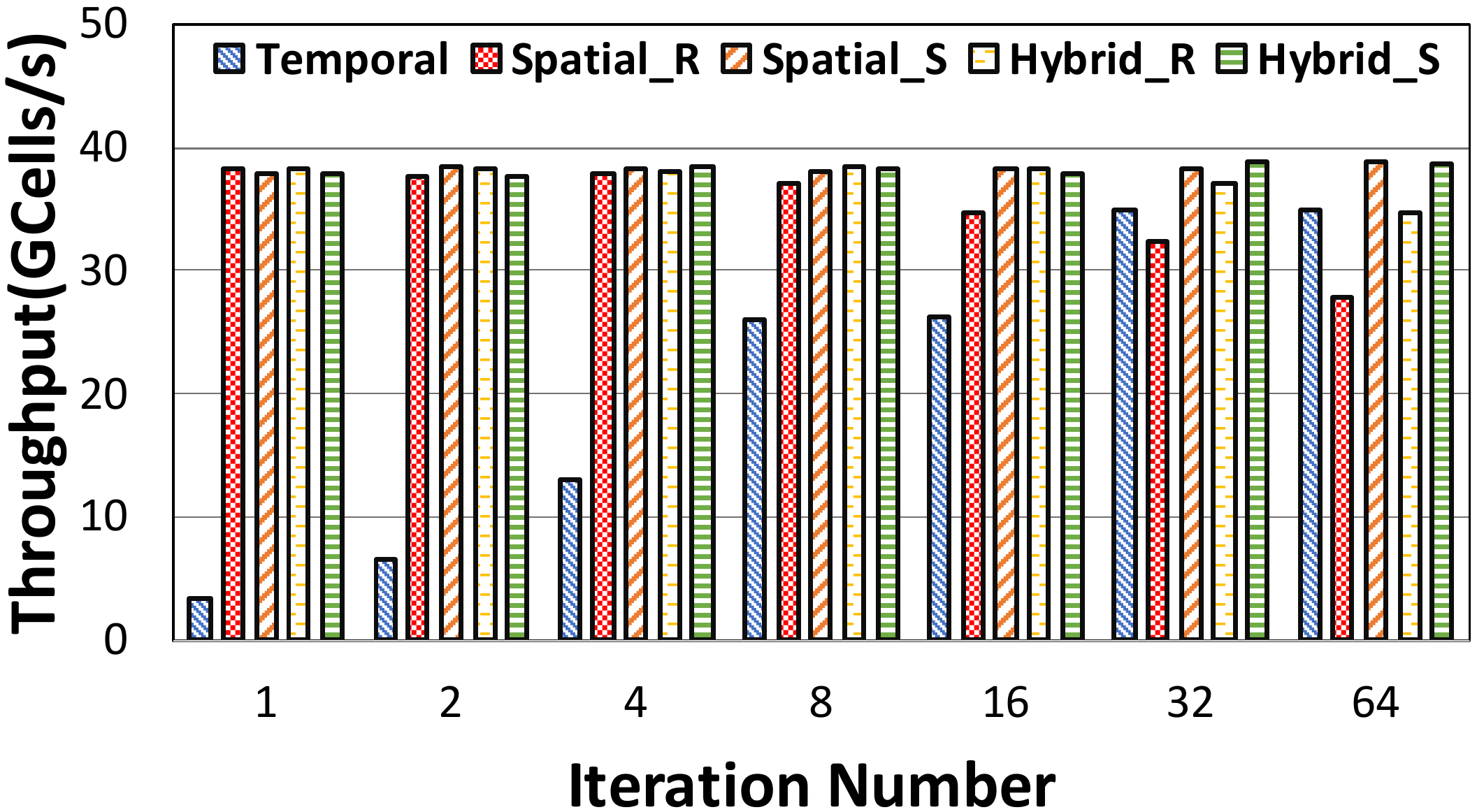}
        \caption{Throughput of SEIDEL2D $4096 \times 4096$}
        \label{fig:d_4096}
    \end{subfigure}
    \caption{\rev{Throughput (GCell/s) comparison of different parallelism optimizations for SEIDEL2D with the number of iterations changing from 1 to 64}} 
    \label{fig:throughput_b}
\end{figure}

\begin{figure}[h]
    \centering
    \begin{subfigure}[b]{0.447\textwidth}
        \centering
        \includegraphics[width=\textwidth]{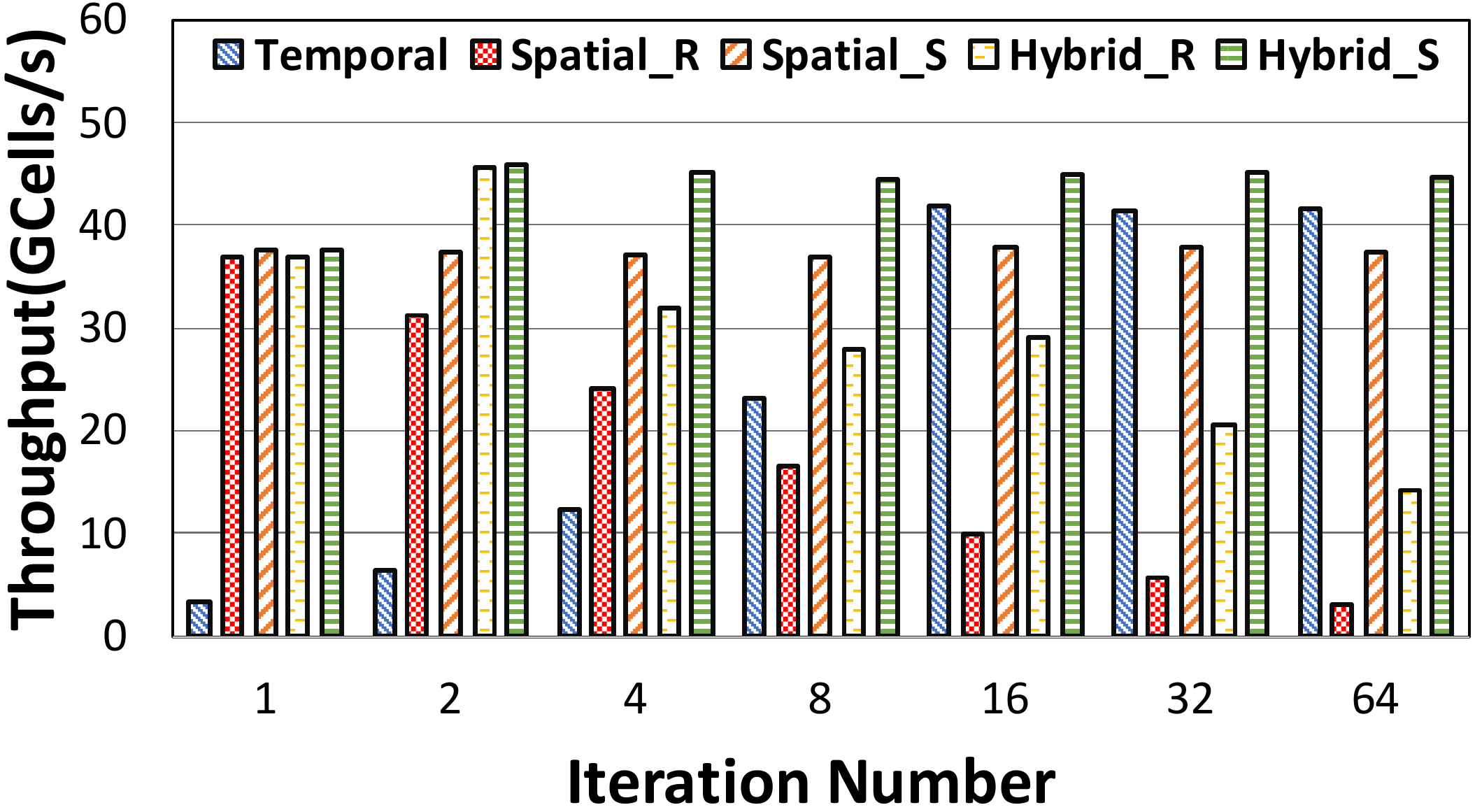}
        \caption{Throughput of DILATE $256 \times 256$}
        \label{fig:e_256}
    \end{subfigure}
    \begin{subfigure}[b]{0.447\textwidth}
        \centering
        \includegraphics[width=\textwidth]{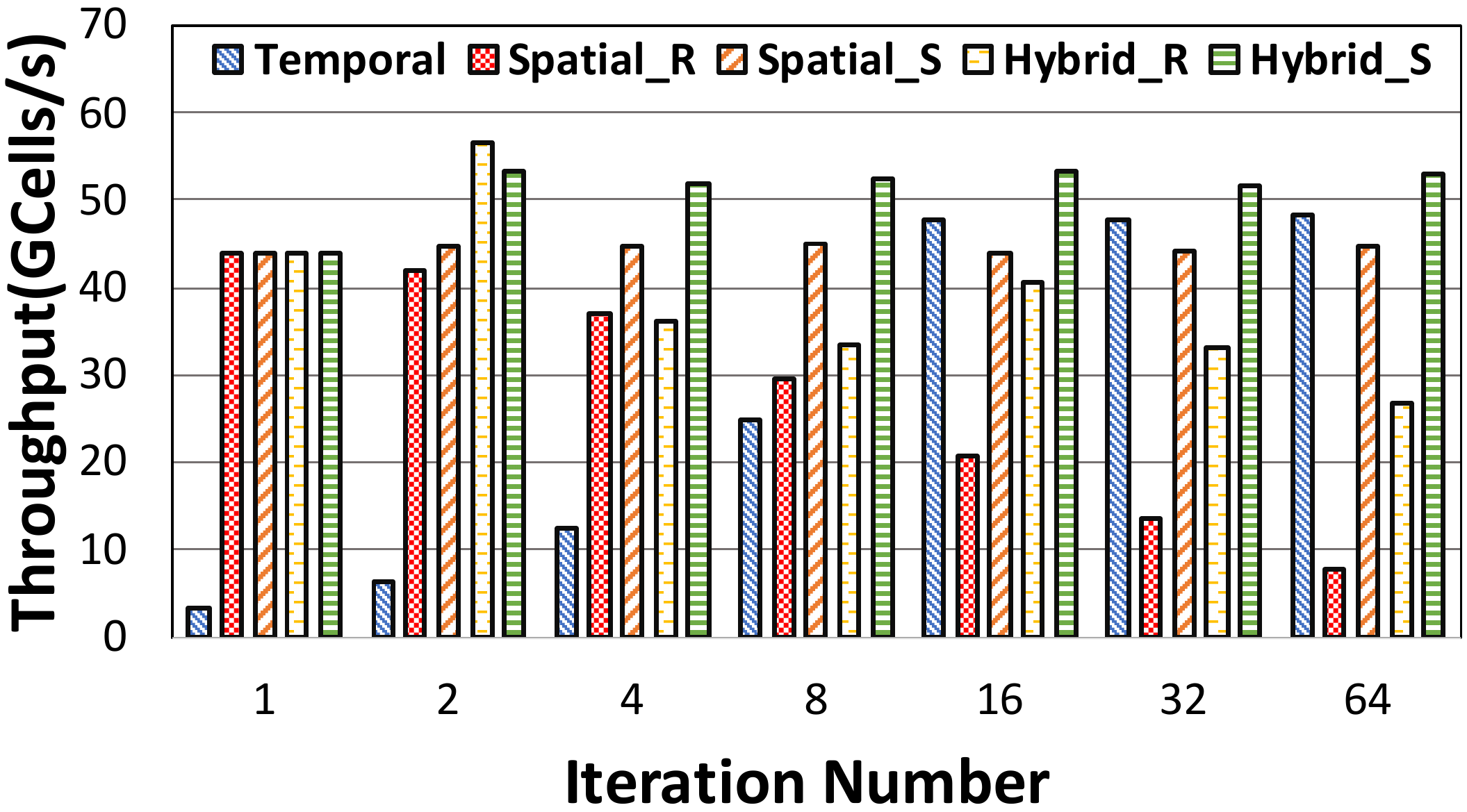}
        \caption{Throughput of DILATE $720 \times 1024$}
        \label{fig:e_720}
    \end{subfigure}
    \begin{subfigure}[b]{0.447\textwidth}
        \centering
        \includegraphics[width=\textwidth]{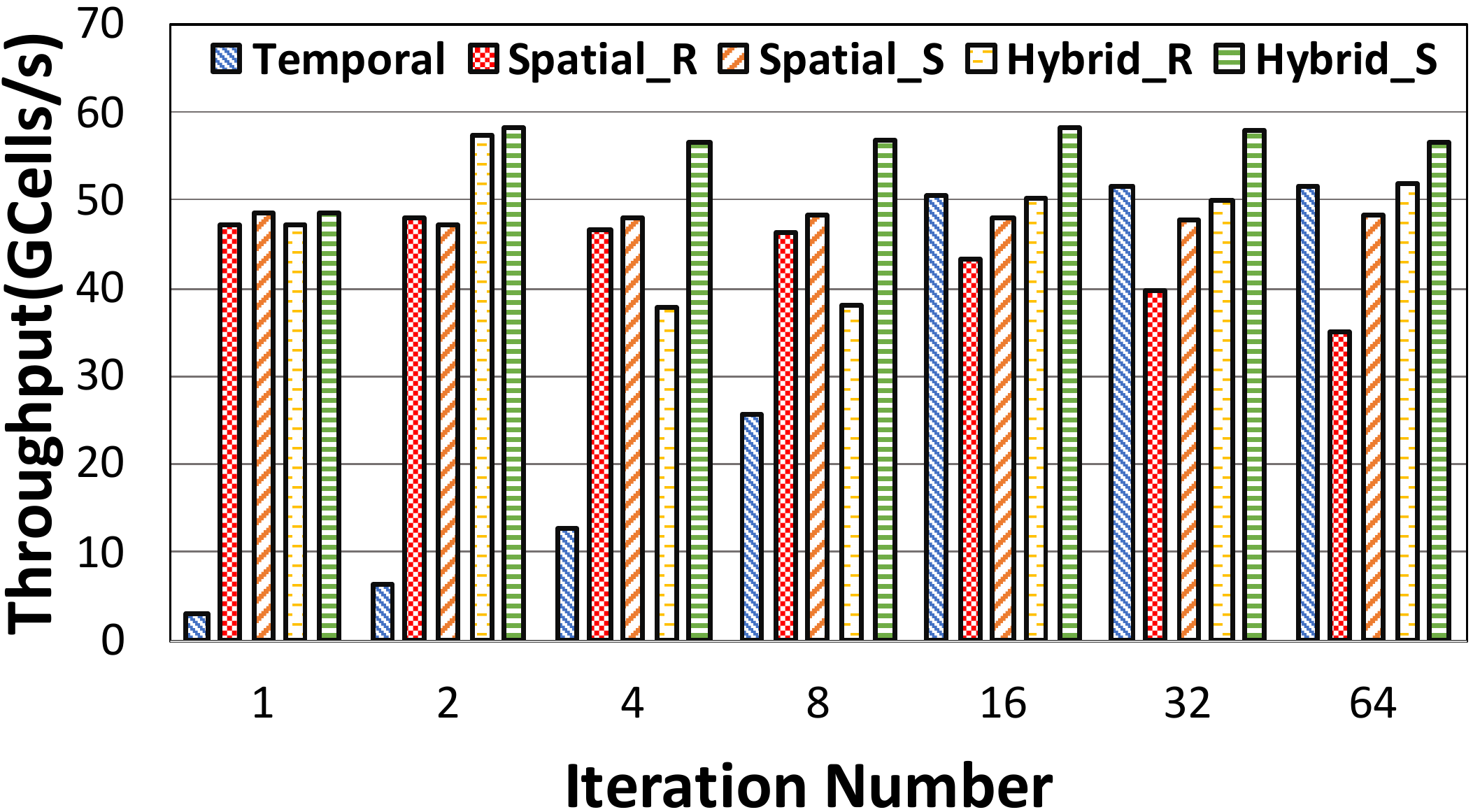}
        \caption{Throughput of DILATE $9720 \times 1024$}
        \label{fig:e_9720}
    \end{subfigure}
    \begin{subfigure}[b]{0.447\textwidth}
        \centering
        \includegraphics[width=\textwidth]{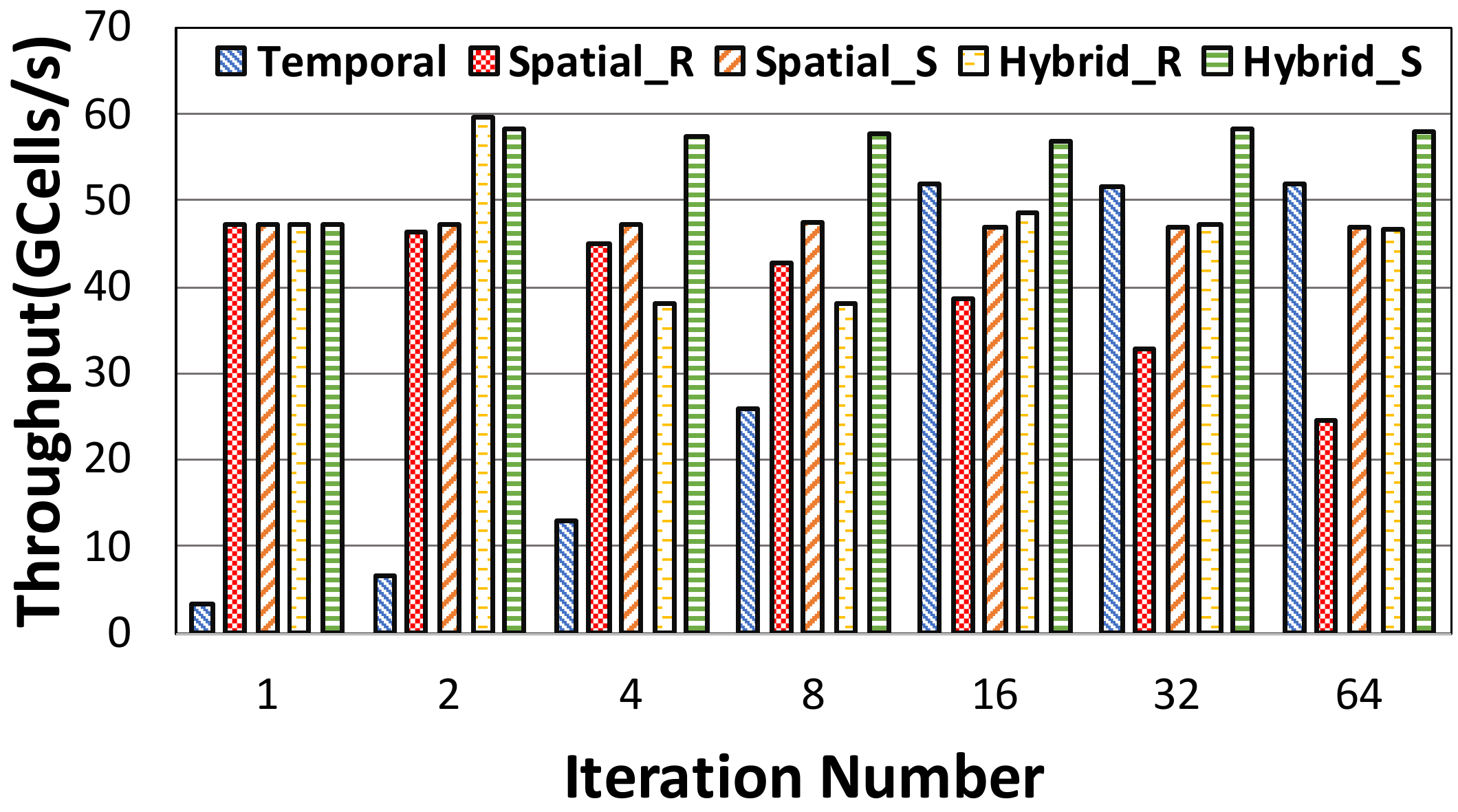}
        \caption{Throughput of DILATE $4096 \times 4096$}
        \label{fig:e_4096}
    \end{subfigure}
    \caption{\rev{Throughput (GCell/s) comparison of different parallelism optimizations for DILATE with the number of iterations changing from 1 to 64}} 
    \label{fig:throughput_c}
\end{figure}

\begin{figure}[h]
    \centering
    \begin{subfigure}[b]{0.445\textwidth}
        \centering
        \includegraphics[width=\textwidth]{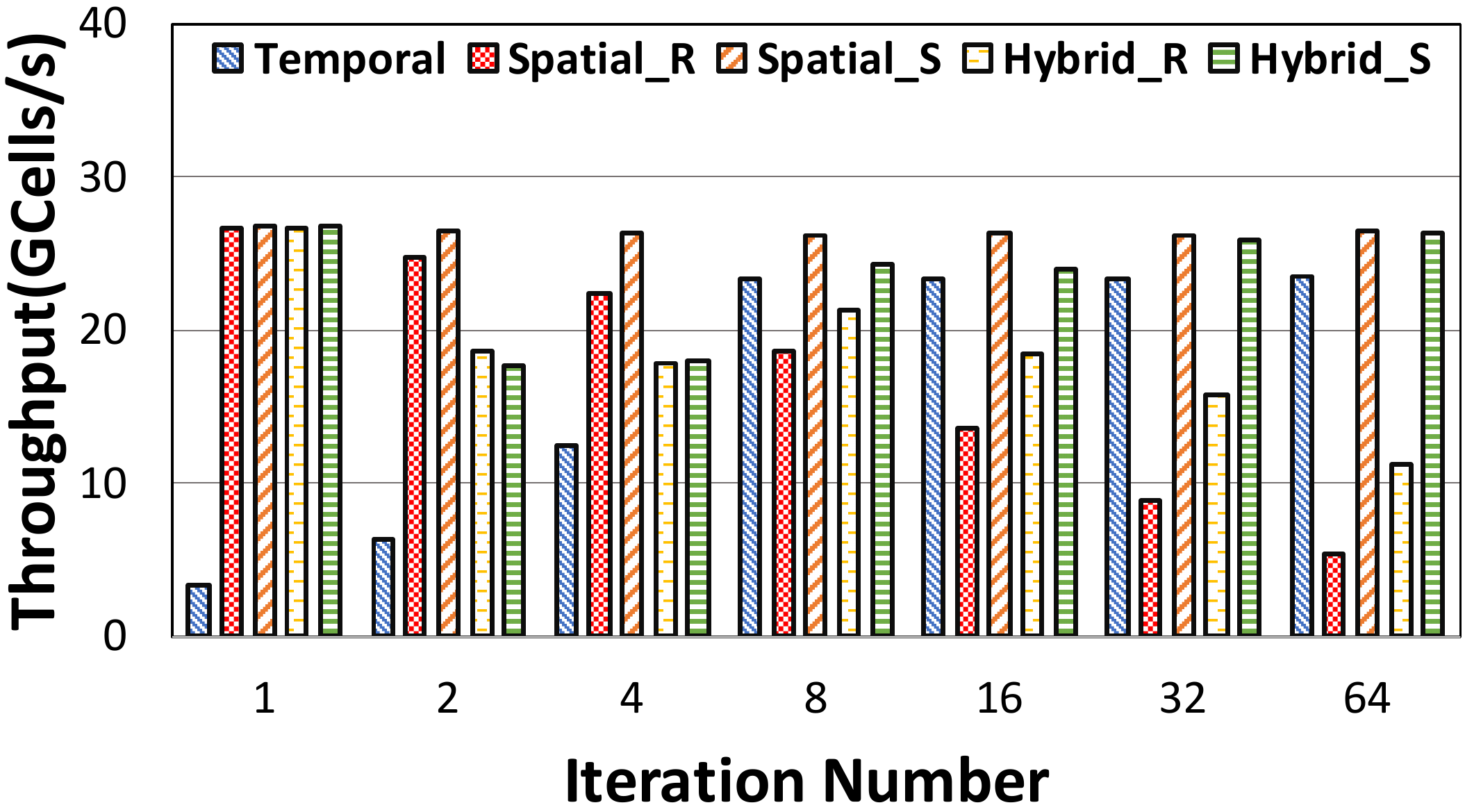}
        \caption{Throughput of HOTSPOT $256 \times 256$}
        \label{fig:f_256}
    \end{subfigure}
    \begin{subfigure}[b]{0.445\textwidth}
        \centering
        \includegraphics[width=\textwidth]{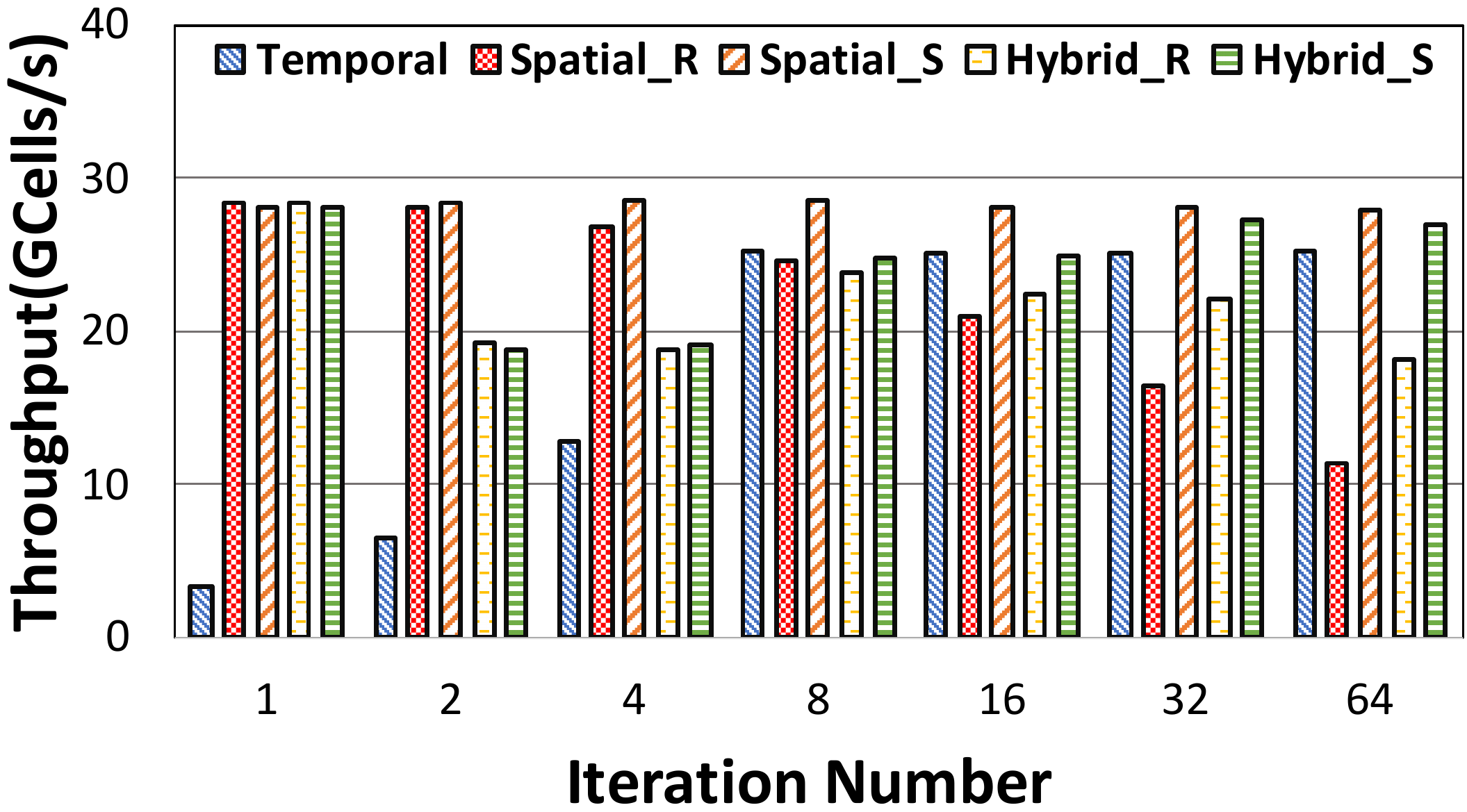}
        \caption{Throughput of HOTSPOT $720 \times 1024$}
        \label{fig:f_720}
    \end{subfigure}
    \begin{subfigure}[b]{0.445\textwidth}
        \centering
        \includegraphics[width=\textwidth]{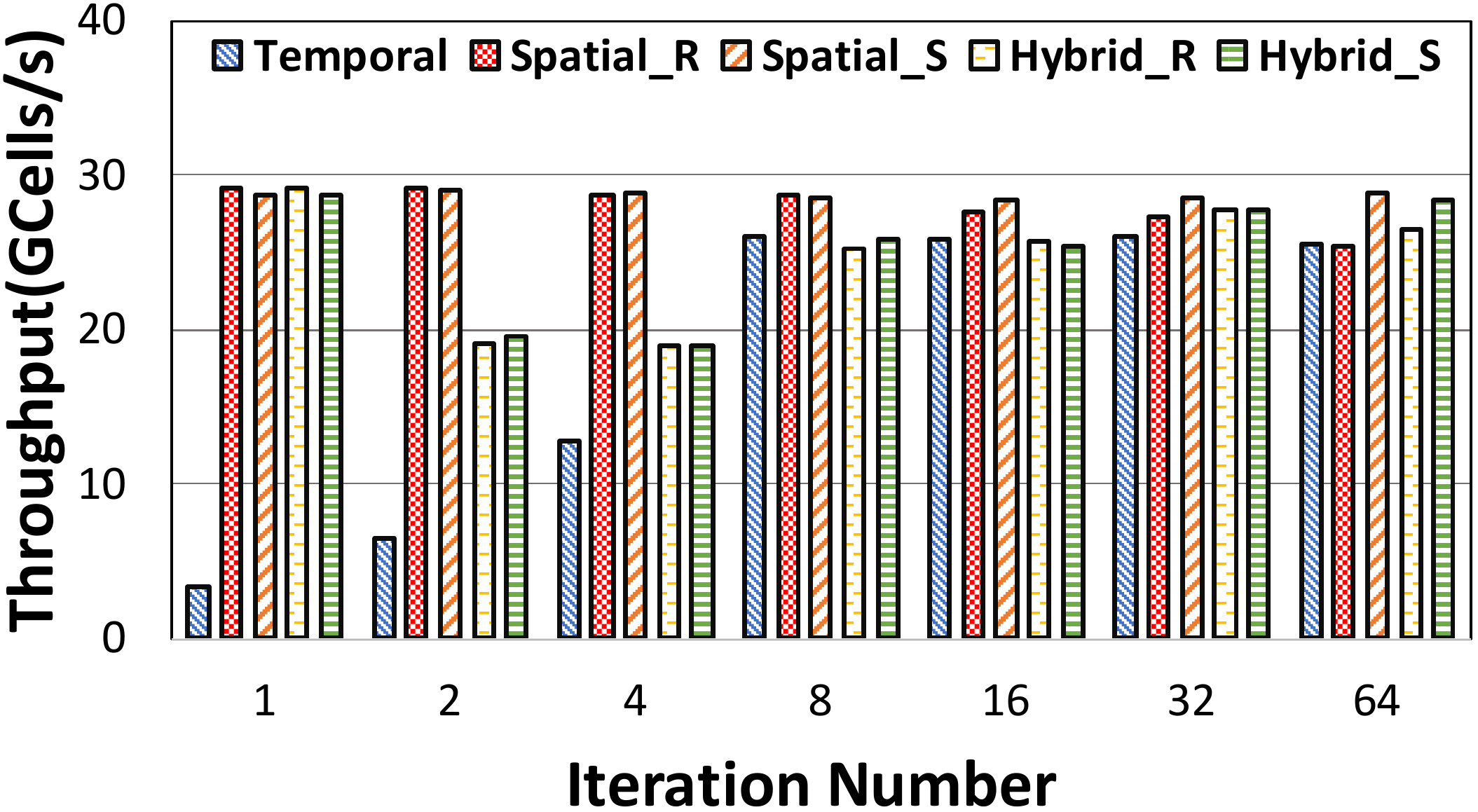}
        \caption{Throughput of HOTSPOT $9720 \times 1024$}
        \label{fig:f_9720}
    \end{subfigure}
    \begin{subfigure}[b]{0.445\textwidth}
        \centering
        \includegraphics[width=\textwidth]{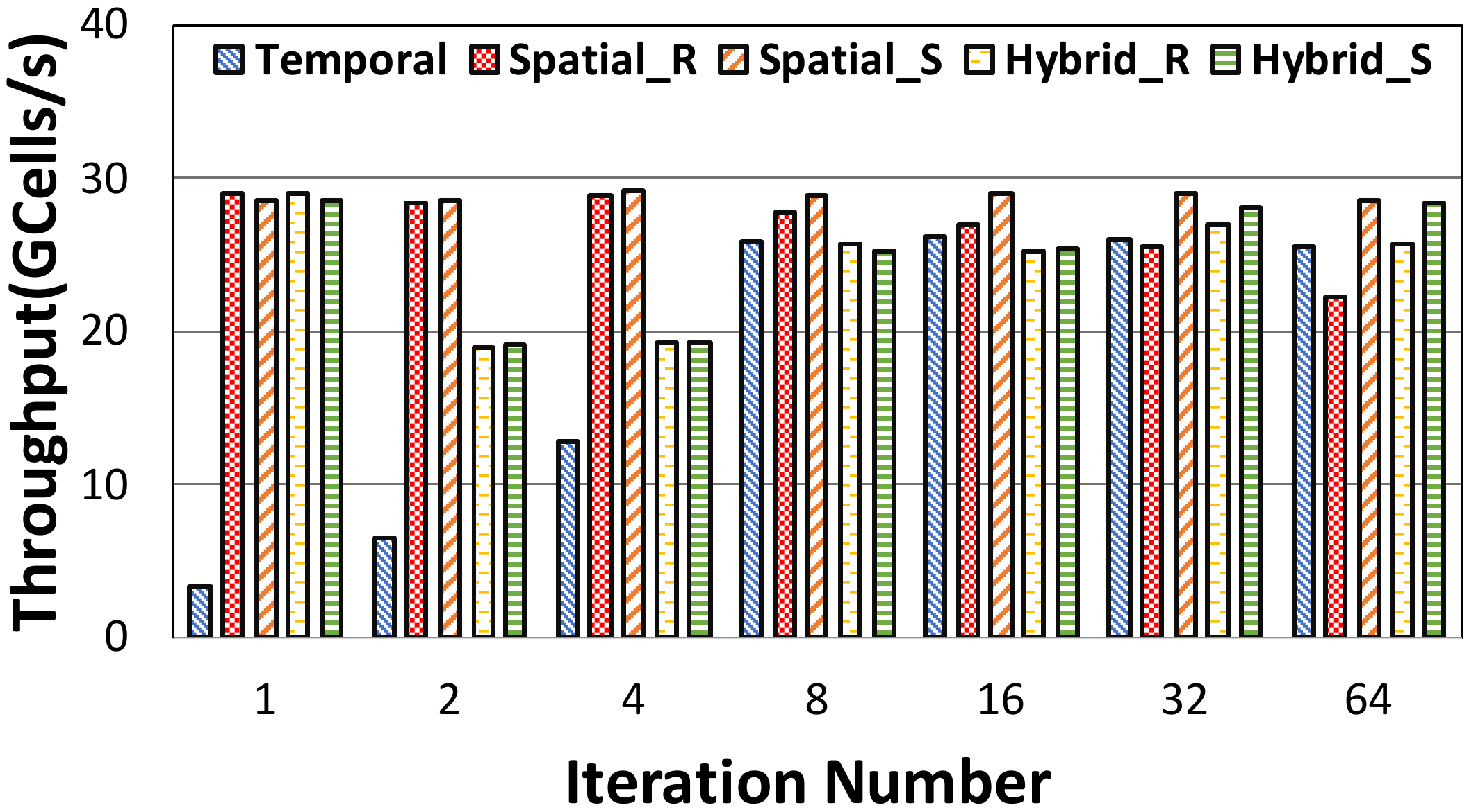}
        \caption{Throughput of HOTSPOT $4096 \times 4096$}
        \label{fig:f_4096}
    \end{subfigure}
    \caption{\rev{Throughput (GCell/s) comparison of different parallelism optimizations for HOTSPOT with the number of iterations changing from 1 to 64}} 
    \label{fig:throughput_d}
\end{figure}

\begin{figure}[h]
    \centering
    \begin{subfigure}[b]{0.447\textwidth}
        \centering
        \includegraphics[width=\textwidth]{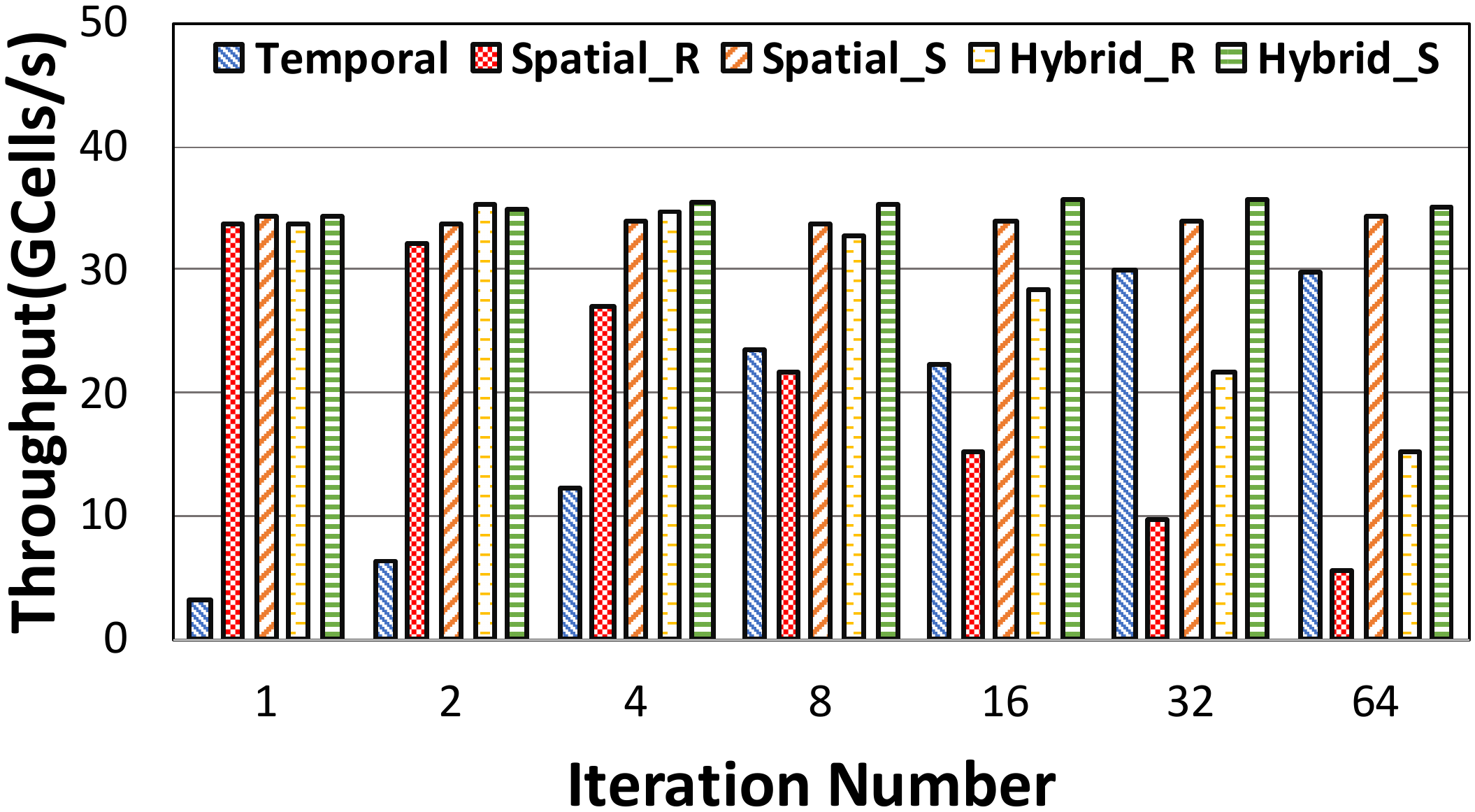}
        \caption{Throughput of HEAT3D $256 \times 16 \times 16$}
        \label{fig:g_256}
    \end{subfigure}
    \begin{subfigure}[b]{0.447\textwidth}
        \centering
        \includegraphics[width=\textwidth]{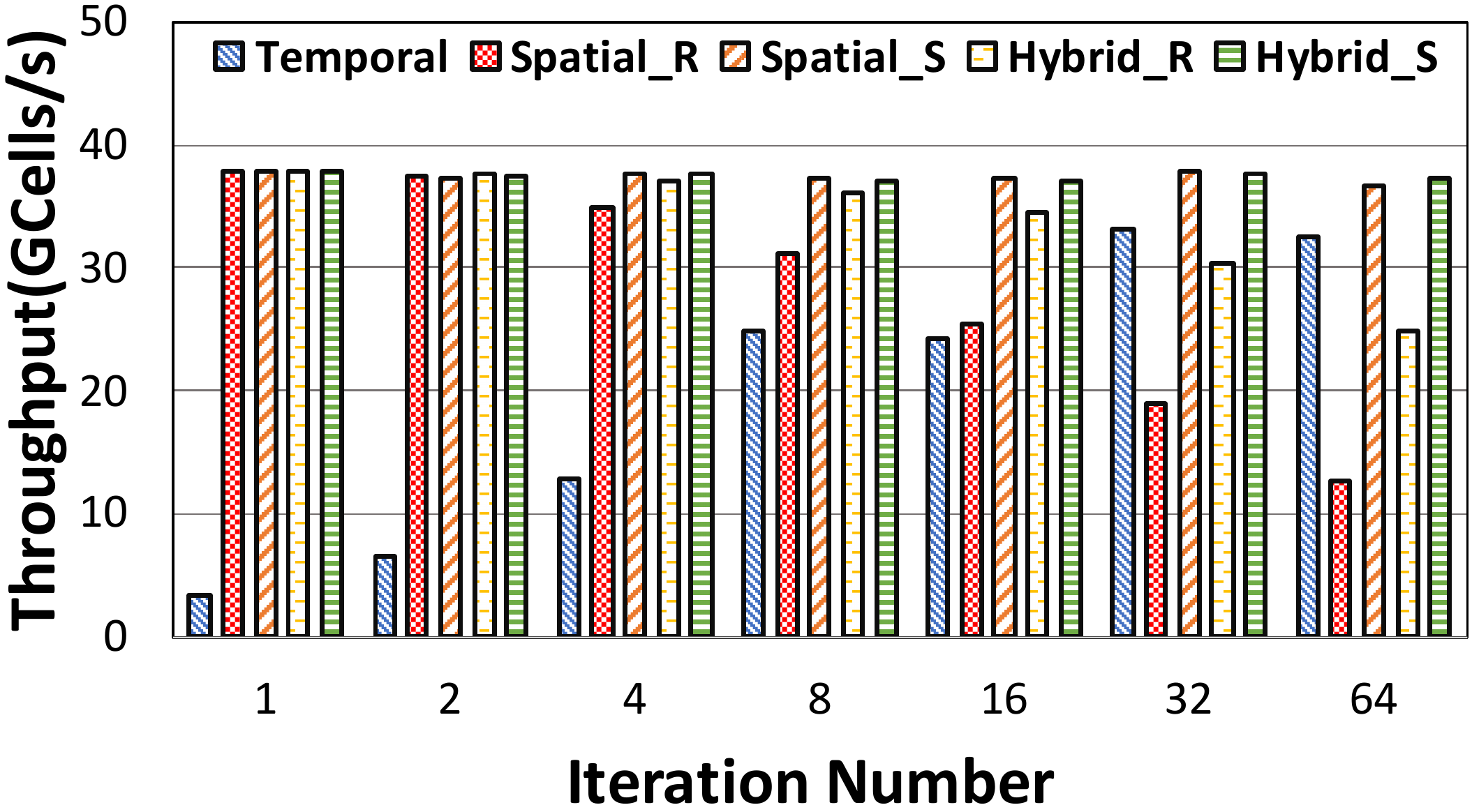}
        \caption{Throughput of HEAT3D $720 \times 32 \times 32$}
        \label{fig:g_720}
    \end{subfigure}
    \begin{subfigure}[b]{0.447\textwidth}
        \centering
        \includegraphics[width=\textwidth]{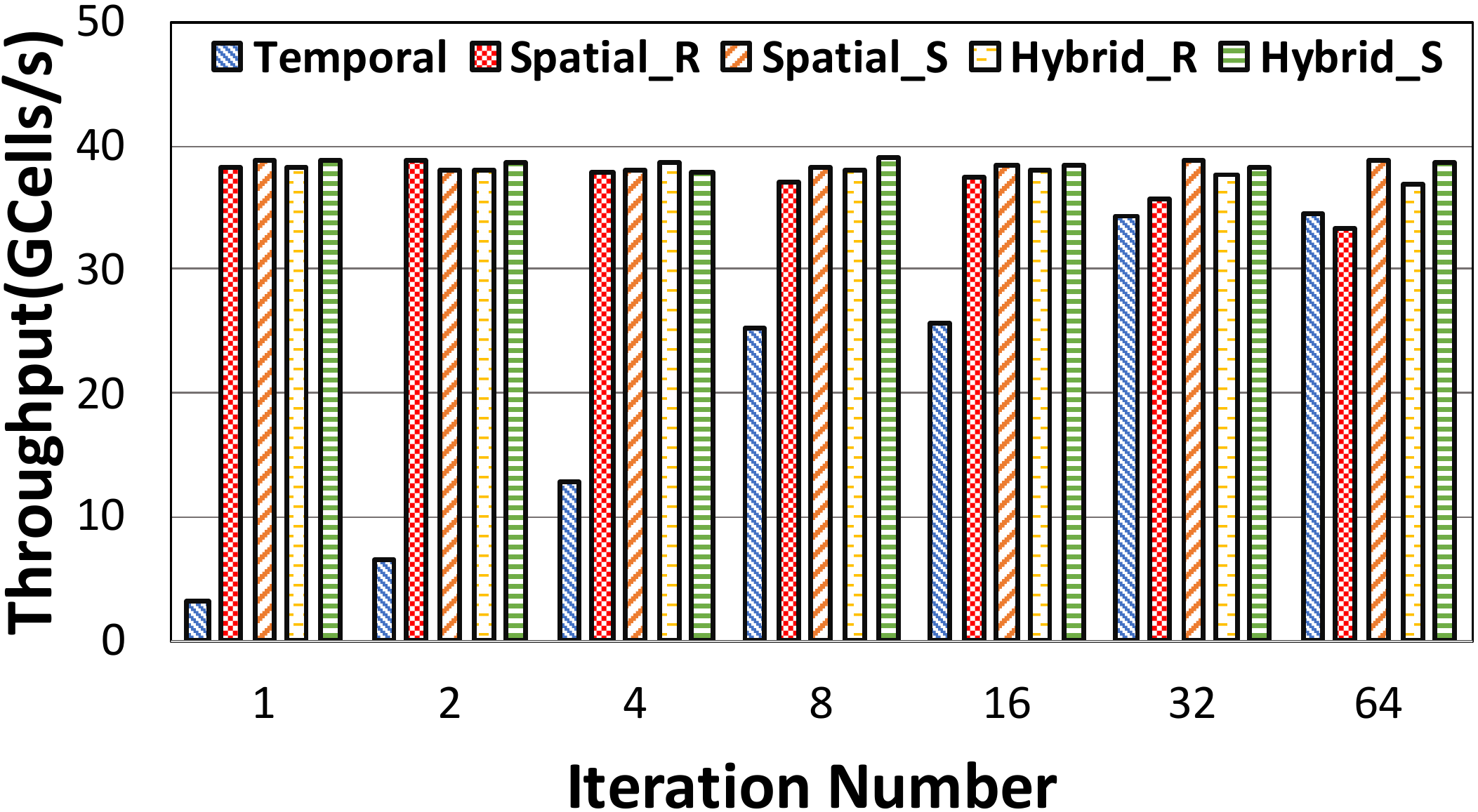}
        \caption{Throughput of HEAT3D $9720 \times 32 \times 32$}
        \label{fig:g_9720}
    \end{subfigure}
    \begin{subfigure}[b]{0.447\textwidth}
        \centering
        \includegraphics[width=\textwidth]{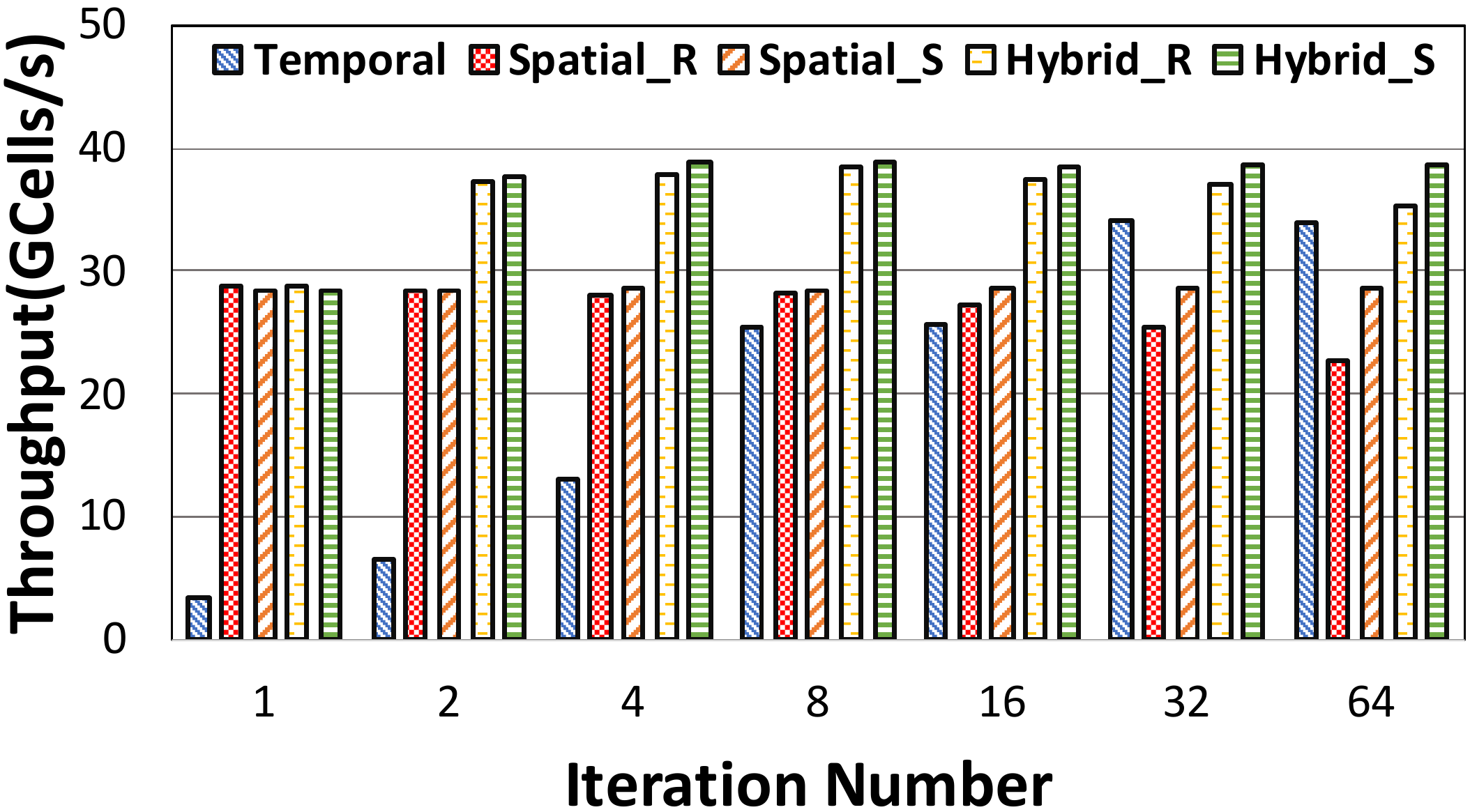}
        \caption{Throughput of HEAT3D $4096 \times 64 \times 64$}
        \label{fig:g_4096}
    \end{subfigure}
    \caption{\rev{Throughput (GCell/s) comparison of different parallelism optimizations for HEAT3D with the number of iterations changing from 1 to 64}} 
    \label{fig:throughput_e}
\end{figure}

\begin{figure}[h]
    \centering
    \begin{subfigure}[b]{0.446\textwidth}
        \centering
        \includegraphics[width=\textwidth]{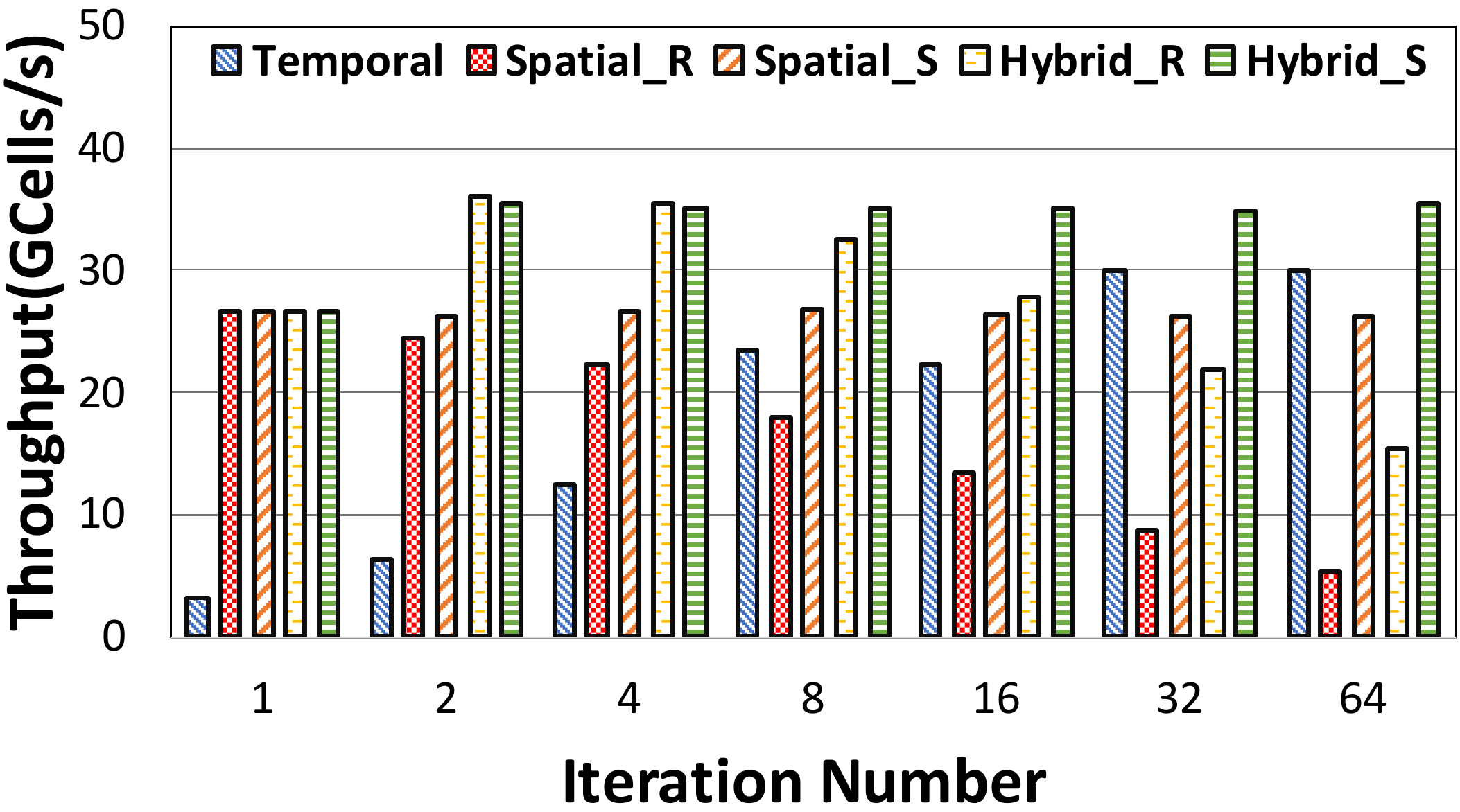}
        \caption{Throughput of SOBEL2D $256 \times 256$}
        \label{fig:h_256}
    \end{subfigure}
    \begin{subfigure}[b]{0.446\textwidth}
        \centering
        \includegraphics[width=\textwidth]{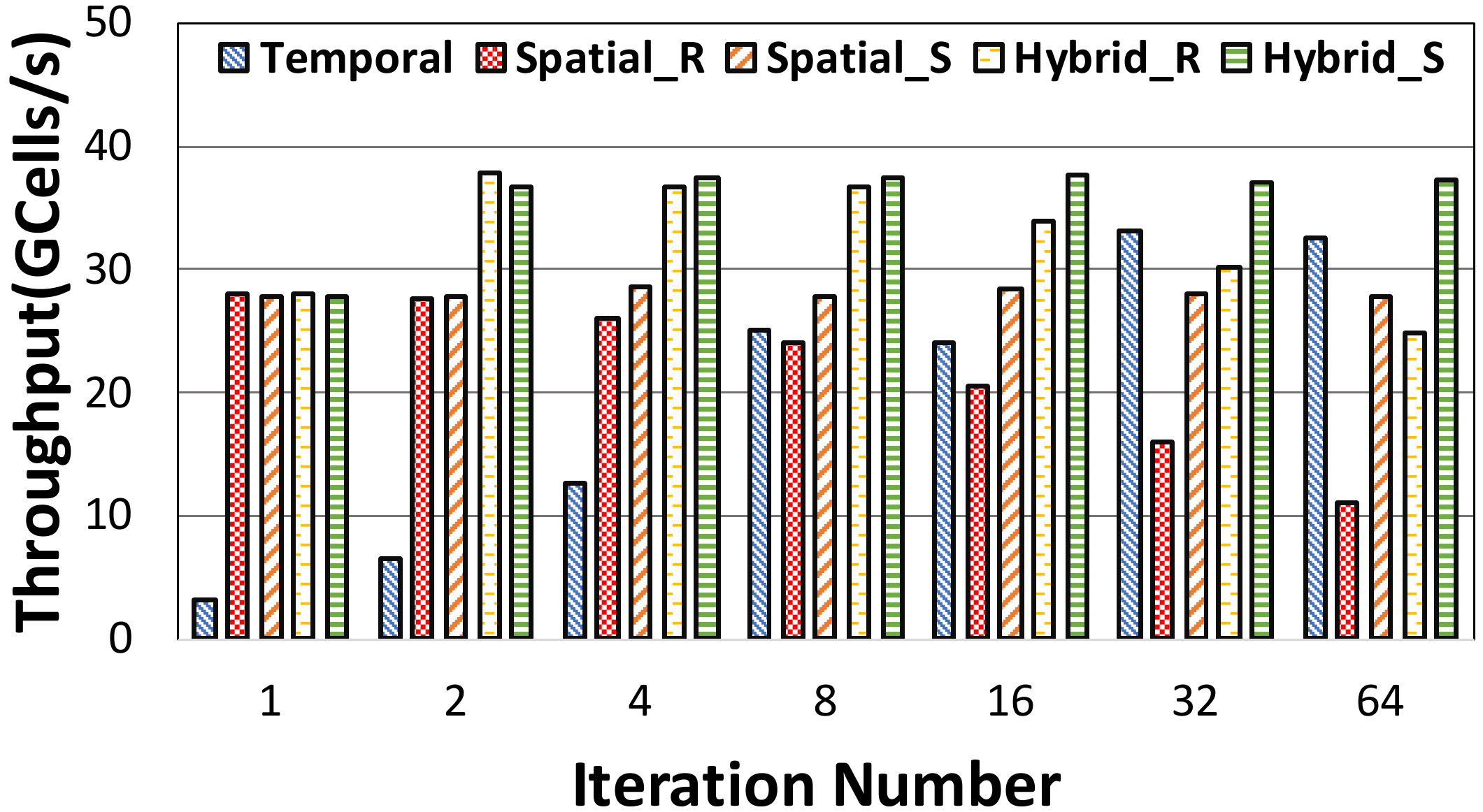}
        \caption{Throughput of SOBEL2D $720 \times 1024$}
        \label{fig:h_720}
    \end{subfigure}
    \begin{subfigure}[b]{0.446\textwidth}
        \centering
        \includegraphics[width=\textwidth]{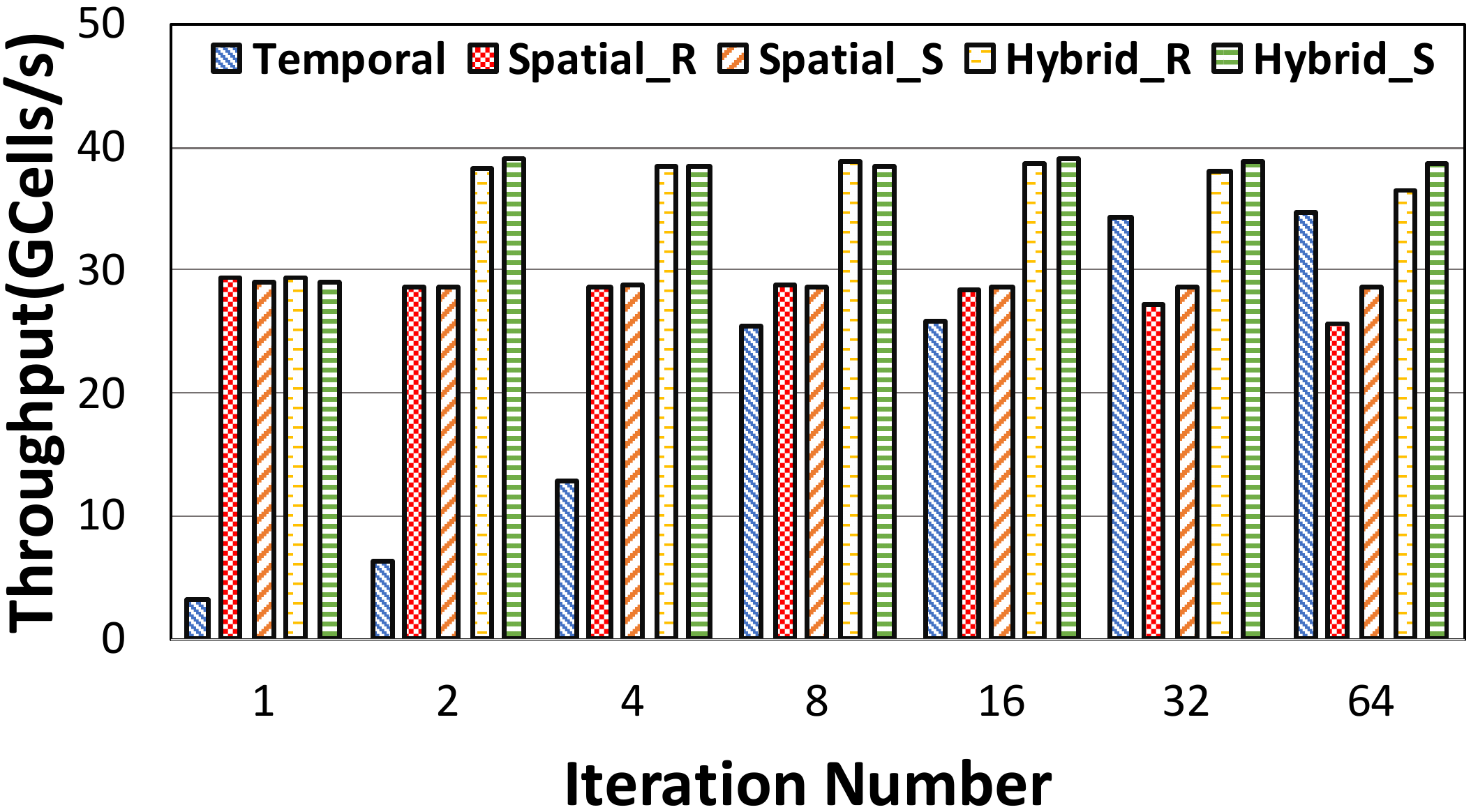}
        \caption{Throughput of SOBEL2D $9720 \times 1024$}
        \label{fig:h_9720}
    \end{subfigure}
    \begin{subfigure}[b]{0.446\textwidth}
        \centering
        \includegraphics[width=\textwidth]{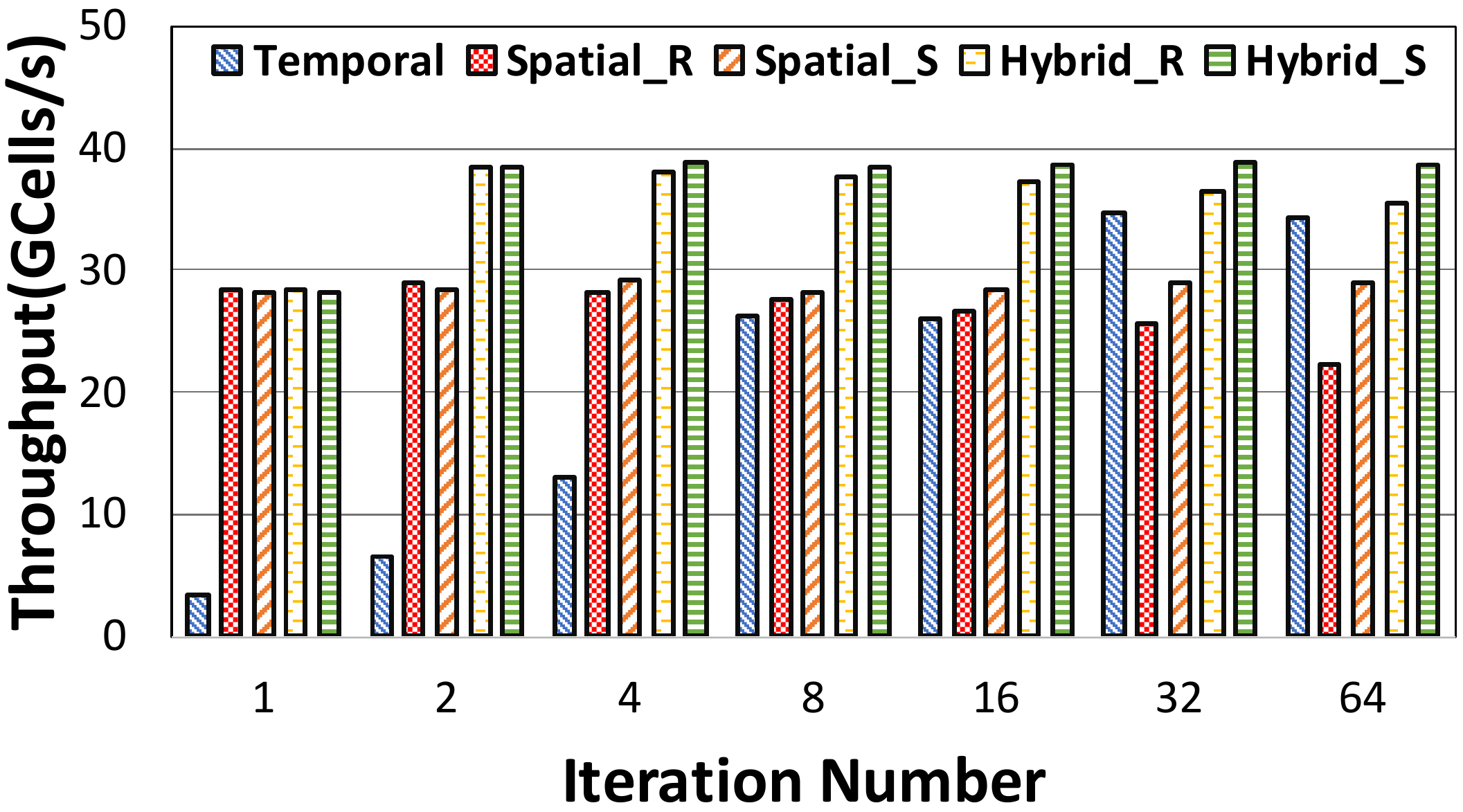}
        \caption{Throughput of SOBEL2D $4096 \times 4096$}
        \label{fig:h_4096}
    \end{subfigure}
    \caption{
\rev{Throughput (GCell/s) comparison of different parallelism optimizations for SOBEL2D with the number of iterations changing from 1 to 64}} 
    \label{fig:throughput_f}
\end{figure}

To evaluate the accuracy of our analytical performance model, we run a wide range of configurations, including different iteration numbers and different parallelism optimizations for each stencil kernel, and compare the model predicted execution time with the  actual measured time of on-board execution. Figure~\ref{fig:accuracy} shows the average (histogram), maximum (top bar), and minimum (bottom bar) error rates of our performance model for each stencil benchmark with different parallelism optimizations. For each histogram, the error rate is averaged across different numbers of stencil iterations from 1 to 64. For all configurations, our performance model has an error rate within 5\% in estimating the performance of our accelerator designs.

\subsubsection{Performance Results of Temporal Parallelism Designs} \label{subsubsec:perf_temporal}
% As shown in Figure~\ref{fig:throughput}, 
\rev{As shown in Figure~\ref{fig:throughput_a} to Figure~\ref{fig:throughput_h}}, the performance of the temporal parallelism designs generally increases with the iteration number, as more stencil iterations are concurrently processed on the FPGA in a dataflow fashion.
This linear performance improvement trend stops when we could not instantiate more temporal stages (i.e., stencil iterations) on the FPGA. For most benchmarks, their maximum number of PEs in temporal parallelism designs are between 
% 8 and 12 
\rev{9 to 15} when their iteration number is large enough, as shown in \rev{Figure 18 to 20}.  %Table~\ref{fig:penumber_64}. 
Therefore, their throughput increases linearly as the iteration number grows from 1 to 8. 
\rev{The two exceptions are JACOBI2D and DILATE. Their linear throughput increase is achieved when iteration ranges from 1 to 16 since their maximum PE numbers are 21 and 18, respectively.}
% The two exceptions are HOTSPOT and SOBEL2D; this linear throughput increase is achieved when the iteration ranges from 1 to 4 since their maximum PE number is less than 8.

When the iteration number is larger than the maximum number of PEs, this performance does not improve linearly with the iteration number; 
%For example, in JACOBI2D and BLUR, their maximum PE counts are 12 and 9, respectively.
the performance is mainly decided by the ratio of iteration number and rounds of FPGA kernel execution. 
% For example, in JACOBI2D, 
\rev{For example, in BLUR,} when the iteration number is 32 and 16, respectively, the maximum number of PEs is 12 in both cases; therefore, the numbers of FPGA kernel runs are 3 and 2, respectively. 
While the work to be done is increased by twice from iteration number 16 to 32, the execution time is only increased by $3/2 = 1.5\times$.
In this way, the throughput of iteration number 32 is larger than that of iteration number 16.

% When the iteration number is larger than the maximum number of PEs, this performance does not improve linearly with the iteration number; 
% %For example, in JACOBI2D and BLUR, their maximum PE counts are 12 and 9, respectively.
% the performance is mainly decided by the ratio of iteration number and maximum number of PEs, which decides the PE underutilization rate in the last round of FPGA kernel execution. For example, in JACOBI2D, comparing when the iteration number is 32 and 16, respectively, the maximum number of PEs is 12 in both cases; therefore, the numbers of FPGA kernel runs are 3 and 2, respectively. In the prior case, for the last round, there are $12 \times 3 - 32 = 4$ PEs idle; in the latter case, for the last round, there are $12 \times 2 - 16 = 8$ PEs idle.
% In this way, the throughput of iteration number 32 is larger than that of iteration number 16.

\subsubsection{Performance Results of Spatial Parallelism Designs} \label{subsubsec:perf_spatial}
\begin{figure}[h]
    \centering
    \begin{subfigure}[b]{0.447\textwidth}
        \centering
        \includegraphics[width=\textwidth]{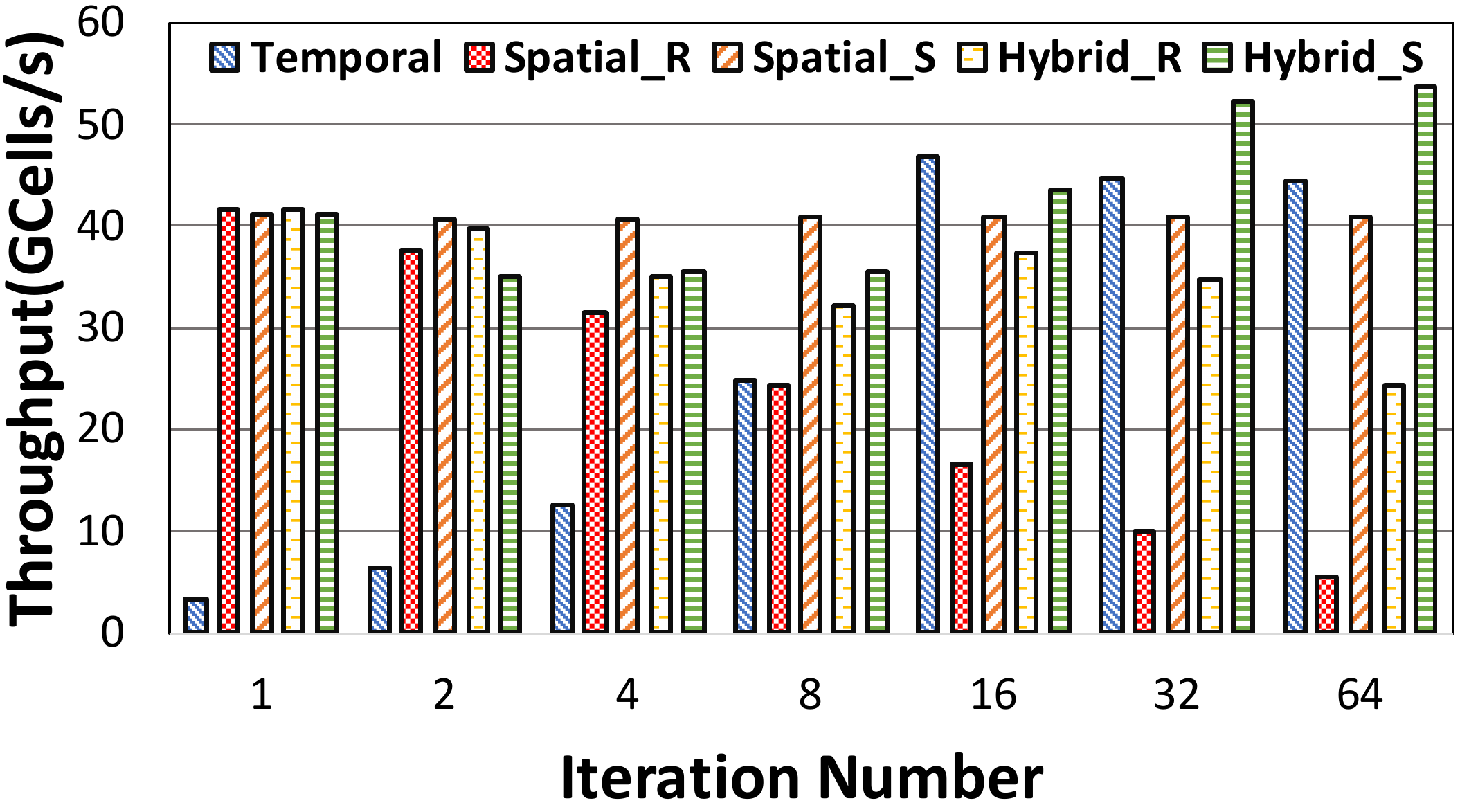}
        \caption{Throughput of JACOBI2D $256 \times 256$}
        \label{fig:a_256}
    \end{subfigure}
    \begin{subfigure}[b]{0.447\textwidth}
        \centering
        \includegraphics[width=\textwidth]{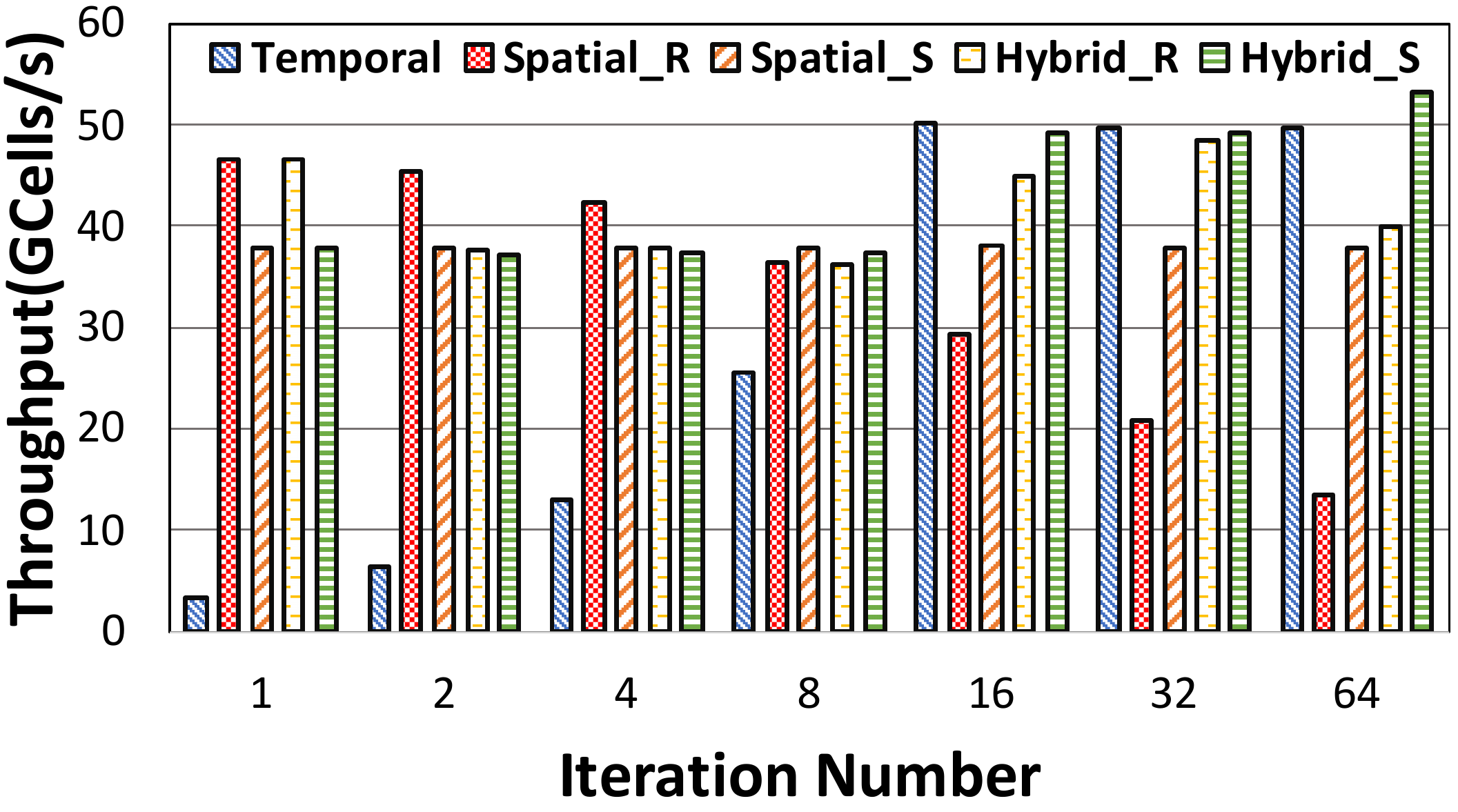}
        \caption{Throughput of JACOBI2D $720 \times 1024$}
        \label{fig:a_720}
    \end{subfigure}
    \begin{subfigure}[b]{0.447\textwidth}
        \centering
        \includegraphics[width=\textwidth]{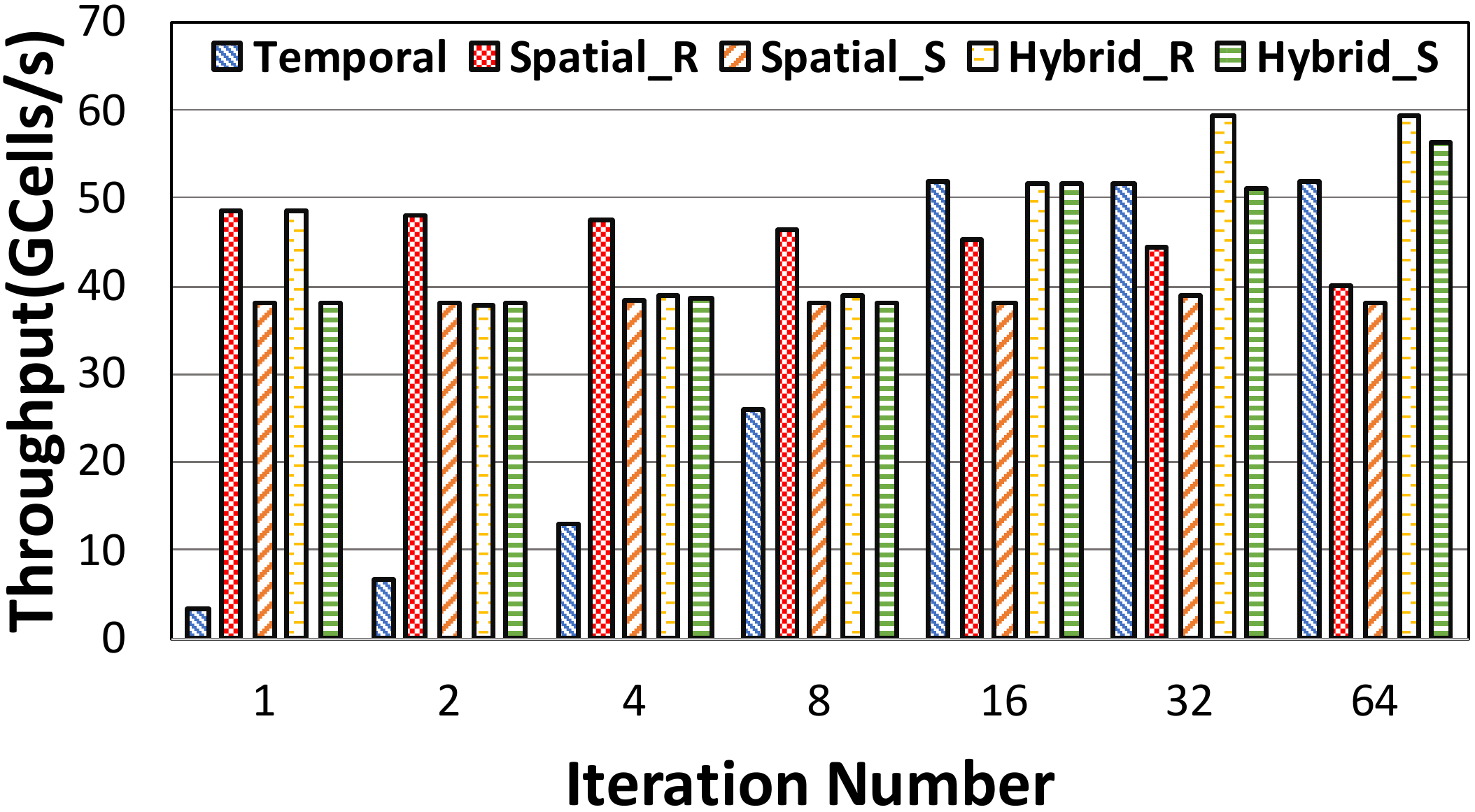}
        \caption{Throughput of JACOBI2D $9720 \times 1024$}
        \label{fig:a_9720}
    \end{subfigure}
    \begin{subfigure}[b]{0.447\textwidth}
        \centering
        \includegraphics[width=\textwidth]{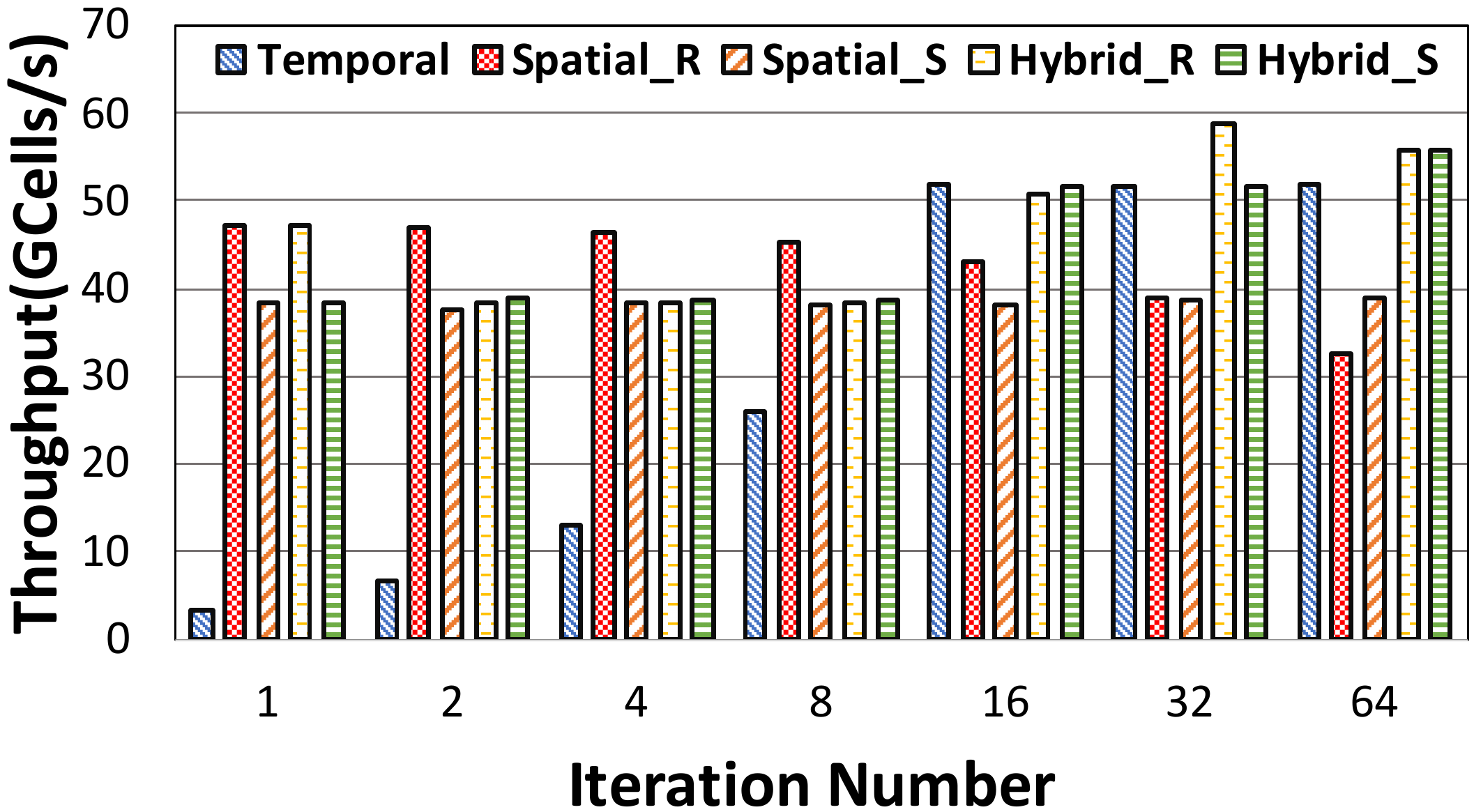}
        \caption{Throughput of JACOBI2D $4096 \times 4096$}
        \label{fig:a_4096}
    \end{subfigure}
    \caption{\rev{Throughput (GCell/s) comparison of different parallelism optimizations for JACOBI2D with the number of iterations changing from 1 to 64}}
    \label{fig:throughput_g}
\end{figure}

\begin{figure}[h]
    \centering
    \begin{subfigure}[b]{0.447\textwidth}
        \centering
        \includegraphics[width=\textwidth]{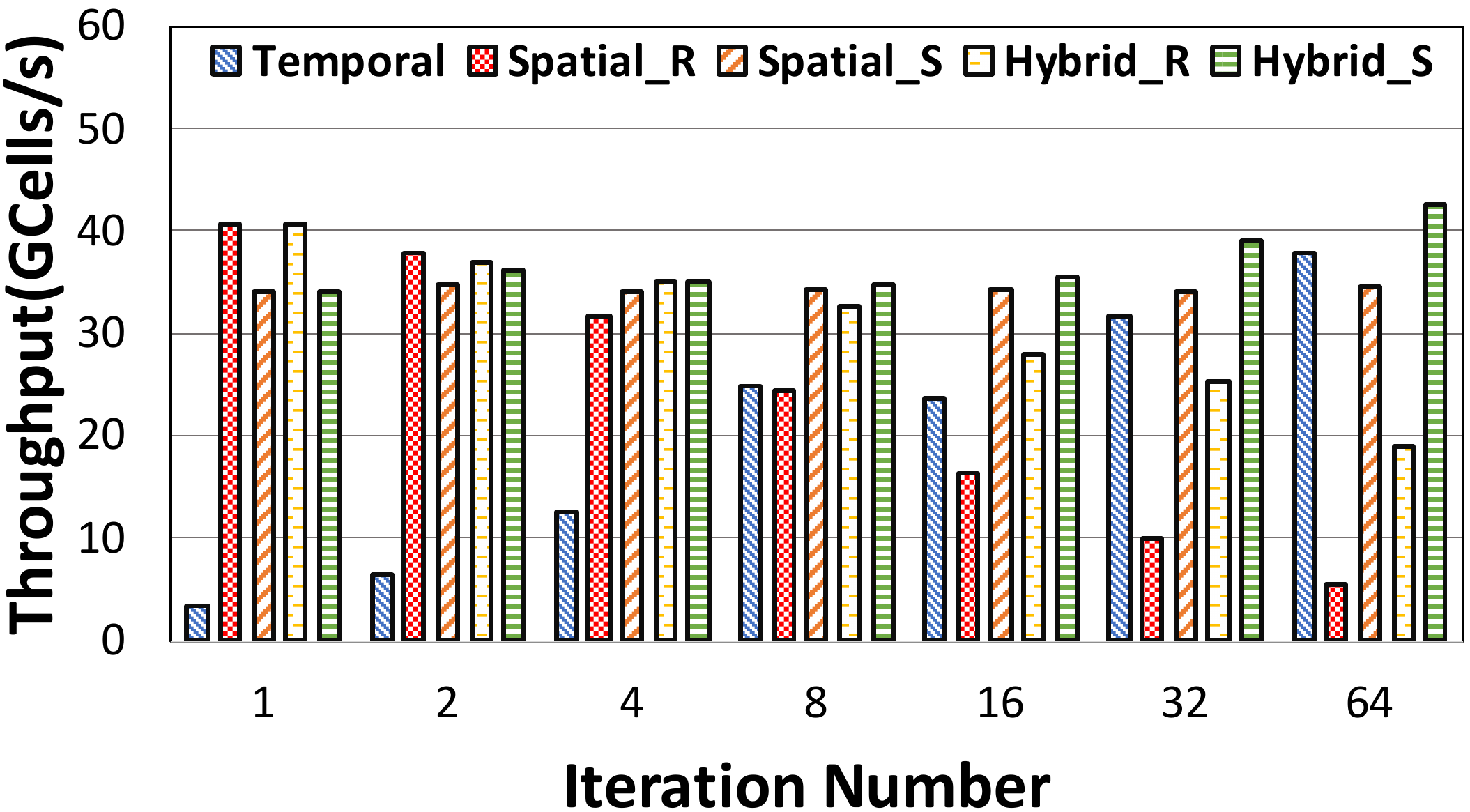}
        \caption{Throughput of JACOBI3D $256 \times 16 \times 16$}
        \label{fig:b_256}
    \end{subfigure}
    \begin{subfigure}[b]{0.447\textwidth}
        \centering
        \includegraphics[width=\textwidth]{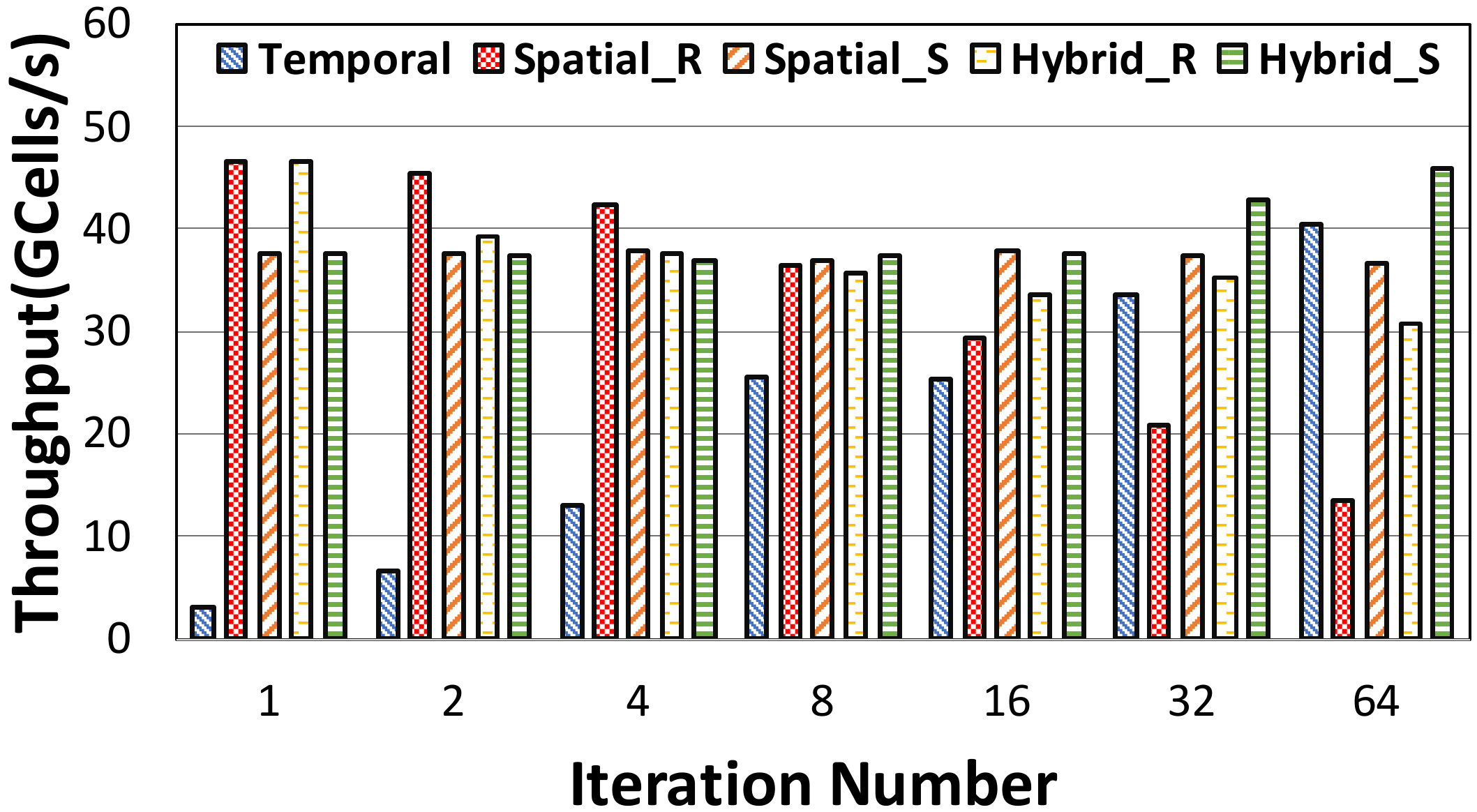}
        \caption{Throughput of JACOBI3D $720 \times 32 \times 32$}
        \label{fig:b_720}
    \end{subfigure}
    \begin{subfigure}[b]{0.447\textwidth}
        \centering
        \includegraphics[width=\textwidth]{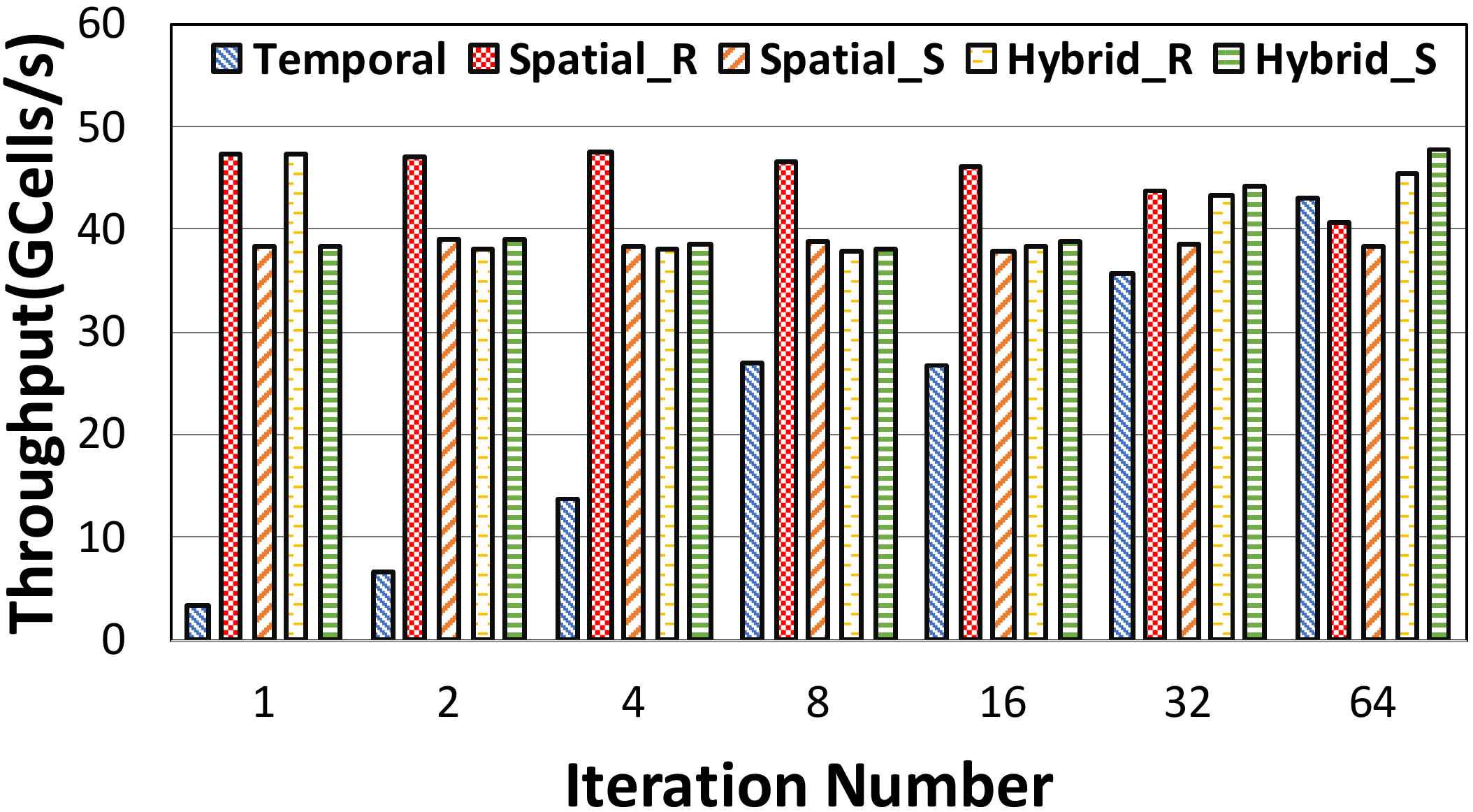}
        \caption{Throughput of JACOBI3D $9720 \times 32 \times 32$}
        \label{fig:b_9720}
    \end{subfigure}
    \begin{subfigure}[b]{0.447\textwidth}
        \centering
        \includegraphics[width=\textwidth]{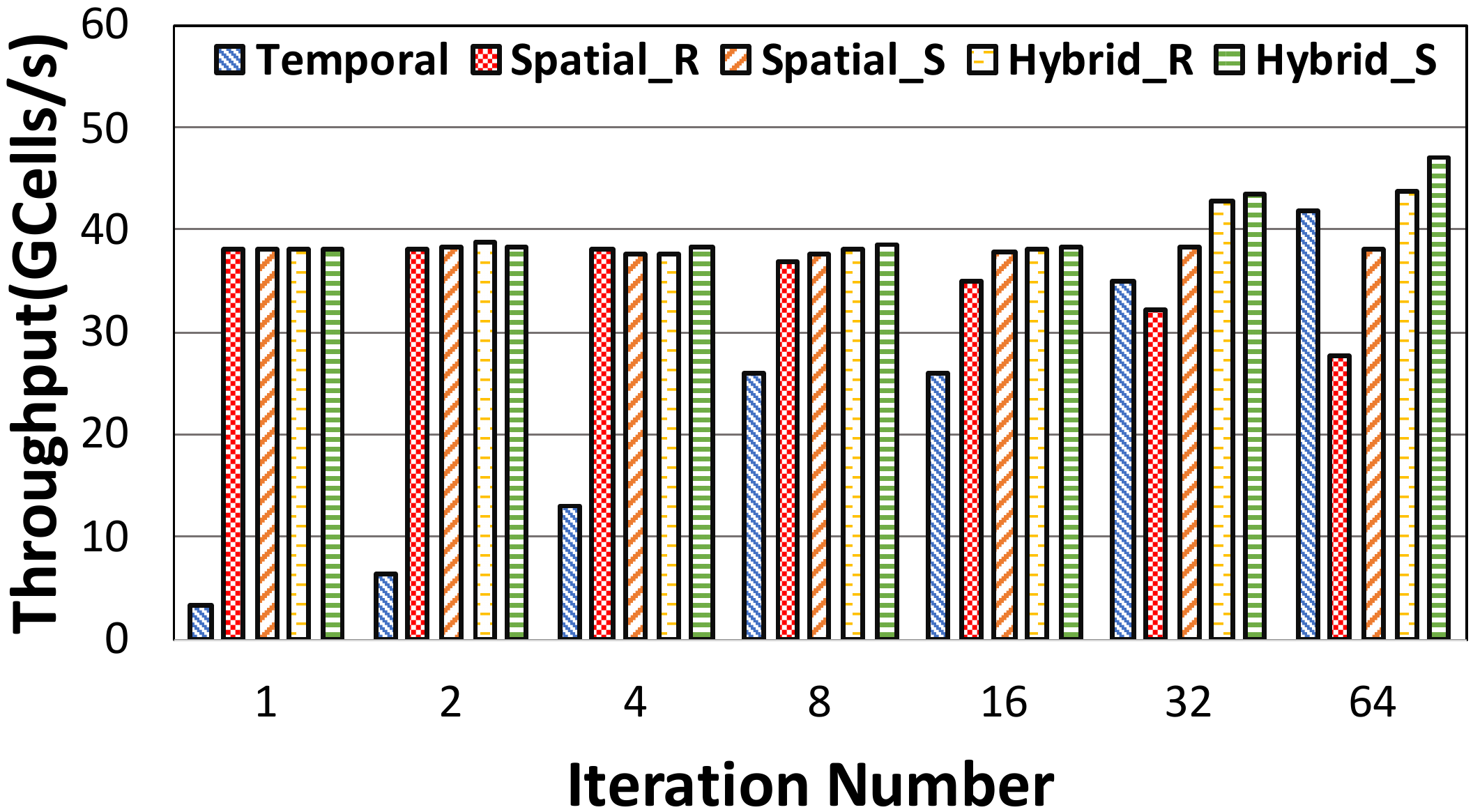}
        \caption{Throughput of JACOBI3D $4096 \times 64 \times 64$}
        \label{fig:b_4096}
    \end{subfigure}
    \caption{\rev{Throughput (GCell/s) comparison of different parallelism optimizations for JACOBI3D with the number of iterations changing from 1 to 64}}
    \label{fig:throughput_h}
\end{figure}

% \xy{0.Accuracy analysis. 1. Parallelism itself 2.Different benchmarks 3. Different Iteration Num}
To better understand the performance difference between the two spatial parallelism design variants presented in Section~\ref{subsec:spatial_parallelism}, $Spatial\_R$ and $Spatial\_S$,  we further analyze the performance trend of these two designs at different iteration numbers\rev{, input sizes} and stencil kernels.
%shows the performance throughput in $GCells/s$ of different parallelism designs among our evaluated stencil benchmarks at different iteration numbers. 
As shown in \rev{Figure~\ref{fig:throughput_a} to Figure~\ref{fig:throughput_h}}, for the $Spatial\_R$ design, its performance generally decreases as the iteration number increases. This is mainly due to the increase of the halo data processing as the iteration number increases. \rev{The performance decrease is worse on smaller input sizes as halo data increase has more significant impact on smaller input sizes. For example, the throughput of $Spatial\_R$ drops faster at $256 \times 256$ and $720 \times 1024$ input sizes in JACOBI2D compared with $9720 \times 1024$ and $4096 \times 4096$ input sizes.} On the other hand, the performance of the $Spatial\_S$ design does not vary with the iteration number. This is because the amount of halo data exchange remains the same as the iteration number increases. These trends align with our performance model in Equations~\ref{eq:spatial_R_model} and~\ref{eq:spatial_S_model}, respectively.

Comparing between these two design variants, when the iteration number is low (i.e., less than 4) and with the same number of PEs, \rev{as shown in Figure~\ref{fig:throughput_a} to~\ref{fig:throughput_f},} $Spatial\_R$ and $Spatial\_S$ achieve about the same throughput. And as the iteration number increases, the $Spatial\_S$ design can maintain its performance and \rev{outperforms the $Spatial\_R$ design especially for smaller input sizes.}
% (slightly) outperforms the $Spatial\_R$ design. 
\rev{A few exceptions are, JACOBI2D (when input size is $720\times1024$, $9720\times1024$ and $4096\times4096$) and JACOBI3D (when input size is $720\times32\times32$, $9720\times32\times32$ and $4096\times64\times64$)},  the $Spatial\_R$ design achieves a better throughput than the $Spatial\_S$ design as $Spatial\_R$ can place more PEs.
This is because,
\rev{border streaming based approach consumes slightly more wires due to the streaming connections than redundant computation based approach to implement border streaming, which affects timing closure, especially when the increase of corss-SLR (i.e., cross-die) connections is approaching FPGA board limit.}
% based on our experiments, occasionally the $Spatial\_S$ design may not be able to instantiate the as many PEs as the $Spatial\_R$ design to satisfy the timing closure requirement.
% Figure~\ref{fig:penumber_256} to~\ref{fig:penumber_4096} 
% % \xy{Update this} 
% shows the maximum number of PEs that we could instantiate for the different parallelism designs when the iteration number is 64 and 2. For the $Spatial\_S$ design in JACOBI2D and JACOBI3D, 
% the additional routing delay from the streaming connections between PEs (which are placed on different SLRs on the U280 FPGA) caused the timing closure issue, and thus less number of PEs could be placed compared to the $Spatial\_R$ design. 

\subsubsection{Performance Results of Hybrid Parallelism Designs} \label{subsubsec:perf_hybrid}
In hybrid parallelism designs, both temporal and spatial parallelisms are exploited. 
The performance from $Hybrid\_R$ and $Hybrid\_S$ parallelism designs reflects a combination of trend from both the temporal parallelism design as described in Section~\ref{subsubsec:perf_temporal} and the spatial designs discussed in Section~\ref{subsubsec:perf_spatial}.

\begin{figure}[h]
    \begin{subfigure}[b]{0.48\textwidth}
        \centering
        \includegraphics[width=\textwidth]{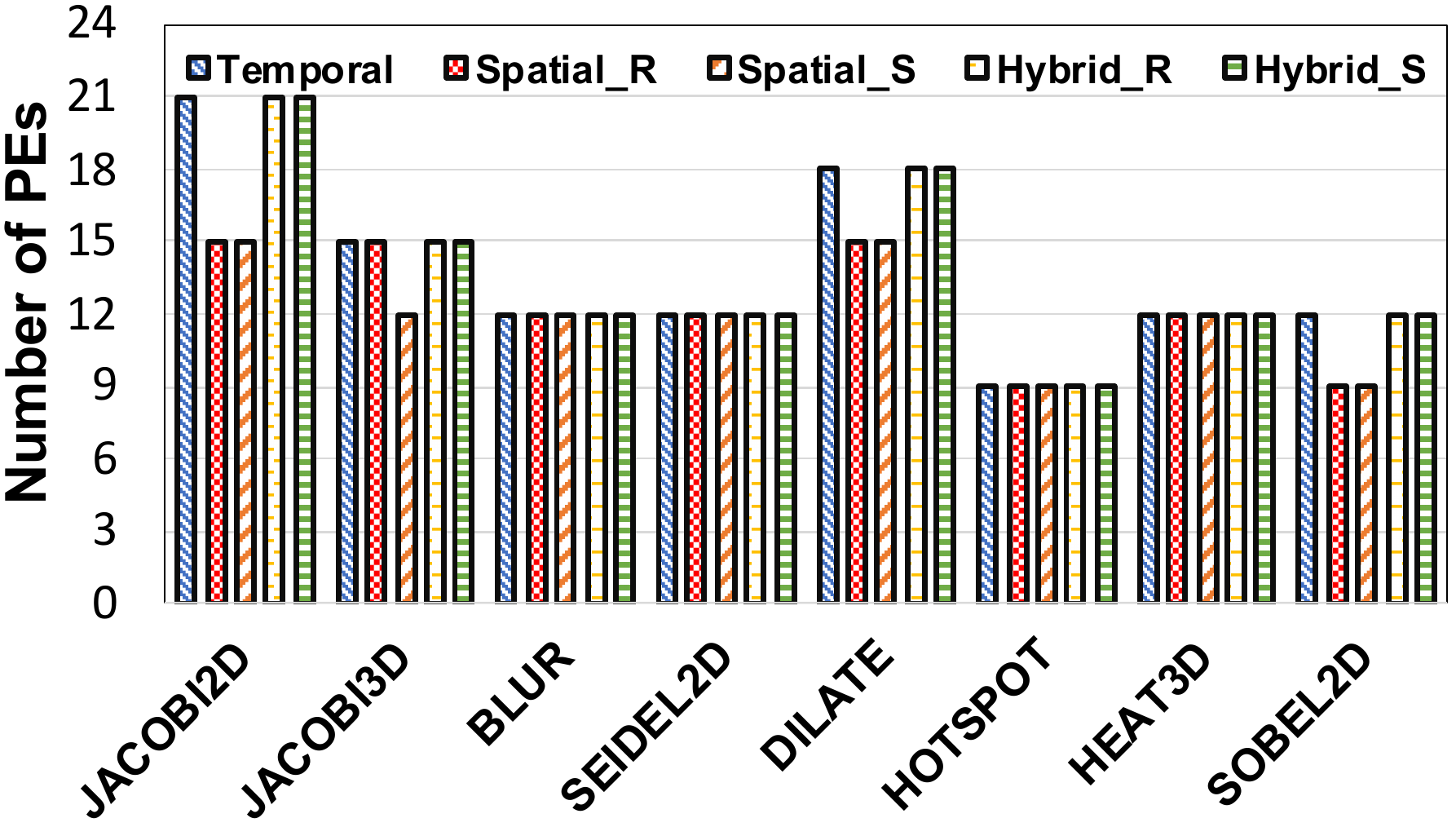}
        \caption{Number of PEs  with the number of iteration = 64}
        \label{fig:penumber_256_64}
    \end{subfigure}
    \hspace{+0.05in}
    \begin{subfigure}[b]{0.48\textwidth}
        \centering
        \includegraphics[width=\textwidth]{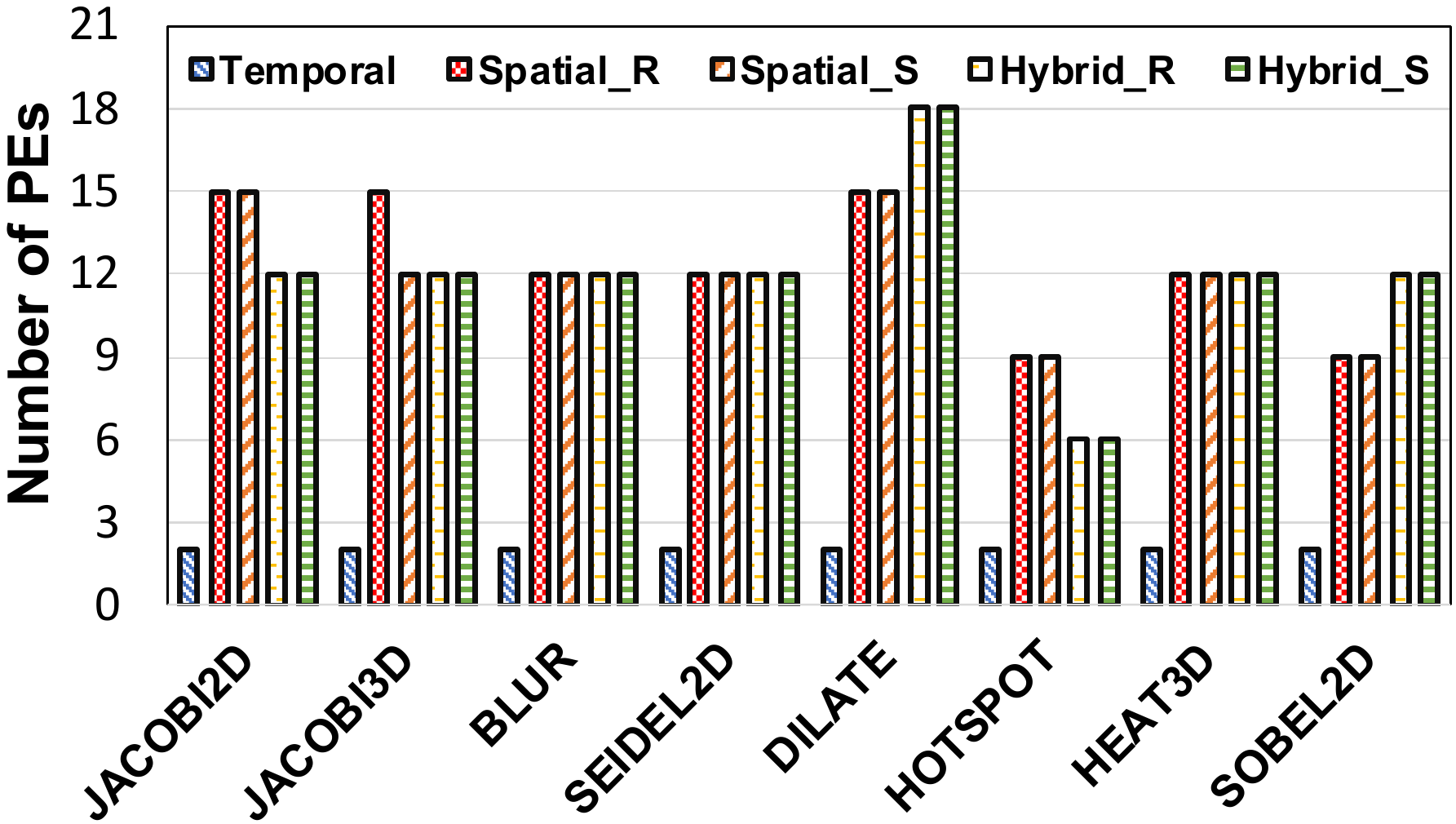}
        \caption{Number of PEs with the number of iteration = 2}
        \label{fig:penumber_256_2}
    \end{subfigure}
    
    \caption{\rev{Total number of PEs for different parallelisms on Alveo U280 with column size = 256}} %\zf{update this figure, enlarge font size}\xy{Updated}}
    \label{fig:penumber_256}
    \vspace{-0.05in}
\end{figure}

\begin{figure}[h]
    \begin{subfigure}[b]{0.48\textwidth}
        \centering
        \includegraphics[width=\textwidth]{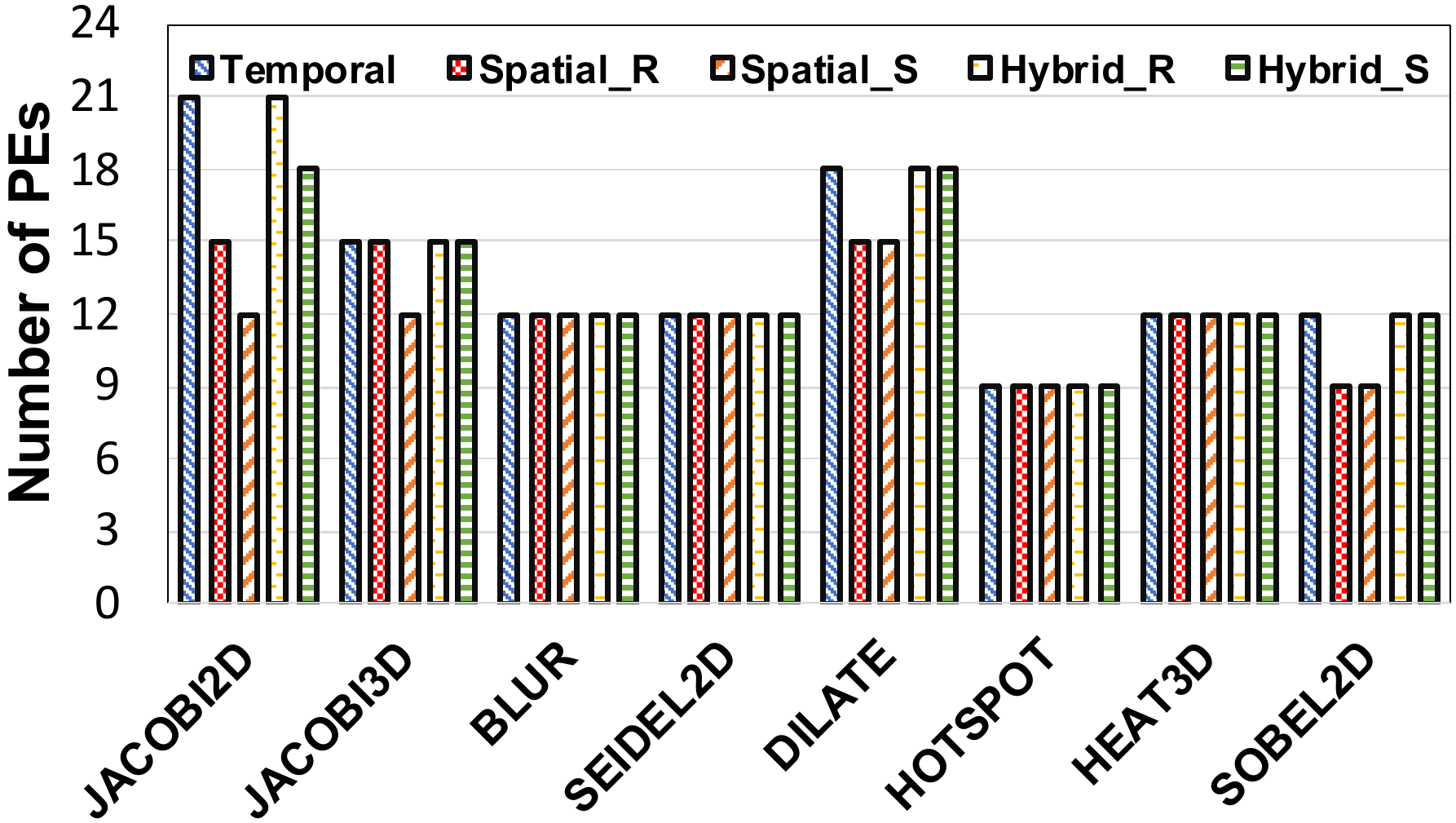}
        \caption{Number of PEs  with the number of iteration = 64}
        \label{fig:penumber_1024_64}
    \end{subfigure}
    \hspace{+0.05in}
    \begin{subfigure}[b]{0.48\textwidth}
        \centering
        \includegraphics[width=\textwidth]{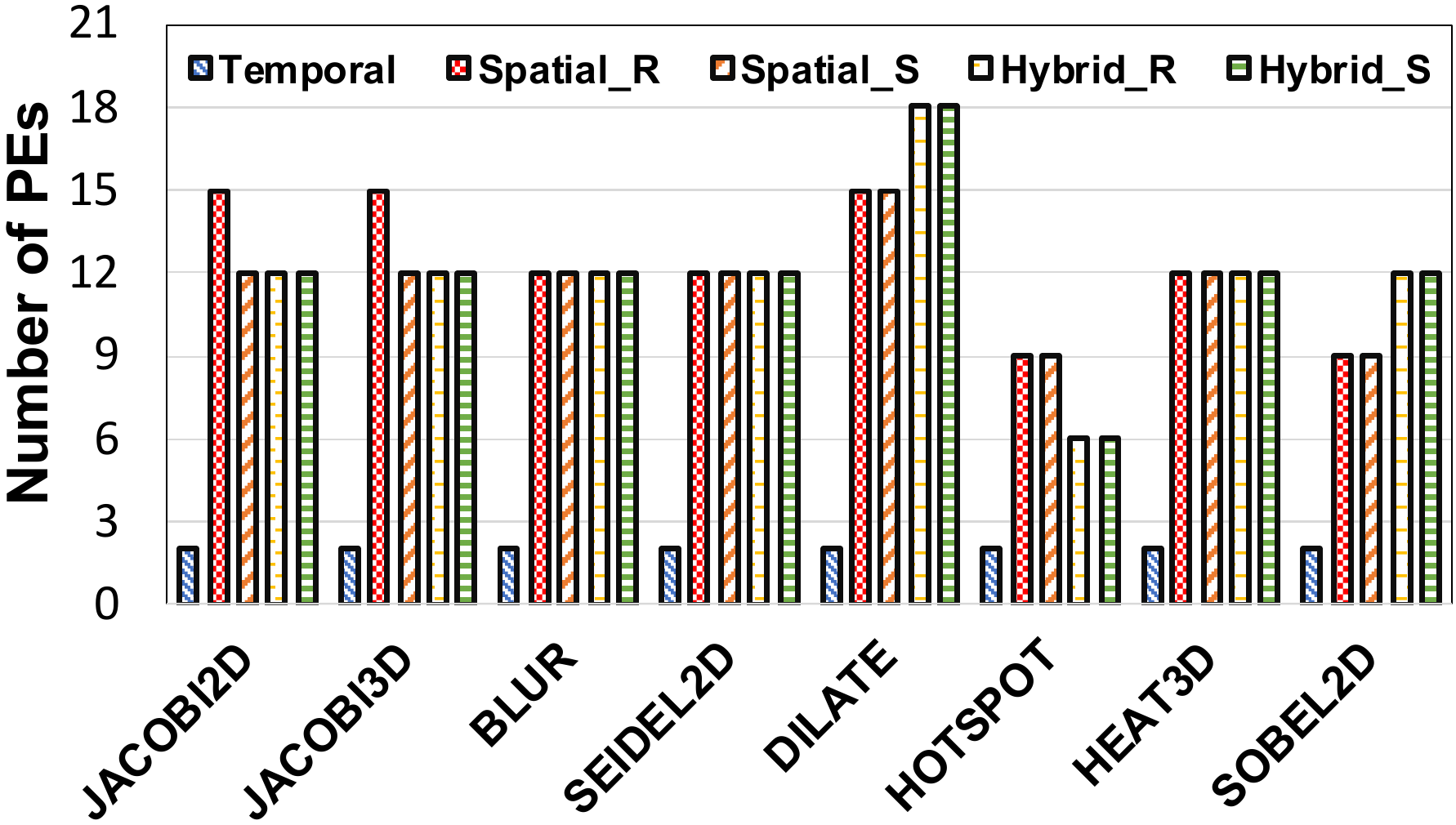}
        \caption{Number of PEs with the number of iteration = 2}
        \label{fig:penumber_1024_2}
    \end{subfigure}
    
    \caption{\rev{Total number of PEs for different parallelisms on Alveo U280 with column size = 1024}} %\zf{update this figure, enlarge font size}\xy{Updated}}
    \label{fig:penumber_1024}
    \vspace{-0.05in}
\end{figure}

\begin{figure}[h]
    \begin{subfigure}[b]{0.48\textwidth}
        \centering
        \includegraphics[width=\textwidth]{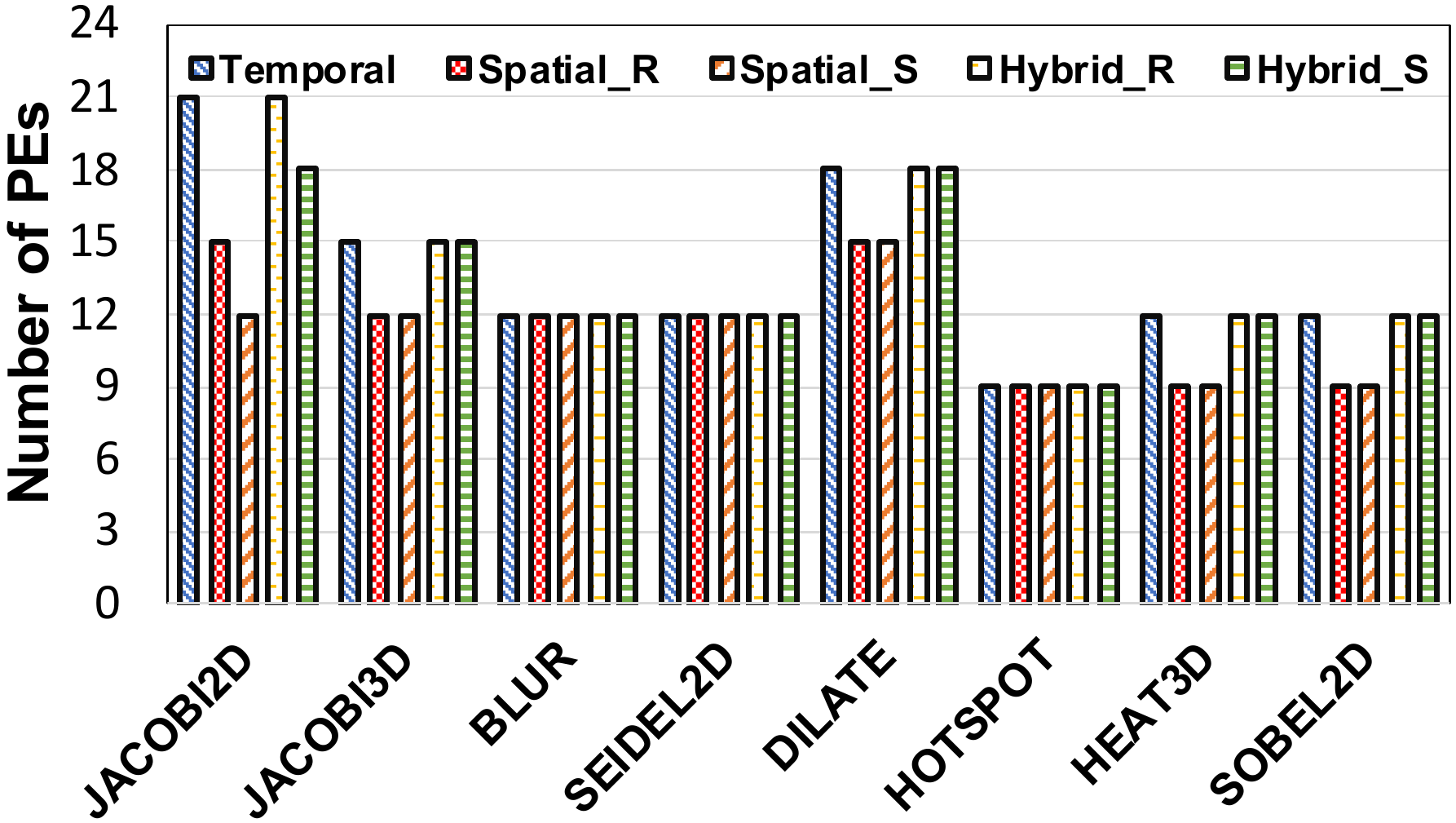}
        \caption{Number of PEs  with the number of iteration = 64}
        \label{fig:penumber_4096_64}
    \end{subfigure}
    \hspace{+0.05in}
    \begin{subfigure}[b]{0.48\textwidth}
        \centering
        \includegraphics[width=\textwidth]{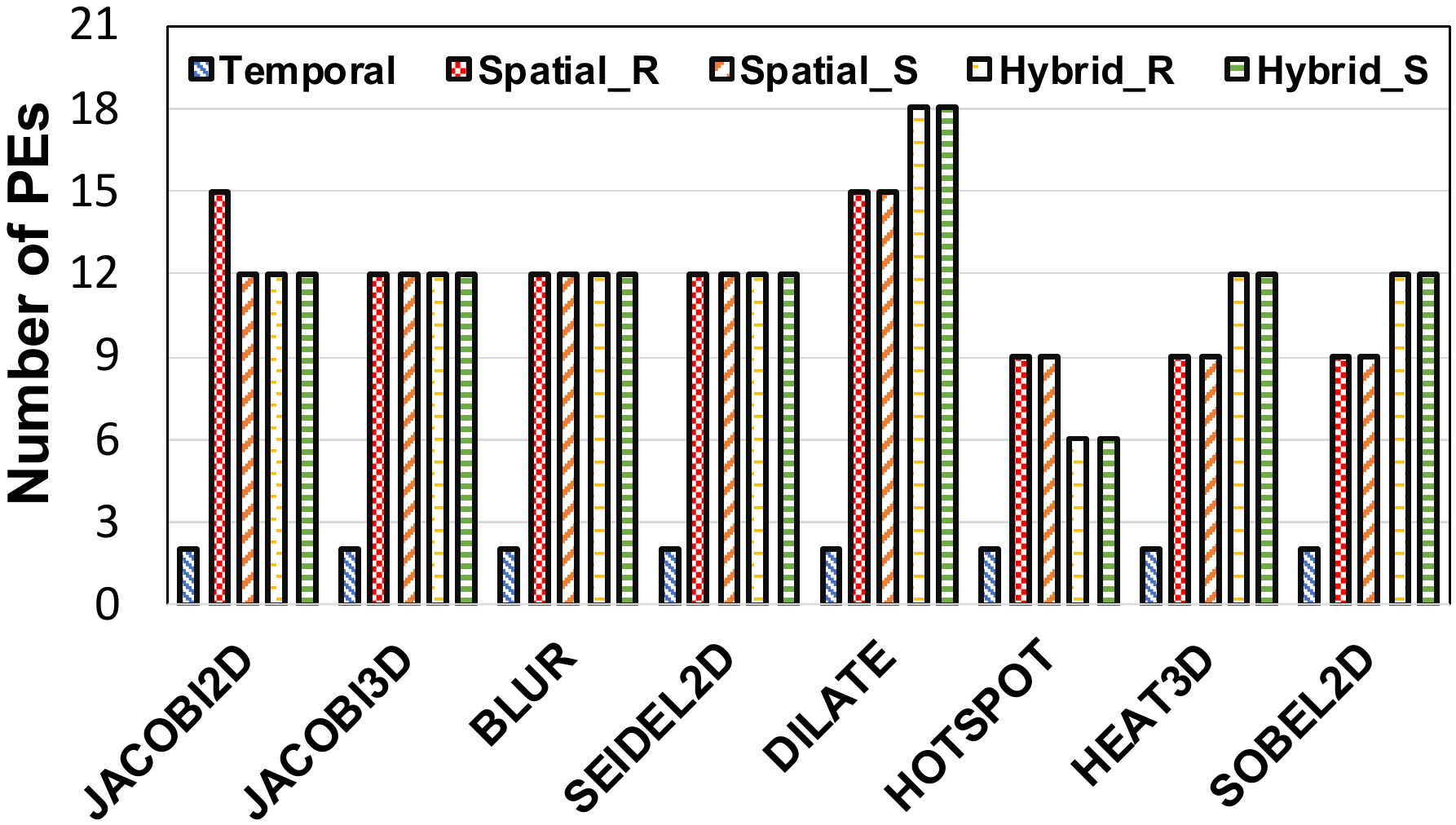}
        \caption{Number of PEs with the number of iteration = 2}
        \label{fig:penumber_4096_2}
    \end{subfigure}
    
    \caption{\rev{Total number of PEs for different parallelisms on Alveo U280 with column size = 4096}} %\zf{update this figure, enlarge font size}\xy{Updated}}
    \label{fig:penumber_4096}
    \vspace{-0.05in}
\end{figure}

\begin{enumerate}
    \item When the iteration number is 1, the hybrid parallelism is the same as spatial parallelism, since each spatial PE group has only one temporal stage. \rev{When iteration number is larger than 1, there are multiple combinations of spatial parallelism degree and temporal parallelism degree. For example, in JACOBI3D at $256 \times 256$, maximum number of PEs can be implemented is 15, as shown in Figure~\ref{fig:penumber_256} to~\ref{fig:penumber_4096}. We choose the degree of spatial parallelism based on the number of SLRs, which is 3 on Alevo U280 board. When iteration number is 2 and 4, 6 spatial PE groups with 2 temporal stages will outperform 3 spatial PE groups with 5 temporal stages even with less PEs. This is because the former can utilize more off-chip memory bandwidth without idle PEs.}

    \item Hybrid parallelism has a similar trend as spatial parallelism with the increase of the iteration number, when the ratio of iteration number and rounds of FPGA kernel execution maintains the same, especially when the iteration number is small. The effect of this ratio is illustrated in Section~\ref{subsubsec:perf_temporal}. For example, in \rev{BLUR, SEIDEL2D and HEAT3D}, the throughput of $Hybrid\_R$ decreases as the iteration number increases in the range from \rev{4} to 64, since the ratio of iteration number and rounds of FPGA kernel execution does not change. And the throughput of $Hybrid\_S$ in the same iteration range stays the same as \rev{the pattern of }$Spatial\_S$ \rev{since they have the same number of PEs}.

    \item However, when iteration number becomes large enough and this ratio of iteration number and rounds of FPGA kernel execution changes, the throughput of hybrid parallelism will have a noticeable change, as temporal parallelism plays a heavier role. \rev{Such pattern is more outstanding at small input size, like $256\times256$.} For example, in JACOBI2D, the throughput of $Hybrid\_R$ decreases slightly when iteration number ranges from \rev{4} to 8, reflecting the characteristics of spatial parallelism. However, there is a big performance boost when the iteration number changes from 8 to 16,  since the ratio changes and temporal parallelism play a heavier role. This ratio change happens again when the iteration number changes from \rev{16} to 64 in JACOBI2D. Such turning points vary with the PE number; for example, \rev{in HOTSPOT}, the turning points of $Hybrid\_S$ are from 4 to 8 and from 16 to 32 since its PE number is 9.
    
\end{enumerate}

\rev{For most benchmarks, $Hybrid\_R$ and $Hybrid\_S$ have the same number of PEs. $Hybrid\_S$ achieves a similar performance to $Hybrid\_R$ at a small iteration number. At a large iteration number, $Hybrid\_S$ outperforms $Hybrid\_R$, because the $Hybrid\_R$ design requires redundant computation for more halo data than $Hybrid\_S$.
Lastly, there is only one case where the border streaming based approach achieves fewer PEs than the redundant computation based approach. Specifically, 
for JACOBI2D at $9270\times1024$ and $4096\times4096$, $Hybrid\_S$ has fewer PEs than $Hybrid\_R$. As a result, when the iteration number is 32, the performance of $Hybrid\_R$ is better than $Hybrid\_S$. However, such advantage is offset by the redundant halo computation overhead for other iteration numbers.}

\subsubsection{\rev{Performance Impact by Different Input Sizes}} \label{subsubsec:perf_input_compare}

\rev{For the four different stencil input sizes, 256x256, 720x1024, 9720x1024, and 4096x4096 (for 2D stencils), we have made the following observations.} 

\rev{First, for the majority of our stencil benchmarks under these input sizes (more specifically, different column sizes), the row buffer resource consumption did not become a bottleneck, as each PE roughly needs to buffer only two rows of data on-chip.} 

\rev{Second, the row sizes do have a performance impact, especially for the redundant computation based spatial parallelism ($Spatial\_R$) and hybrid parallelism ($Hybrid\_R$). With a smaller row size (e.g., 256), when the iteration count becomes larger, the performance of $Spatial\_R$ decreases significantly as the redundant computation adds a very significant overhead. Therefore, the border streaming based spatial parallelism ($Spatial\_S$) is a better choice. While with a larger row size (e.g., 9720), such overhead is much smaller and the difference between $Spatial\_R$ and $Spatial\_S$ is marginal. A similar performance impact is observed for $Hybrid\_R$.}

\rev{Third, in general, the overall throughput for the small 256x256 input size is relatively lower than those with larger input sizes. The reasons are twofold. First, the execution time of extra halo regions for a smaller input size occupies a high execution time percentage, i.e., the overhead is bigger. Second, with the smaller input size, the memory burst size for each HBM bank is relatively small, thus leading to lower off-chip memory bandwidth utilization.}

\subsubsection{Performance Comparison between Temporal, Spatial, and Hybrid Parallelisms} \label{subsubsec:perf_overall_compare}

\rev{Overall, temporal parallelism achieves the lowest performance amongst all parallelism variants. When the iteration count is low, e.g., 1 or 2, temporal parallelism cannot efficiently exploit the HBM memory bandwidth. Even when the iteration count is as large as 64, temporal parallelism also may not give the best performance since the iteration count may not be evenly divisible by the temporal stages instantiated on hardware. Take JACOBI2D as an example, there are 21 temporal stages on the hardware as shown in Figure~\ref{fig:penumber_4096_64}. When its iteration count is 64, it needs to execute the hardware ceil (64/21) = 4 rounds. In the last round, there is only one last iteration (64 - 21$\times$3 = 1) that needs to be executed; 20 temporal stages on hardware are under-utilized.}

\rev{For the remaining parallelism variants, spatial and hybrid, boarder streaming based approach generally achieves better performance than the redundant computation based method as detailed above in Section~\ref{subsubsec:perf_spatial} and Section~\ref{subsubsec:perf_hybrid}. However, depending on stencil kernel, iteration number, and input sizes, the best parallelism may vary.}

\rev{First, there are cases where $Spatial\_S$ and $Hybrid\_S$ achieve a similar performance and are the best among all parallelisms, specifically for BLUR, SEIDEL2D, and HEAT3D kernels. The reason is that both parallelisms have 12 PEs and can fully utilize them under different iteration numbers and input sizes. Specifically for $Hybrid\_S$, when the iteration number is 2, the degree of spatial parallelism is 6 and the degree of temporal parallelism is 2; when the iteration number larger than 2, the degree of spatial parallelism is 3 and the degree of temporal parallelism is 4. Therefore, all 12 PEs can be fully utilized with different iteration numbers.}

% $Hybrid\_S$ > $Spatial\_S$
\rev{Second, there are cases where $Hybrid\_S$ outperforms $Spatial\_S$ and $Hybrid\_S$ is the best among all parallelisms, specifically for DILATE, SOBEL2D, JACOBI2D with large iteration number, and JACOBI3D with large iteration number. There are two reasons behind this: 1) for DILATE and JACOBI2D, due to the HBM bank (i.e., bandwidth) restriction, $Spatial\_S$ has fewer PEs than $Hybrid\_S$; 2) for SOBEL2D and JACOBI3D, $Spatial\_S$ has fewer PEs than $Hybrid\_S$ as it is harder to pass the timing closure.}

% $Spatial\_S$ > $Hybrid\_S$
\rev{Third, there is also one case where $Spatial\_S$ achieves a better performance than $Hybrid\_S$ and is the best, specifically for HOTSPOT at a small iteration number. In fact, both $Spatial\_S$ and $Hybrid\_S$ have 9 PEs in this case. However, in $Hybrid\_S$, the degree of spatial parallelism is 3 and the degree of temporal parallelism is 3, which cannot be evenly divided by the iteration number. As a result, some PEs are underutilized in $Hybrid\_S$, leading to a lower performance than $Spatial\_S$.}

% $Spatial\_R$ > $Hybrid\_S$
\rev{Lastly, there are two exceptional cases where $Spatial\_R$ performs better than $Hybrid\_S$ and is the best, specifically for JACOBI2D and JACOBI3D when the iteration number is small and the number of input rows is large. This is because the $Hybrid\_S$ significantly under-utilizes the number of PEs when the iteration count is small, especially when iteration count is 2 or 4. For example, $Spatial\_R$ of JACOBI3D can utilize all 15 PEs when the iteration number is 2, while $Hybrid\_S$ can only utilizes 12 PEs with the best configuration of 6 spatial PE groups and 2 temporal PEs in each group.}

\subsubsection{The Best Parallelism Configurations and Their Resource Utilization}
% Each benchmark, best parallelism
%HBM number/ resource /Frequecny

% Different benchmarks & Different iteration number have different best parallelism
% limited by HBM number or resource utilization.
% Why - How - What's new

\begin{table}[]
\caption{\rev{Configuration of the best parallelism on Alveo U280, for the input size of 9720x1024}}
\label{tab:configuration}
\small
\begin{tabular}{|c|ccccc|ccccc|}
\hline
\multirow{2}{*}{} & \multicolumn{5}{c|}{Iteration = 64}                                                                                                & \multicolumn{5}{c|}{Iteration   = 2}                                                                                               \\ \cline{2-11} 
                  & \multicolumn{1}{c|}{Parallelism} & \multicolumn{1}{c|}{Frequency} & \multicolumn{1}{c|}{k}  & \multicolumn{1}{c|}{s} & \begin{tabular}[c]{@{}c@{}}\#HBM\\ banks\end{tabular} & \multicolumn{1}{c|}{Parallelism} & \multicolumn{1}{c|}{Frequency} & \multicolumn{1}{c|}{k}  & \multicolumn{1}{c|}{s} & \begin{tabular}[c]{@{}c@{}}\#HBM\\ banks\end{tabular} \\ \hline
JACOBI2D          & \multicolumn{1}{c|}{Hybrid\_S}   & \multicolumn{1}{c|}{250 MHz}   & \multicolumn{1}{c|}{3}  & \multicolumn{1}{c|}{7} & 6           & \multicolumn{1}{c|}{Spatial\_R}  & \multicolumn{1}{c|}{233 MHz}   & \multicolumn{1}{c|}{15} & \multicolumn{1}{c|}{1} & 30          \\ \hline
JACOBI3D          & \multicolumn{1}{c|}{Hybrid\_S}  & \multicolumn{1}{c|}{250 MHz}   & \multicolumn{1}{c|}{3} & \multicolumn{1}{c|}{5} & 6          & \multicolumn{1}{c|}{Spatial\_R}  & \multicolumn{1}{c|}{226 MHz}   & \multicolumn{1}{c|}{15} & \multicolumn{1}{c|}{1} & 30          \\ \hline
BLUR              & \multicolumn{1}{c|}{Hybrid\_S}  & \multicolumn{1}{c|}{249 MHz}   & \multicolumn{1}{c|}{3} & \multicolumn{1}{c|}{4} & 6          & \multicolumn{1}{c|}{Spatial\_R}  & \multicolumn{1}{c|}{229 MHz}   & \multicolumn{1}{c|}{12} & \multicolumn{1}{c|}{1} & 24          \\ \hline
SEIDEL2D          & \multicolumn{1}{c|}{Hybird\_S}   & \multicolumn{1}{c|}{225 MHz}   & \multicolumn{1}{c|}{3}  & \multicolumn{1}{c|}{4} & 6           & \multicolumn{1}{c|}{Spatial\_R}  & \multicolumn{1}{c|}{225 MHz}   & \multicolumn{1}{c|}{12} & \multicolumn{1}{c|}{1} & 24          \\ \hline
DILATE            & \multicolumn{1}{c|}{Hybrid\_S}   & \multicolumn{1}{c|}{250 MHz}   & \multicolumn{1}{c|}{3}  & \multicolumn{1}{c|}{6} & 6           & \multicolumn{1}{c|}{Hybrid\_S}  & \multicolumn{1}{c|}{250 MHz}   & \multicolumn{1}{c|}{6} & \multicolumn{1}{c|}{2} & 12          \\ \hline
HOTSPOT           & \multicolumn{1}{c|}{\revsnd{Hybrid\_S}}  & \multicolumn{1}{c|}{250 MHz}   & \multicolumn{1}{c|}{3}  & \multicolumn{1}{c|}{3} & 9           & \multicolumn{1}{c|}{Spatial\_S}  & \multicolumn{1}{c|}{250 MHz}   & \multicolumn{1}{c|}{9}  & \multicolumn{1}{c|}{1} & 27          \\ \hline
HEAT3D            & \multicolumn{1}{c|}{Hybrid\_S}  & \multicolumn{1}{c|}{225 MHz}   & \multicolumn{1}{c|}{3} & \multicolumn{1}{c|}{4} & 6          & \multicolumn{1}{c|}{Spatial\_R}  & \multicolumn{1}{c|}{230 MHz}   & \multicolumn{1}{c|}{12} & \multicolumn{1}{c|}{1} & 24          \\ \hline
SOBEL2D           & \multicolumn{1}{c|}{Hybrid\_S}   & \multicolumn{1}{c|}{250 MHz}   & \multicolumn{1}{c|}{3}  & \multicolumn{1}{c|}{4} & 6           & \multicolumn{1}{c|}{Hybrid\_S}  & \multicolumn{1}{c|}{250 HHz}   & \multicolumn{1}{c|}{3}  & \multicolumn{1}{c|}{4} & 6          \\ \hline
\end{tabular}
\end{table}

As discussed above, the best parallelism optimization varies with the stencil benchmark and the number of iterations. Table~\ref{tab:configuration} summarizes the best parallelism configuration for each benchmark \rev{for the input size of 9720$\times$1024}, when the number of iterations is 64 and 2, respectively. 
When the number of iterations is 64, \rev{ $Hybrid\_S$ achieves the best performance for all benchmarks as it is not affected by the redundant halo computation overhead.}
Note that one advantage of hybrid parallelism over spatial parallelism is that it requires much less off-chip bandwidth (shown as the number of HBM banks in Table ~\ref{tab:configuration}).
When the number of iterations is 2, 
\rev{spatial parallelism achieves the best performance for most benchmarks for most of the benchmarks;} 
% and there is not much parallelism along the temporal dimension; depending on the benchmark, 
both $Spatial\_R$ or $Spatial\_S$ achieve a similar performance. \rev{There are some exceptions, DILATE and SOBLE2D, where $Hybrid\_S$ achieves the best performance. This is because their $Spatial\_R$ and $Spatial\_S$ parallelism implements less number of PEs due to the limitation of the available HBM banks and timing closure issues, respectively.} 

For the best parallelism configurations, \rev{the degree of spatial parallelism} ($k$) and the number of temporal stages ($s$) are also included in Table~\ref{tab:configuration}. These number also vary between benchmarks, which again highlights the importance of an automation framework to compile the high-level DSL to the optimized FPGA design. All of our designs achieve a clock frequency of at least 225 MHz to fully utilize bandwidth of each HBM bank.

\begin{figure}[t]
    \begin{subfigure}[b]{0.48\textwidth}
        \centering
        \includegraphics[width=\textwidth]{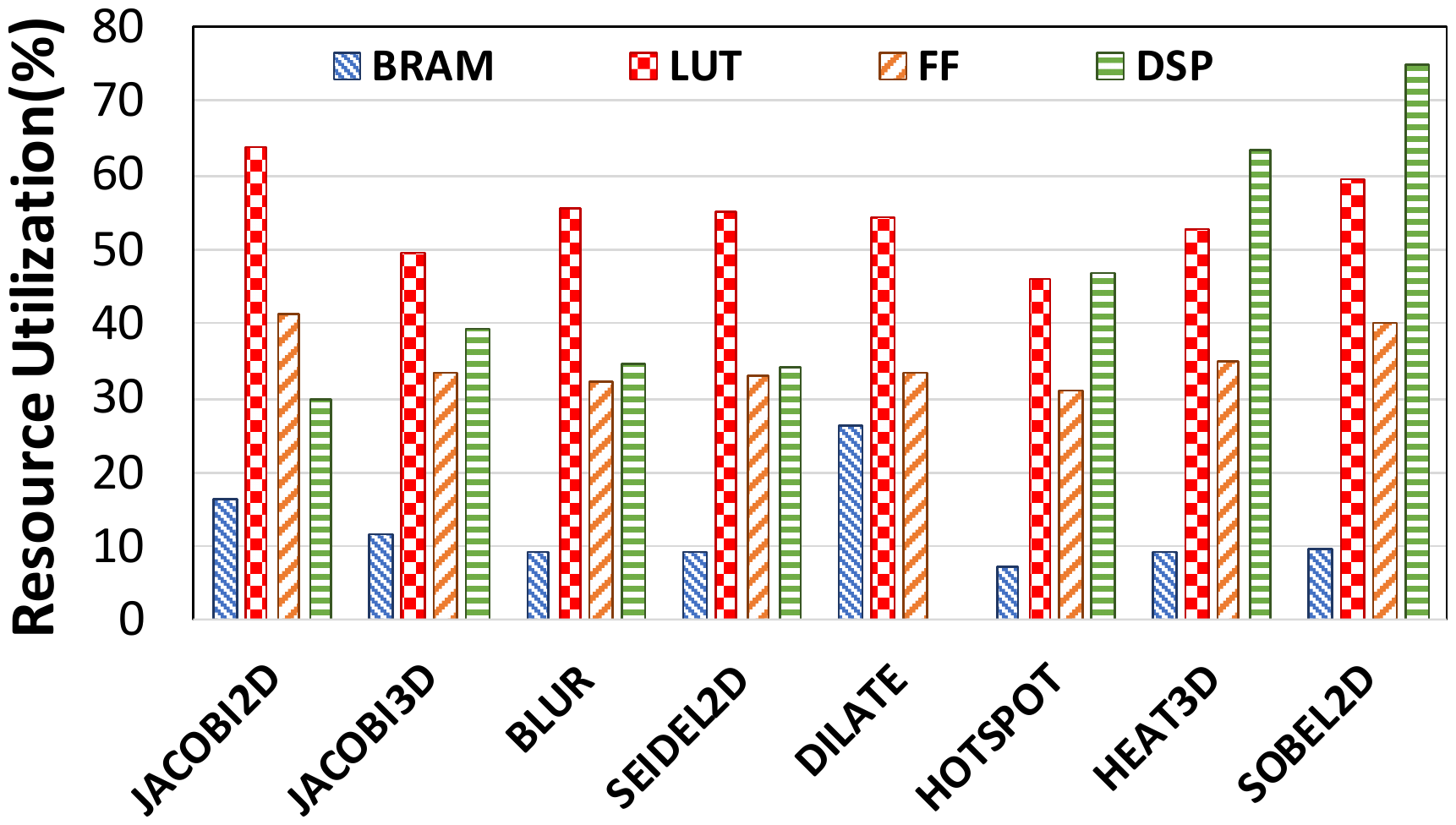}
        \caption{Resource utilization of the best parallelism with the number of iteration = 64}
        \label{fig:resource_64}
    \end{subfigure} 
    \hspace{+0.1in}
    \begin{subfigure}[b]{0.48\textwidth}
        \centering
        \includegraphics[width=\textwidth]{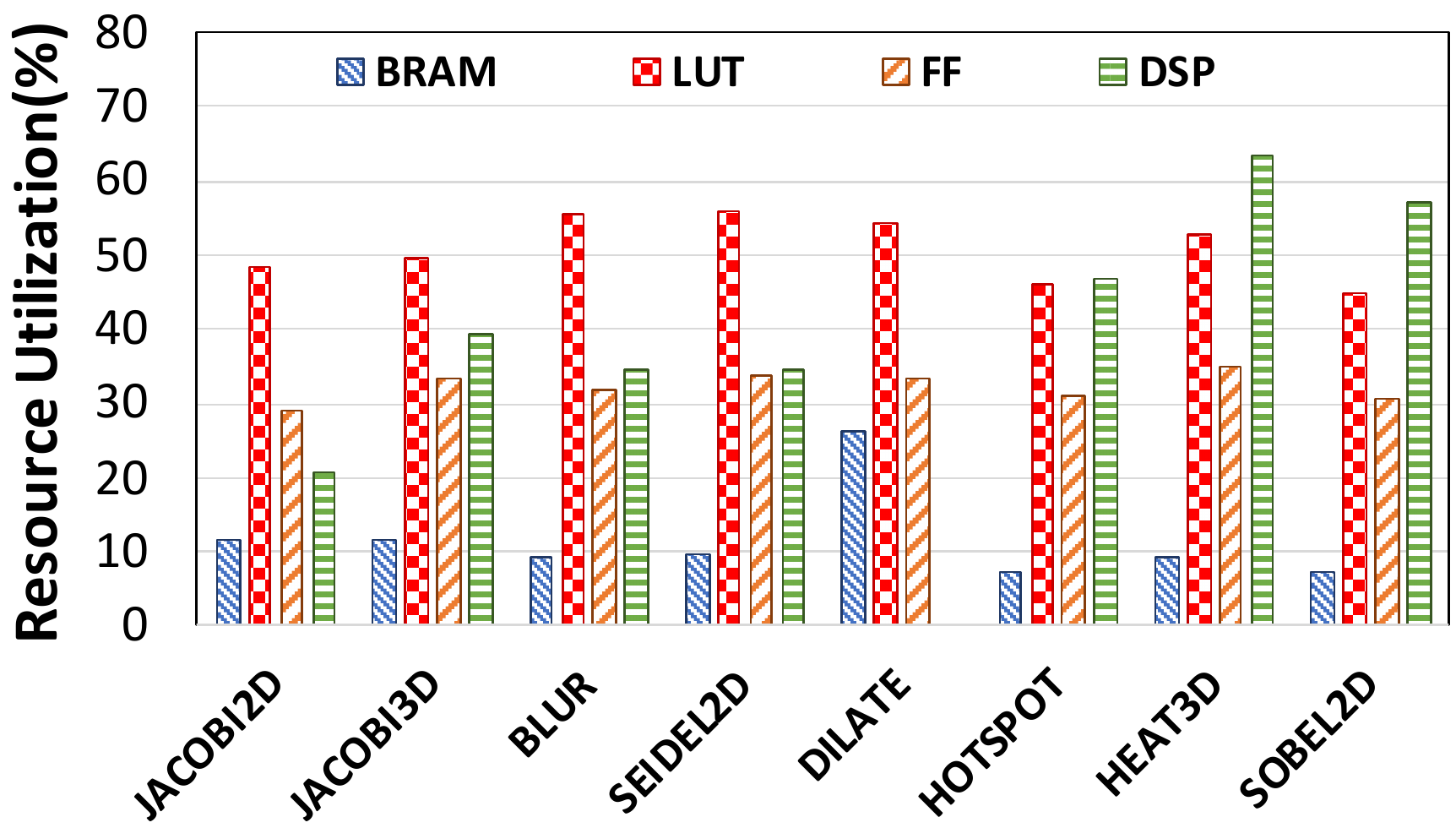}
        \caption{Resource utilization of the best parallelism with the number of iteration = 2}
        \label{fig:resource_2}
    \end{subfigure}
    
    \caption{\rev{Resource utilization of the best parallelism configuration on Alveo U280, for the input size of 9720x1024}}
    \label{fig:best_resource}
    \vspace{-0.05in}
\end{figure}

Finally, we also show the utilization of on-chip resources and off-chip HBM banks for the best parallelism configurations in Figure~\ref{fig:best_resource} and Table~\ref{tab:configuration}, respectively. 
The bottleneck resource changes as the computation intensity increases. As shown in Figure~\ref{fig:best_resource}, for benchmarks with lower computation intensity, such as JACOBI2D, JACOBI3D, BLUR, SEIDEL2D, and DILATE, LUT has the highest resource utilization rate compared with other resources.
For benchmarks with higher computation intensity, such as HOTSPOT, HEAT3D, and SOBEL2D, DSP is the bottleneck to scale up to more PEs.
% For some benchmarks such as JACOBI2D and JACOBI3D, their performance (number of PEs) are limited by the number of available HBM banks (off-chip bandwidth). %For other benchmarks whose resource utilization is far from reaching the limitation, the timing closure problem prevents them from instantiating more PEs. In future work, we plan to further investigate timing and floorplanning optimizations to place more PEs and improve the performance for \toolName~designs.

%The PE number of each parallelism is limited by not only the available on-chip hardware resource, but also the number of HBM banks. PE number of some benchmarks, like BLUR and SEIDEL2D, is limited by hardware resource (LUT). When iteration number is 2, while JACOBI2D and JACOBI3D not fully utilize all resource with $Spatial_R$ parallelism, they almost consume all HBM banks. In this way, both benchmarks can hardly incorporate more PEs.

%Figure ~\ref{fig:best_resource} shows the resource utilization of best parallelism of each benchmark when the iteration number is 64 and 2. 

% As shown in Table~\ref{tab:configuration}, different benchmarks may have different parallelisms with the same iteration number. Different iteration numbers 

% 1. Figure and table show that different benchmarks and different iteration number have different best parallelisms.
% 2. The PE number is limited by HBM number or resource utilization.
% 3. Most can reach 225MHz, only HEAT3D to be further optimized.

\subsection{\rev{Comparison to Prior Work}}

\rev{As discussed in Section~\ref{subsec:single-pe}, SODA~\cite{SODA18} is less efficient than our \toolName~temporal parallelism implementation due to the additional on-chip line buffer usage. To conduct a fair comparison between SODA and \toolName~, we integrate SODA with TAPA/AutoBridge~\cite{tapa,Autobridge21} to address the major resource inefficiency, and rerun all the experiments on the same HBM-based U280 FPGA. The on-chip line buffer to buffer the input data from off-chip memory, shown in Figure~\ref{fig:resource}, is also removed in this integration, since TAPA replaces the resource-inefficient AXI interface with a lightweight streaming interface. As a result, both SODA and \toolName~temporal parallelism implementation achieve the same performance.
%, as discussed in Sec~\ref{subsec:result_multiPE_parallisim}. 
However, SODA only supports temporal parallelism, and does not support other types of parallelisms that we have explored in this paper; therefore, its performance is sub-optimal when the iteration count is small or the iteration count cannot be evenly divided by the temporal stages in the hardware design. Compared to SODA, \toolName~achieves better throughput with an average of at least \rev{3.74}$\times$ speedup across all configurations. The highest speedup over temporal parallelism is reached in JACOBI3D when iteration number is 1, where redundant computation based spatial parallelism can reach \rev{15.73}$\times$ speedup.}

\rev{The stencil accelerator design proposed in~\cite{SpatialTemporyBlocking18} only supports temporal parallelism and its throughput is measured when the iteration count is super large. 
For temporal parallelism, the throughput is determined by the bandwidth of a single memory bank. Since their FPGA uses DDR4, which has a higher bandwidth (19.2GB/s theoretical bandwidth) than that of a single HBM bank (14.4GB/s theoretical bandwidth) in our results, their reported GCell/s is higher than ours. However, for the given HBM bank, our implementation already achieves the best performance that the HBM bank can achieve. More importantly, we have explored different parallelisms and can automatically generate the design with the best parallelism.}

\section{Conclusion}\label{sec:concl}

In this paper we propose a scalable and automatic stencil acceleration framework on modern HBM-based FPGAs called~\toolName. 
In terms of the accelerator design architecture, \toolName~ employs a multi-PE approach to exploit temporal and spatial parallelisms for better scalability.
Each single PE design is optimized for on-chip data reuse, off-chip memory access, and the on-chip buffer usage.
For design automation, \toolName~provides a high-level DSL for domain experts to configure and define the stencil operation. Then a code generator automatically explores the design space based on our analytical performance model and generates an optimized stencil accelerator design with the best parallelism optimization.
Experimental results across a wide range of stencil benchmarks show that our \toolName~can achieve 

speedup on the HBM-based Xilinx Alveo U280 FPGA, compared to state-of-the-art automatic stencil acceleration framework SODA~\cite{SODA18} that only exploits temporal parallelism. Finally, we plan to open source our tool in the near future.

% For future work, 
% %we are working on integrating spatial blocking to further improve the flexibility of our design in exploiting the spatial parallelism. Second, to , 
% we plan to incorporate the more floorplanning optimizations for streaming-based applications introduced in Autobridge~\cite{Autobridge21} to improve the timing closure of \toolName~designs and place more PEs on the FPGA. Moreover, we plan to conduct a quantitative comparison of~\toolName~to latest stencil acceleration studies. 

\newpage
\bibliographystyle{ACM-Reference-Format}
\bibliography{reference}

%%% -*-BibTeX-*-
%%% Do NOT edit. File created by BibTeX with style
%%% ACM-Reference-Format-Journals [18-Jan-2012].

\begin{thebibliography}{31}

%%% ====================================================================
%%% NOTE TO THE USER: you can override these defaults by providing
%%% customized versions of any of these macros before the \bibliography
%%% command.  Each of them MUST provide its own final punctuation,
%%% except for \shownote{}, \showDOI{}, and \showURL{}.  The latter two
%%% do not use final punctuation, in order to avoid confusing it with
%%% the Web address.
%%%
%%% To suppress output of a particular field, define its macro to expand
%%% to an empty string, or better, \unskip, like this:
%%%
%%% \newcommand{\showDOI}[1]{\unskip}   % LaTeX syntax
%%%
%%% \def \showDOI #1{\unskip}           % plain TeX syntax
%%%
%%% ====================================================================

\ifx \showCODEN    \undefined \def \showCODEN     #1{\unskip}     \fi
\ifx \showDOI      \undefined \def \showDOI       #1{#1}\fi
\ifx \showISBNx    \undefined \def \showISBNx     #1{\unskip}     \fi
\ifx \showISBNxiii \undefined \def \showISBNxiii  #1{\unskip}     \fi
\ifx \showISSN     \undefined \def \showISSN      #1{\unskip}     \fi
\ifx \showLCCN     \undefined \def \showLCCN      #1{\unskip}     \fi
\ifx \shownote     \undefined \def \shownote      #1{#1}          \fi
\ifx \showarticletitle \undefined \def \showarticletitle #1{#1}   \fi
\ifx \showURL      \undefined \def \showURL       {\relax}        \fi
% The following commands are used for tagged output and should be
% invisible to TeX
\providecommand\bibfield[2]{#2}
\providecommand\bibinfo[2]{#2}
\providecommand\natexlab[1]{#1}
\providecommand\showeprint[2][]{arXiv:#2}

\bibitem[\protect\citeauthoryear{Alobaid, Baraki, and Epple}{Alobaid
  et~al\mbox{.}}{2014}]%
        {CFD14}
\bibfield{author}{\bibinfo{person}{Falah Alobaid}, \bibinfo{person}{Nabil
  Baraki}, {and} \bibinfo{person}{Bernd Epple}.}
  \bibinfo{year}{2014}\natexlab{}.
\newblock \showarticletitle{Investigation into improving the efficiency and
  accuracy of CFD/DEM simulations}.
\newblock \bibinfo{journal}{\emph{Particuology}}  \bibinfo{volume}{16}
  (\bibinfo{year}{2014}), \bibinfo{pages}{41--53}.
\newblock


\bibitem[\protect\citeauthoryear{Cattaneo, Natale, Sicignano, Sciuto, and
  Santambrogio}{Cattaneo et~al\mbox{.}}{2015}]%
        {TemporalTACO16}
\bibfield{author}{\bibinfo{person}{Riccardo Cattaneo},
  \bibinfo{person}{Giuseppe Natale}, \bibinfo{person}{Carlo Sicignano},
  \bibinfo{person}{Donatella Sciuto}, {and} \bibinfo{person}{Marco~Domenico
  Santambrogio}.} \bibinfo{year}{2015}\natexlab{}.
\newblock \showarticletitle{On How to Accelerate Iterative Stencil Loops: A
  Scalable Streaming-Based Approach}.
\newblock \bibinfo{journal}{\emph{ACM Trans. Archit. Code Optim.}}
  \bibinfo{volume}{12}, \bibinfo{number}{4}, Article \bibinfo{articleno}{53}
  (\bibinfo{date}{dec} \bibinfo{year}{2015}), \bibinfo{numpages}{26}~pages.
\newblock


\bibitem[\protect\citeauthoryear{Chi and Cong}{Chi and Cong}{2020}]%
        {StencilComputeReuse20}
\bibfield{author}{\bibinfo{person}{Yuze Chi} {and} \bibinfo{person}{Jason
  Cong}.} \bibinfo{year}{2020}\natexlab{}.
\newblock \showarticletitle{Exploiting Computation Reuse for Stencil
  Accelerators}. In \bibinfo{booktitle}{\emph{Proceedings of the 57th
  ACM/EDAC/IEEE Design Automation Conference}}. Article
  \bibinfo{articleno}{184}, \bibinfo{numpages}{6}~pages.
\newblock


\bibitem[\protect\citeauthoryear{Chi, Cong, Wei, and Zhou}{Chi
  et~al\mbox{.}}{2018}]%
        {SODA18}
\bibfield{author}{\bibinfo{person}{Yuze Chi}, \bibinfo{person}{Jason Cong},
  \bibinfo{person}{Peng Wei}, {and} \bibinfo{person}{Peipei Zhou}.}
  \bibinfo{year}{2018}\natexlab{}.
\newblock \showarticletitle{SODA: Stencil with Optimized Dataflow
  Architecture}. In \bibinfo{booktitle}{\emph{2018 IEEE/ACM International
  Conference on Computer-Aided Design (ICCAD)}}. \bibinfo{pages}{1--8}.
\newblock


\bibitem[\protect\citeauthoryear{Chi, Guo, Lau, Choi, Wang, and Cong}{Chi
  et~al\mbox{.}}{2021}]%
        {tapa}
\bibfield{author}{\bibinfo{person}{Yuze Chi}, \bibinfo{person}{Licheng Guo},
  \bibinfo{person}{Jason Lau}, \bibinfo{person}{Young-kyu Choi},
  \bibinfo{person}{Jie Wang}, {and} \bibinfo{person}{Jason Cong}.}
  \bibinfo{year}{2021}\natexlab{}.
\newblock \showarticletitle{Extending High-Level Synthesis for Task-Parallel
  Programs}. In \bibinfo{booktitle}{\emph{2021 IEEE 29th Annual International
  Symposium on Field-Programmable Custom Computing Machines (FCCM)}}.
  \bibinfo{pages}{204--213}.
\newblock
\urldef\tempurl%
\url{https://doi.org/10.1109/FCCM51124.2021.00032}
\showDOI{\tempurl}


\bibitem[\protect\citeauthoryear{Cong, Fang, Lo, Wang, Xu, and Zhang}{Cong
  et~al\mbox{.}}{2018}]%
        {RodiniaHLS}
\bibfield{author}{\bibinfo{person}{Jason Cong}, \bibinfo{person}{Zhenman Fang},
  \bibinfo{person}{Michael Lo}, \bibinfo{person}{Hanrui Wang},
  \bibinfo{person}{Jingxian Xu}, {and} \bibinfo{person}{Shaochong Zhang}.}
  \bibinfo{year}{2018}\natexlab{}.
\newblock \showarticletitle{Understanding Performance Differences of FPGAs and
  GPUs}. In \bibinfo{booktitle}{\emph{2018 IEEE 26th Annual International
  Symposium on Field-Programmable Custom Computing Machines (FCCM)}}.
  \bibinfo{pages}{93--96}.
\newblock
\urldef\tempurl%
\url{https://doi.org/10.1109/FCCM.2018.00023}
\showDOI{\tempurl}


\bibitem[\protect\citeauthoryear{Cooke, Fowers, Hunt, and Stitt}{Cooke
  et~al\mbox{.}}{2013}]%
        {SlidingWindow}
\bibfield{author}{\bibinfo{person}{Patrick Cooke}, \bibinfo{person}{Jeremy
  Fowers}, \bibinfo{person}{Lee Hunt}, {and} \bibinfo{person}{Greg Stitt}.}
  \bibinfo{year}{2013}\natexlab{}.
\newblock \showarticletitle{A High-Performance, Low-Energy FPGA Accelerator for
  Correntropy-Based Feature Tracking (Abstract Only)}. In
  \bibinfo{booktitle}{\emph{Proceedings of the ACM/SIGDA International
  Symposium on Field Programmable Gate Arrays}} (Monterey, California, USA)
  \emph{(\bibinfo{series}{FPGA '13})}. \bibinfo{publisher}{Association for
  Computing Machinery}, \bibinfo{address}{New York, NY, USA},
  \bibinfo{pages}{278}.
\newblock


\bibitem[\protect\citeauthoryear{Datta, Kamil, Williams, Oliker, Shalf, and
  Yelick}{Datta et~al\mbox{.}}{2009}]%
        {small-iterations}
\bibfield{author}{\bibinfo{person}{Kaushik Datta}, \bibinfo{person}{Shoaib
  Kamil}, \bibinfo{person}{Samuel Williams}, \bibinfo{person}{Leonid Oliker},
  \bibinfo{person}{John Shalf}, {and} \bibinfo{person}{Katherine Yelick}.}
  \bibinfo{year}{2009}\natexlab{}.
\newblock \showarticletitle{Optimization and Performance Modeling of Stencil
  Computations on Modern Microprocessors}.
\newblock \bibinfo{journal}{\emph{SIAM Rev.}} \bibinfo{volume}{51},
  \bibinfo{number}{1} (\bibinfo{year}{2009}), \bibinfo{pages}{129--159}.
\newblock
\showISSN{00361445, 10957200}
\urldef\tempurl%
\url{http://www.jstor.org/stable/20454196}
\showURL{%
\tempurl}


\bibitem[\protect\citeauthoryear{Dejanović, Vaderna, Milosavljević, and
  Vuković}{Dejanović et~al\mbox{.}}{2017}]%
        {textX}
\bibfield{author}{\bibinfo{person}{I. Dejanović}, \bibinfo{person}{R.
  Vaderna}, \bibinfo{person}{G. Milosavljević}, {and} \bibinfo{person}{Ž.
  Vuković}.} \bibinfo{year}{2017}\natexlab{}.
\newblock \showarticletitle{TextX: A Python tool for Domain-Specific Languages
  implementation}.
\newblock \bibinfo{journal}{\emph{Knowledge-Based Systems}}
  \bibinfo{volume}{115} (\bibinfo{year}{2017}), \bibinfo{pages}{1--4}.
\newblock
\showISSN{0950-7051}
\urldef\tempurl%
\url{https://doi.org/10.1016/j.knosys.2016.10.023}
\showDOI{\tempurl}


\bibitem[\protect\citeauthoryear{Du and Yamaguchi}{Du and Yamaguchi}{2020}]%
        {StencilHBM20}
\bibfield{author}{\bibinfo{person}{Changdao Du} {and} \bibinfo{person}{Yoshiki
  Yamaguchi}.} \bibinfo{year}{2020}\natexlab{}.
\newblock \showarticletitle{High-Level Synthesis Design for Stencil
  Computations on FPGA with High Bandwidth Memory}.
\newblock \bibinfo{journal}{\emph{Electronics}} \bibinfo{volume}{9},
  \bibinfo{number}{8} (\bibinfo{year}{2020}).
\newblock


\bibitem[\protect\citeauthoryear{Escobedo and Lin}{Escobedo and Lin}{2018}]%
        {GraphBasedFPGA18}
\bibfield{author}{\bibinfo{person}{Juan Escobedo} {and}
  \bibinfo{person}{Mingjie Lin}.} \bibinfo{year}{2018}\natexlab{}.
\newblock \showarticletitle{Graph-Theoretically Optimal Memory Banking for
  Stencil-Based Computing Kernels}. In \bibinfo{booktitle}{\emph{Proceedings of
  the 2018 ACM/SIGDA International Symposium on Field-Programmable Gate
  Arrays}} (Monterey, CALIFORNIA, USA) \emph{(\bibinfo{series}{FPGA '18})}.
  \bibinfo{publisher}{Association for Computing Machinery},
  \bibinfo{address}{New York, NY, USA}, \bibinfo{pages}{199–208}.
\newblock


\bibitem[\protect\citeauthoryear{Faramarzi, Rajan, and Christensen}{Faramarzi
  et~al\mbox{.}}{2013}]%
        {ImageProcessing13}
\bibfield{author}{\bibinfo{person}{Esmaeil Faramarzi}, \bibinfo{person}{Dinesh
  Rajan}, {and} \bibinfo{person}{Marc~P. Christensen}.}
  \bibinfo{year}{2013}\natexlab{}.
\newblock \showarticletitle{Unified Blind Method for Multi-Image
  Super-Resolution and Single/Multi-Image Blur Deconvolution}.
\newblock \bibinfo{journal}{\emph{IEEE Transactions on Image Processing}}
  \bibinfo{volume}{22}, \bibinfo{number}{6} (\bibinfo{year}{2013}),
  \bibinfo{pages}{2101--2114}.
\newblock


\bibitem[\protect\citeauthoryear{Firmansyah, Wijayanto, and
  Yamaguchi}{Firmansyah et~al\mbox{.}}{2018}]%
        {StencilSoC18}
\bibfield{author}{\bibinfo{person}{Iman Firmansyah}, \bibinfo{person}{Yusuf~Nur
  Wijayanto}, {and} \bibinfo{person}{Yoshiki Yamaguchi}.}
  \bibinfo{year}{2018}\natexlab{}.
\newblock \showarticletitle{2D Stencil Computation on Cyclone V SoC FPGA using
  OpenCL}. In \bibinfo{booktitle}{\emph{2018 International Conference on Radar,
  Antenna, Microwave, Electronics, and Telecommunications (ICRAMET)}}.
  \bibinfo{pages}{121--124}.
\newblock


\bibitem[\protect\citeauthoryear{Guo, Chi, Wang, Lau, Qiao, Ustun, Zhang, and
  Cong}{Guo et~al\mbox{.}}{2021}]%
        {Autobridge21}
\bibfield{author}{\bibinfo{person}{Licheng Guo}, \bibinfo{person}{Yuze Chi},
  \bibinfo{person}{Jie Wang}, \bibinfo{person}{Jason Lau},
  \bibinfo{person}{Weikang Qiao}, \bibinfo{person}{Ecenur Ustun},
  \bibinfo{person}{Zhiru Zhang}, {and} \bibinfo{person}{Jason Cong}.}
  \bibinfo{year}{2021}\natexlab{}.
\newblock \showarticletitle{AutoBridge: Coupling Coarse-Grained Floorplanning
  and Pipelining for High-Frequency HLS Design on Multi-Die FPGAs}. In
  \bibinfo{booktitle}{\emph{The 2021 ACM/SIGDA International Symposium on
  Field-Programmable Gate Arrays}}. \bibinfo{publisher}{Association for
  Computing Machinery}, \bibinfo{pages}{81–92}.
\newblock


\bibitem[\protect\citeauthoryear{Guo, Maidee, Zhou, Lavin, Wang, Chi, Qiao,
  Kaviani, Zhang, and Cong}{Guo et~al\mbox{.}}{2022}]%
        {guo2022rapidstream}
\bibfield{author}{\bibinfo{person}{Licheng Guo}, \bibinfo{person}{Pongstorn
  Maidee}, \bibinfo{person}{Yun Zhou}, \bibinfo{person}{Chris Lavin},
  \bibinfo{person}{Jie Wang}, \bibinfo{person}{Yuze Chi},
  \bibinfo{person}{Weikang Qiao}, \bibinfo{person}{Alireza Kaviani},
  \bibinfo{person}{Zhiru Zhang}, {and} \bibinfo{person}{Jason Cong}.}
  \bibinfo{year}{2022}\natexlab{}.
\newblock \showarticletitle{RapidStream: Parallel Physical Implementation of
  FPGA HLS Designs}. In \bibinfo{booktitle}{\emph{Proceedings of the 2022
  ACM/SIGDA International Symposium on Field-Programmable Gate Arrays}}.
  \bibinfo{pages}{1--12}.
\newblock


\bibitem[\protect\citeauthoryear{Holewinski, Pouchet, and
  Sadayappan}{Holewinski et~al\mbox{.}}{2012}]%
        {GPUStencil12}
\bibfield{author}{\bibinfo{person}{Justin Holewinski},
  \bibinfo{person}{Louis-No\"{e}l Pouchet}, {and} \bibinfo{person}{P.
  Sadayappan}.} \bibinfo{year}{2012}\natexlab{}.
\newblock \showarticletitle{High-Performance Code Generation for Stencil
  Computations on GPU Architectures}. In \bibinfo{booktitle}{\emph{Proceedings
  of the 26th ACM International Conference on Supercomputing}}.
  \bibinfo{pages}{311–320}.
\newblock


\bibitem[\protect\citeauthoryear{Kamalakkannan, Mudalige, Reguly, and
  Fahmy}{Kamalakkannan et~al\mbox{.}}{2021}]%
        {HighLevelStencil21}
\bibfield{author}{\bibinfo{person}{Kamalavasan Kamalakkannan},
  \bibinfo{person}{Gihan~R. Mudalige}, \bibinfo{person}{István~Z. Reguly},
  {and} \bibinfo{person}{Suhaib~A. Fahmy}.} \bibinfo{year}{2021}\natexlab{}.
\newblock \showarticletitle{High-Level FPGA Accelerator Design for
  Structured-Mesh-Based Explicit Numerical Solvers}. In
  \bibinfo{booktitle}{\emph{2021 IEEE International Parallel and Distributed
  Processing Symposium (IPDPS)}}. \bibinfo{pages}{1087--1096}.
\newblock


\bibitem[\protect\citeauthoryear{Kyparissas and Dollas}{Kyparissas and
  Dollas}{2020}]%
        {CellularAutomata21}
\bibfield{author}{\bibinfo{person}{Nikolaos Kyparissas} {and}
  \bibinfo{person}{Apostolos Dollas}.} \bibinfo{year}{2020}\natexlab{}.
\newblock \showarticletitle{Large-Scale Cellular Automata on FPGAs: A New
  Generic Architecture and a Framework}.
\newblock \bibinfo{journal}{\emph{ACM Trans. Reconfigurable Technol. Syst.}}
  \bibinfo{volume}{14}, \bibinfo{number}{1}, Article \bibinfo{articleno}{5}
  (\bibinfo{date}{dec} \bibinfo{year}{2020}), \bibinfo{numpages}{32}~pages.
\newblock


\bibitem[\protect\citeauthoryear{Matsumura, Zohouri, Wahib, Endo, and
  Matsuoka}{Matsumura et~al\mbox{.}}{2020}]%
        {GPUAutoStencil20}
\bibfield{author}{\bibinfo{person}{Kazuaki Matsumura},
  \bibinfo{person}{Hamid~Reza Zohouri}, \bibinfo{person}{Mohamed Wahib},
  \bibinfo{person}{Toshio Endo}, {and} \bibinfo{person}{Satoshi Matsuoka}.}
  \bibinfo{year}{2020}\natexlab{}.
\newblock \showarticletitle{AN5D: Automated Stencil Framework for High-Degree
  Temporal Blocking on GPUs}. \bibinfo{pages}{199–211}.
\newblock


\bibitem[\protect\citeauthoryear{Natale, Stramondo, Bressana, Cattaneo, Sciuto,
  and Santambrogio}{Natale et~al\mbox{.}}{2016}]%
        {TemporalICCAD17}
\bibfield{author}{\bibinfo{person}{Giuseppe Natale}, \bibinfo{person}{Giulio
  Stramondo}, \bibinfo{person}{Pietro Bressana}, \bibinfo{person}{Riccardo
  Cattaneo}, \bibinfo{person}{Donatella Sciuto}, {and}
  \bibinfo{person}{Marco~D. Santambrogio}.} \bibinfo{year}{2016}\natexlab{}.
\newblock \showarticletitle{A polyhedral model-based framework for dataflow
  implementation on FPGA devices of Iterative Stencil Loops}. In
  \bibinfo{booktitle}{\emph{2016 IEEE/ACM International Conference on
  Computer-Aided Design (ICCAD)}}. \bibinfo{pages}{1--8}.
\newblock


\bibitem[\protect\citeauthoryear{Nguyen, Satish, Chhugani, Kim, and
  Dubey}{Nguyen et~al\mbox{.}}{2010}]%
        {CPUGPUStencil10}
\bibfield{author}{\bibinfo{person}{Anthony Nguyen}, \bibinfo{person}{Nadathur
  Satish}, \bibinfo{person}{Jatin Chhugani}, \bibinfo{person}{Changkyu Kim},
  {and} \bibinfo{person}{Pradeep Dubey}.} \bibinfo{year}{2010}\natexlab{}.
\newblock \showarticletitle{3.5-D Blocking Optimization for Stencil
  Computations on Modern CPUs and GPUs}. In \bibinfo{booktitle}{\emph{SC '10:
  Proceedings of the 2010 ACM/IEEE International Conference for High
  Performance Computing, Networking, Storage and Analysis}}.
  \bibinfo{pages}{1--13}.
\newblock


\bibitem[\protect\citeauthoryear{Reggiani, Del~Sozzo, Conficconi, Natale,
  Moroni, and Santambrogio}{Reggiani et~al\mbox{.}}{2021}]%
        {HDLMultiFPGA21}
\bibfield{author}{\bibinfo{person}{Enrico Reggiani}, \bibinfo{person}{Emanuele
  Del~Sozzo}, \bibinfo{person}{Davide Conficconi}, \bibinfo{person}{Giuseppe
  Natale}, \bibinfo{person}{Carlo Moroni}, {and} \bibinfo{person}{Marco~D.
  Santambrogio}.} \bibinfo{year}{2021}\natexlab{}.
\newblock \showarticletitle{Enhancing the Scalability of Multi-FPGA Stencil
  Computations via Highly Optimized HDL Components}.
\newblock \bibinfo{journal}{\emph{ACM Trans. Reconfigurable Technol. Syst.}}
  \bibinfo{volume}{14}, \bibinfo{number}{3}, Article \bibinfo{articleno}{15}
  (\bibinfo{date}{aug} \bibinfo{year}{2021}), \bibinfo{numpages}{33}~pages.
\newblock


\bibitem[\protect\citeauthoryear{Singh, Diamantopoulos, Hagleitner, Gomez-Luna,
  Stuijk, Mutlu, and Corporaal}{Singh et~al\mbox{.}}{2020}]%
        {NERO20}
\bibfield{author}{\bibinfo{person}{Gagandeep Singh}, \bibinfo{person}{Dionysios
  Diamantopoulos}, \bibinfo{person}{Christoph Hagleitner},
  \bibinfo{person}{Juan Gomez-Luna}, \bibinfo{person}{Sander Stuijk},
  \bibinfo{person}{Onur Mutlu}, {and} \bibinfo{person}{Henk Corporaal}.}
  \bibinfo{year}{2020}\natexlab{}.
\newblock \showarticletitle{NERO: A Near High-Bandwidth Memory Stencil
  Accelerator for Weather Prediction Modeling}. In
  \bibinfo{booktitle}{\emph{2020 30th International Conference on
  Field-Programmable Logic and Applications (FPL)}}. \bibinfo{pages}{9--17}.
\newblock


\bibitem[\protect\citeauthoryear{Waidyasooriya and Hariyama}{Waidyasooriya and
  Hariyama}{2019}]%
        {MultiFPGAStencil19}
\bibfield{author}{\bibinfo{person}{Hasitha~Muthumala Waidyasooriya} {and}
  \bibinfo{person}{Masanori Hariyama}.} \bibinfo{year}{2019}\natexlab{}.
\newblock \showarticletitle{Multi-FPGA Accelerator Architecture for Stencil
  Computation Exploiting Spacial and Temporal Scalability}.
\newblock \bibinfo{journal}{\emph{IEEE Access}}  \bibinfo{volume}{7}
  (\bibinfo{year}{2019}), \bibinfo{pages}{53188--53201}.
\newblock


\bibitem[\protect\citeauthoryear{Wang and Chandramowlishwaran}{Wang and
  Chandramowlishwaran}{2020}]%
        {CPUStencil20}
\bibfield{author}{\bibinfo{person}{Hengjie Wang} {and} \bibinfo{person}{Aparna
  Chandramowlishwaran}.} \bibinfo{year}{2020}\natexlab{}.
\newblock \showarticletitle{Pencil: A Pipelined Algorithm for Distributed
  Stencils}. In \bibinfo{booktitle}{\emph{SC20: International Conference for
  High Performance Computing, Networking, Storage and Analysis}}.
  \bibinfo{pages}{1--16}.
\newblock


\bibitem[\protect\citeauthoryear{Wang and Liang}{Wang and Liang}{2017}]%
        {ShuoWangDAC17}
\bibfield{author}{\bibinfo{person}{Shuo Wang} {and} \bibinfo{person}{Yun
  Liang}.} \bibinfo{year}{2017}\natexlab{}.
\newblock \showarticletitle{A comprehensive framework for synthesizing stencil
  algorithms on FPGAs using OpenCL model}. In \bibinfo{booktitle}{\emph{2017
  54th ACM/EDAC/IEEE Design Automation Conference (DAC)}}.
  \bibinfo{pages}{1--6}.
\newblock


\bibitem[\protect\citeauthoryear{Wolfram}{Wolfram}{2018}]%
        {CellularAutomata18}
\bibfield{author}{\bibinfo{person}{Stephen Wolfram}.}
  \bibinfo{year}{2018}\natexlab{}.
\newblock \bibinfo{booktitle}{\emph{Computation Theory of Cellular Automata}}.
\newblock \bibinfo{pages}{159--202}.
\newblock
\showISBNx{9780429494093}


\bibitem[\protect\citeauthoryear{Xilinx}{Xilinx}{2020a}]%
        {alveoU280}
\bibfield{author}{\bibinfo{person}{Xilinx}.} \bibinfo{year}{2020}\natexlab{a}.
\newblock \bibinfo{title}{Alveo U280 Data Center Accelerator Cards Data Sheet}.
\newblock
\newblock
\urldef\tempurl%
\url{https://www.xilinx.com/support/documentation/data\_sheets/ds963-u280.pdf}
\showURL{%
\tempurl}
\newblock
\shownote{Last accessed July 28, 2020.}


\bibitem[\protect\citeauthoryear{Xilinx}{Xilinx}{2020b}]%
        {vitis}
\bibfield{author}{\bibinfo{person}{Xilinx}.} \bibinfo{year}{2020}\natexlab{b}.
\newblock \bibinfo{title}{Vitis Unified Software Platform}.
\newblock
\newblock
\urldef\tempurl%
\url{https://www.xilinx.com/products/design-tools/vitis/vitis-platform.html\#development}
\showURL{%
\tempurl}
\newblock
\shownote{Last accessed Nov 26, 2021.}


\bibitem[\protect\citeauthoryear{Zohouri, Podobas, and Matsuoka}{Zohouri
  et~al\mbox{.}}{2018a}]%
        {SpatialTemporyBlocking18}
\bibfield{author}{\bibinfo{person}{Hamid~Reza Zohouri}, \bibinfo{person}{Artur
  Podobas}, {and} \bibinfo{person}{Satoshi Matsuoka}.}
  \bibinfo{year}{2018}\natexlab{a}.
\newblock \showarticletitle{Combined Spatial and Temporal Blocking for
  High-Performance Stencil Computation on FPGAs Using OpenCL}. In
  \bibinfo{booktitle}{\emph{Proceedings of the 2018 ACM/SIGDA International
  Symposium on Field-Programmable Gate Arrays}}. \bibinfo{pages}{153–162}.
\newblock


\bibitem[\protect\citeauthoryear{Zohouri, Podobas, and Matsuoka}{Zohouri
  et~al\mbox{.}}{2018b}]%
        {HighOrderFPGAStencil18}
\bibfield{author}{\bibinfo{person}{Hamid~Reza Zohouri}, \bibinfo{person}{Artur
  Podobas}, {and} \bibinfo{person}{Satoshi Matsuoka}.}
  \bibinfo{year}{2018}\natexlab{b}.
\newblock \showarticletitle{High-Performance High-Order Stencil Computation on
  FPGAs Using OpenCL}. In \bibinfo{booktitle}{\emph{2018 IEEE International
  Parallel and Distributed Processing Symposium Workshops (IPDPSW)}}.
  \bibinfo{pages}{123--130}.
\newblock


\end{thebibliography}
\end{document}